\documentclass[%
 reprint,
 amsmath,amssymb,
 aps,
]{revtex4-1}

\usepackage{graphicx}% Include figure files
\usepackage{dcolumn}% Align table columns on decimal point
\usepackage{bm}% bold math
\begin{document}

\preprint{APS/123-QED}

\title{Braneworld non-minimal inflation with induced gravity}% Force line breaks with \\

\author{Kourosh Nozari}
 \homepage{knozari@umz.ac.ir}
\affiliation{Research Institute for Astronomy and Astrophysics of
Maragha (RIAAM),\\
P. O. Box 55134-441, Maragha, Iran}

\author{Narges Rashidi}
\homepage{n.rashidi@umz.ac.ir}%
\affiliation{Department of Physics, Faculty of Basic Sciences,\\
University of Mazandaran,\\
P. O. Box 47416-95447, Babolsar, IRAN}

\date{\today}% It is always \today, today,
             %  but any date may be explicitly specified

\begin{abstract}
We study cosmological inflation on a warped DGP braneworld where
inflaton field is non-minimally coupled to induced gravity on the
brane. We present a detailed calculation of the perturbations and
inflation parameters both in Jordan and Einstein frame. We analyze
the parameters space of the model fully to justify about the
viability of the model in confrontation with recent observational
data. We compare the results obtained in these two frames also in
order to judge which frame gives more acceptable results in
comparison with observational data.
\begin{description}
\item[PACS numbers]
98.80.Cq,\, 98.80.-k,\, 04.50.-h
\item[Key Words]
Braneworld Inflation, Induced Gravity, Scalar-Tensor Gravity
\end{description}
\end{abstract}

\maketitle

%\tableofcontents

\section{Introduction}

Although the standard big bang cosmology has great successes in
confrontation with observation, it suffers from some shortcomings
such as the flatness, horizon and relics problems. It has been shown
that an accelerating stage during the early time evolution of the
universe with $\ddot{a}>0$ ($p<-\rho /3$) has the capability to
solve these problems. This is the early time inflationary stage. The
inflation also provides a mechanism for production of density
perturbations needed to seed the formation of structures in the
universe. It has been shown that a simple scalar field (usually
dubbed \textit{inflaton}) whose energy dominates the universe and
whose potential energy dominates over the kinetic term (the
\textit{slow-roll} conditions) gives the required inflation
\cite{Gut81,Lin82,Alb82,Lin90,Lid00a,Lid97,Rio02,Lyt09}. Despite the
great successes of the inflation paradigm, there are several
problems with no concrete solutions: natural realization of
inflation in a fundamental theory, cosmological constant and dark
energy problem, unexpected low power spectrum at large scales and
egregious running of the spectral index are some of these problems
\cite{Bra05}. Another unsolved problem in the spirit of the
inflationary scenario is that we don't know how to integrate it with
ideas of the particle physics. For example, we would like to
identify the inflaton, the scalar field that drives inflation, with
one of the known fields of particle physics. Also, it is important
that the inflaton potential emerges naturally from underlying
fundamental theory \cite{Lid97}.

Braneworld scenarios open new windows to address at least part of
these difficulties \cite{Lid04,Buc04}. One of the various braneworld
scenarios, is the model proposed by Dvali, Gabadadze and Porrati
(DGP). This setup is based on a modification of the gravitational
theory in an induced gravity perspective
\cite{Dva00,Dva01a,Dva01b,Lue06}. This induced gravity term in the
brane part of the action, leads to deviations from the standard
4-dimensional gravity over large distances. In the DGP model, the
bulk is a flat Minkowski spacetime, but a reduced gravity term
appears on the brane without tension. Some aspects of the braneworld
inflation in the pure DGP setup are studied in \cite{Laz04,Cor08}.
Maeda, Mizuno and Torii have constructed a braneworld scenario which
combines the Randall-Sundrum II (RS II) \cite{Ran99} and DGP models
\cite{Mae03}. In this combination, an induced curvature term appears
on the brane in the RS II model. This model has been called the
\textit{warped} DGP braneworld in literatures
\cite{Cai04,Zha06,Noz07a,Noz11}. Some aspects of the inflation on
the warped DGP setup are studied in Refs.
\cite{Cai04,Zha06,Noz07a,Noz11}.

We note that in a braneworld setup, the induced gravity on the brane
arises as a result of quantum corrections. For instance, in the
Randall-Sundrum II braneworld scenario quantum corrections arise due
to induced coupling between brane matter and the bulk gravitons. The
induced gravity leads to the appearance of terms proportional to the
4-dimensional Ricci scalar in the brane part of the action. While
the RS model gives high-energy modifications to general relativity,
the DGP braneworld produces a low energy modification that leads to
late-time acceleration of brane universe even in the absence of dark
energy. The RS II braneworld scenario modifies certainly the high
energy, ultra-violet (UV) sector of the general relativity. Also the
DGP gravity is essentially a low-energy, infra-red (IR) modification
of the general relativity. Since the warped DGP scenario contains
both UV and IR modifications simultaneously, inflation in a warped
DGP setup is physically more reasonable than the pure RS II or DGP
case. An important issue we are interested in this paper, is that
whether high-energy inflation is subjected to the \emph{induced
gravity} effect. If the induced gravity correction takes the
dominant role, then there is no RS-type high-energy regime in the
early universe and we recover the DGP model. From another
perspective, as the energy scale of inflation grows, the induced
gravity correction acts to limit the growth of amplitude
\emph{relative to the 4D case} \cite{Lan00,Lan07,Lop04,Kal05}.
Although induced gravity is an IR modification of General Relativity
and it seems that these modifications have nothing to do with
inflation, however the mentioned points are important enough to be
the reason for study of the warped DGP-braneworld inflation. We note
also that as has been shown in \cite{Laz04}, brane assisted
inflation may be equally successful beyond general relativity. It
has been proved that this is the case in the RS and DGP models
provided certain conditions hold. Since we considered the normal
branch of solutions, as has been shown in \cite{Laz04} the
conditions for the occurrence of inflation are less restrictive.

On the other hand, considering a braneworld setup has the advantage
that bulk fields such as Radions (for stability purposes) can have
projection(s) on the brane that is a suitable candidate for inflaton
field on the brane. The projection of the bulk inflaton on the brane
behaves just like an ordinary inflaton field in four dimensions in
the low energy regime. While the origin of inflaton field in
standard 4D case is not so trivial, in a braneworld picture we can
imagine this field as a projection of bulk field(s). This may help
to reduce at least part of lacuna of standard scenario. We note also
that as has been shown in \cite{Buc04}, inflation in warped de
Sitter string theory geometries bypasses the difficulties of
computing corrections to $\eta$ slow-roll parameter relative to the
effective four dimensional perspective.

Since inflaton can interact with other fields such as the
gravitational sector of the theory, in the spirit of scalar-tensor
theories, we can consider a non-minimal coupling (NMC) of the
inflaton field with intrinsic (Ricci) curvature on the brane.
Braneworld model with scalar field minimally or non-minimally
coupled to gravity have been studied extensively (see \cite{Noz07b}
and references therein). We note that generally the introduction of
the NMC is not just a matter of taste. The NMC is instead forced
upon us in many situations of physical and cosmological interest.
There are compelling reasons to include an explicit non-minimal
coupling in the action. For instance, non-minimal coupling arises at
the quantum level when quantum corrections to the scalar field
theory are considered. Even if for the classical, unperturbed theory
this non-minimal coupling vanishes, it is necessary for the
renormalizability of the scalar field theory in curved space. In
most theories used to describe inflationary scenarios, it turns out
that a non-vanishing value of the coupling constant cannot be
avoided
\cite{Far96,Far00,Far99,Spo84,Fut89,Sal89,Fak90,Sch05,Mak91,Fak92,Lib98,Hwa99,Hwa98,Tsu99a,Tsu99b,Tsu00a,Chi00,Tsu00b,
Gun01,Koh05,Mar06,Boj06,Bau08,Eas09,Her10,Pal10,Pal11}.
Nevertheless, incorporation of an explicit non-minimal coupling has
disadvantage that it is harder to realize inflation even with
potentials that are known to be inflationary in the minimal theory
\cite{Far96,Far00,Far99}. Using the conformal equivalence between
gravity theories with minimally and non-minimally coupled scalar
fields, for any inflationary model based on a minimally-coupled
scalar field, it is possible to construct infinitely many
conformally related models with a non-minimal coupling
\cite{Spo84,Fut89,Sal89,Fak90,Sch05,Mak91,Fak92,Lib98,Hwa99,Hwa98,Tsu99a,Tsu99b,Tsu00a,Chi00,Tsu00b,
Gun01,Koh05,Mar06,Boj06,Bau08,Eas09,Her10,Pal10,Pal11,Kai95,Chi08}.
However, an important question then arises: are these conformally
related frames really equivalent from physics viewpoint? This issue
has been considered by several authors
\cite{Cap97,Nan98,Fla04,Bha07,Far07,Noz09,Cap10a,Cap10b,Qui11a,Qui11b}
and as a part of our primary goal, we are going to address this
issue from a detailed comparison of the inflationary parameters in
these two (Einstein and Jordan) frames.

Based on the mentioned preliminaries, in this paper we study
cosmological inflation on a warped DGP braneworld where inflaton
field is non-minimally coupled to induced gravity on the brane. We
present a detailed calculation of the perturbations and inflation
parameters both in Jordan and Einstein frame by adopting quadratic
and quartic potentials. We analyze the parameter spaces of the
models with details to have a comparison between two frames and also
in order to constraint these models in confrontation with recent
observational data.

\section{Braneworld inflation with induced gravity in Jordan frame}

The action of a warped DGP model in which a single scalar field is
non-minimally coupled to induced gravity on the brane can be written
in the following form

\begin{widetext}
\begin{equation}
S=\frac{1}{2\kappa_{5}^{2}}\int d^{5}x\sqrt{-g^{(5)}}\bigg[R^{(5)}-
2\Lambda_{5}\bigg]+\int_{brane} d^{4} x
\sqrt{-q}\bigg[\frac{1}{2\kappa_{4}^{2}}R+\frac{f(\varphi)}{2}\,R-\lambda-
\frac{1}{2}q^{\mu\nu}\partial_{\mu}\varphi\partial_{\nu}\varphi-V(\varphi)\bigg]
 \label{eq:wideeq}
\end{equation}
\end{widetext}

where $\kappa_{5}^{2}$\, is the five dimensional gravitational
constant, \,$R$\, is the induced Ricci scalar on the brane,
\,$R^{(5)}$ is 5-dimensional Ricci scalar, \,$\lambda$\, is the
brane tension and $\Lambda_{5}$ is the bulk cosmological constant.
Also $q$ is the trace of the brane metric, $q_{\mu\nu}$. We remind
that the mentioned action results in pure DGP model
\cite{Dva00,Dva01a,Dva01b} if $\lambda=0$ and $\Lambda_{5}=0$, and
pure RSII model \cite{Ran99} if $\mu=0$ where $\mu$ is a mass scale
which may correspond to the 4D Planck mass \cite{Mae03}. Also
$f(\varphi)$ shows an explicit non-minimal coupling of the scalar
field with induced gravity on the brane. We note that the fields and
their interactions on the brane at the classical level will be
determined by the bulk physics through boundary conditions on the
brane. For instance, if $\Phi$ is assumed to be a bulk scalar field,
as has been shown in \cite{Him01,Him03,Yok01,Sas04,Noz12}, the
effective field on the brane will be $\varphi=\sqrt{r_{c}}\Phi$ and
$V(\varphi)=\frac{r_{c}}{2}V(\frac{\Phi}{\sqrt{r_{c}}})$ through
junction conditions on the brane. Also as we will show (see Eq. (6)
below), $\Lambda_{5}=-\frac{\kappa_{5}^{4}}{6}\lambda^{2}\,$. So,
these parameters cannot be freely adjusted and are influenced by
bulk physics.

The generalized cosmological dynamics in this setup is given by the
following Friedmann equation
\begin{eqnarray}
H^{2}=\frac{\kappa_{4}^{2}}{3}\rho_{\varphi}+\frac{\kappa_{4}^{2}}{3}\lambda
+\frac{2\kappa_{4}^{4}}{\kappa_{5}^{4}}\hspace{4.3cm}\nonumber\\
\pm\frac{2\kappa_{4}^{2}}{\kappa_{5}^{2}}\,\sqrt{\frac{\kappa_{4}^{4}}{\kappa_{5}^{4}}+
\frac{\kappa_{4}^{2}}{3}\rho_{\varphi}+
\frac{\kappa_{4}^{2}}{3}\lambda-\frac{\Lambda_{5}}{6}-\frac{\mathcal{C}}{a^{4}}}.\hspace{0.5cm}
 \label{eq:wideeq}
\end{eqnarray}
where $\rho_{\varphi}$\,, the energy-density corresponding to the
non-minimally coupled scalar field is defined as follows
\begin{equation}
\rho_{\varphi}=\frac{1}{2}\dot{\varphi}^{2}+V(\varphi)-6f'(\varphi)H\dot{\varphi},
\end{equation}
and the corresponding pressure is given by
\begin{equation}
p_{\varphi}=\frac{1}{2}\dot{\varphi}^{2}-V(\varphi)+2f'(\varphi)\ddot{\varphi}+4f'(\varphi)H\dot{\varphi}
+2f''(\varphi)\dot{\varphi}^{2}.
\end{equation}
We note that in this paper a prime represents the derivative with
respect to the scalar field and a dot marks derivative with respect
to the cosmic time. Now let's to introduce the effective
cosmological constant on the brane as
\begin{equation}
\Lambda_{eff}=\kappa_{4}^{2}\lambda+\frac{6\kappa_{4}^{4}}{\kappa_{5}^{4}}
\pm\frac{\sqrt{6}\kappa_{4}^{4}}{\kappa_{5}^{4}}
\sqrt{\Big(2\kappa_{4}^{2}\lambda-\Lambda_{5}\Big)\frac{\kappa_{5}^{4}}{\kappa_{4}^{4}}+6}\,.
\end{equation}
Since we are interested in the inflationary dynamics driven by a
scalar field with a self-interacting potential, we put the effective
cosmological constant equal to zero. In this way, we find
\begin{equation}
\Lambda_{5}=-\frac{\kappa_{5}^{4}}{6}\lambda^{2}\,.
\end{equation}
So, we can rewrite the Friedmann equation (2) as follows
\begin{eqnarray}
H^{2}=\frac{\kappa_{4}^{2}}{3}\rho_{\varphi}+\frac{\kappa_{4}^{2}}{3}\lambda
+\frac{2\kappa_{4}^{4}}{\kappa_{5}^{4}}\hspace{4.3cm}\nonumber\\
\pm\frac{2\kappa_{4}^{2}}{\kappa_{5}^{2}}\,\sqrt{\frac{\kappa_{4}^{4}}{\kappa_{5}^{4}}+
\frac{\kappa_{4}^{2}}{3}\rho_{\varphi}+
\frac{\kappa_{4}^{2}}{3}\lambda-\frac{\kappa_{5}^{4}}{36}\lambda^{2}-\frac{\mathcal{C}}{a^{4}}}\,.\hspace{0.5cm}
\end{eqnarray}
Also, the second Friedmann equation is
\begin{equation}
\dot{H}=\frac{\kappa_{4}^{2}}{6H} \dot{\rho}_{\varphi}
\pm\frac{\kappa_{4}^{2}}{\kappa_{5}^{2}}\frac{\frac{\kappa_{4}^{2}}{6H}\dot{\rho}_{\varphi}
+\frac{2\mathcal{C}}{a^{4}}}{\sqrt{\frac{\kappa_{4}^{4}}{\kappa_{5}^{4}}+
\frac{\kappa_{4}^{2}}{3}\rho_{\varphi}+
\frac{\kappa_{4}^{2}}{3}\lambda-\frac{\kappa_{5}^{4}}{36}\lambda^{2}-\frac{\mathcal{C}}{a^{4}}}}\,\,.
\end{equation}
Variation of the action (1) with respect to the scalar field gives
the following equation of motion
\begin{equation}
\ddot{\varphi}+3H\dot{\varphi}-\frac{1}{2}f'(\varphi)R+\frac{dV}{d\varphi}=0\,.
\end{equation}
In the slow-roll approximation, where $\dot{\varphi}^{2}\ll
V(\varphi)$ and $\ddot{\varphi}\ll|3H\dot{\varphi}|$, energy density
and equation of motion for scalar field take the following forms
respectively
\begin{equation}
\rho_{\varphi}\simeq V(\varphi)-6f'(\varphi)H\dot{\varphi}\,,
\end{equation}
\begin{equation}
3H\dot{\varphi}-\frac{1}{2}f'(\varphi)R+\frac{dV}{d\varphi}\simeq
0\,.
\end{equation}
Also, the Friedmann equation now takes the following form
\begin{widetext}
\begin{equation}
H^{2}\simeq\frac{\kappa_{4}^{2}}{3}V-\frac{\kappa_{4}^{2}}{3}f'^{2}R+\frac{2\kappa_{4}^{2}}{3}f'V'
+\frac{\kappa_{4}^{2}}{3}\lambda
+\frac{2\kappa_{4}^{4}}{\kappa_{5}^{4}}
\pm\frac{2\kappa_{4}^{2}}{\kappa_{5}^{2}}\,\sqrt{\frac{\kappa_{4}^{4}}{\kappa_{5}^{4}}+
\frac{\kappa_{4}^{2}}{3}V(\varphi)-\frac{\kappa_{4}^{2}}{3}f'^{2}R+\frac{2\kappa_{4}^{2}}{3}f'V'+
\frac{\kappa_{4}^{2}}{3}\lambda-\frac{\kappa_{5}^{4}}{36}\lambda^{2}-\frac{\mathcal{C}}{a^{4}}}\,.
\end{equation}
\end{widetext}
Now, we define the slow-roll parameters as follows
\begin{equation}
\epsilon\equiv -\frac{\dot{H}}{H^{2}}\,\,,
\end{equation}
\begin{equation}
\eta\equiv -\frac{1}{H}\frac{\ddot{H}}{\dot{H}}\,\,.
\end{equation}
In the slow-roll approximation and by using equation (12) we find
\begin{equation}
\epsilon \simeq \frac{1}{2\kappa_{4}^{2}}\frac{V'^2}{V^2}\times
{\cal{A}}(\varphi)\,,
\end{equation}
and
\begin{equation}
\eta \simeq \frac{1}{\kappa_{4}^{2}}\frac{V''}{V}\times
{\cal{B}}(\varphi)\,,
\end{equation}
where by definition
\begin{widetext}
\begin{eqnarray}
{\cal{A}}(\varphi)=\Bigg(\frac{1}{V'}
-\frac{f'R}{2V'^{2}}\Bigg)\Bigg(V'-2f'f''R+2f''V'+2f'V''\Bigg)\hspace{7.5cm}\nonumber\\
\times\frac{1\pm\frac{\kappa_{4}^{2}}{\kappa_{5}^{2}}
\frac{1-\frac{\mathcal{C}}{a^{4}}\frac{36H^{2}}{\kappa_{4}^{2}\Big(V'
-\frac{f'R}{2}\Big)\Big(V'-2f'f''R+2f''V'+2f'V''\Big)}}{\sqrt{\frac{\kappa_{4}^{4}}{\kappa_{5}^{4}}+
\frac{\kappa_{4}^{2}}{3}V-\frac{\kappa_{4}^{2}}{3}f'^{2}R+\frac{2\kappa_{4}^{2}}{3}f'V'+
\frac{\kappa_{4}^{2}}{3}\lambda-\frac{\kappa_{5}^{4}}{36}\lambda^{2}
-\frac{\mathcal{C}}{\hat{a}^{4}}}}}{\Bigg[1
+\frac{\lambda}{V}-\frac{f'^{2}R}{V}+\frac{2f'V'}{V}+\frac{6\kappa_{4}^{2}}{\kappa_{5}^{4}V}
\pm\frac{6}{\kappa_{5}^{2}V}\,\sqrt{\frac{\kappa_{4}^{4}}{\kappa_{5}^{4}}+
\frac{\kappa_{4}^{2}}{3}V-\frac{\kappa_{4}^{2}}{3}f'^{2}R+\frac{2\kappa_{4}^{2}}{3}f'V'+
\frac{\kappa_{4}^{2}}{3}\lambda-\frac{\kappa_{5}^{4}}{36}\lambda^{2}
-\frac{\mathcal{C}}{a^{4}}}\,\,\Bigg]^2},
\end{eqnarray}
\end{widetext}
and
\begin{widetext}
\begin{eqnarray}
{\cal{B}}(\varphi)=\Bigg(1
-\frac{f''R}{2V''}\Bigg)\hspace{12.5cm}\nonumber\\
\times\Bigg\{\frac{1}{1
+\frac{\lambda}{V}-\frac{f'^{2}R}{V}+\frac{2f'V'}{V}+\frac{6\kappa_{4}^{2}}{\kappa_{5}^{4}V}
\pm\frac{6}{\kappa_{5}^{2}V}\,\sqrt{\frac{\kappa_{4}^{4}}{\kappa_{5}^{4}}+
\frac{\kappa_{4}^{2}}{3}V-\frac{\kappa_{4}^{2}}{3}f'^{2}R+\frac{2\kappa_{4}^{2}}{3}f'V'+
\frac{\kappa_{4}^{2}}{3}\lambda-\frac{\kappa_{5}^{4}}{36}\lambda^{2}
-\frac{\mathcal{C}}{a^{4}}}}\Bigg\}\,.
\end{eqnarray}
\end{widetext}
As we will show, these parameters which reflect the braneworld and
non-minimal nature of our model, in the large field regime intensify
the increment of the slow-roll parameters. Inflation can be attained
only if $\{\epsilon,\eta\}<1$; once one of these parameters reaches
unity, the inflation phase terminates. We note that
${\cal{A}}(\varphi)$ and ${\cal{B}}(\varphi)$ are contributions
originating from braneworld nature of the setup and also the
non-minimal coupling of the scalar field and induced gravity on the
brane.

The number of e-folds during inflation is given by
\begin{equation}
N=\int_{t_{i}}^{t_{f}} H dt\,,
\end{equation}
which in the slow-roll approximation can be written as
\begin{equation}
N\simeq\int_{\varphi_{i}}^{\varphi_{f}}
3H^{2}\frac{1}{\frac{1}{2}f'R-V'} d\varphi\,,
\end{equation}
where $\varphi_{i}$ denotes the value of $\varphi$ when the universe
scale observed today crosses the Hubble horizon during inflation and
$\varphi_{f}$ is the value of $\varphi$ when the universe exits the
inflationary phase. For a warped DGP model with non-minimally
coupled scalar field on the brane, this quantity in Jordan frame
becomes
\begin{widetext}
\begin{eqnarray}
N=\int_{\varphi_{hc}}^{\varphi_{f}}
\Bigg(\frac{3V}{V'}\Bigg)\Bigg(\frac{V'}{\frac{1}{2}f'R-V'}\Bigg)\Bigg[\frac{\kappa_{4}^{2}}{3}
+\frac{1}{V}\bigg(\frac{\kappa_{4}^{2}}{3}\lambda-\frac{\kappa_{4}^{2}f'^{2}R}{3}+\frac{2\kappa_{4}^{2}f'V'}{3}
+\frac{2\kappa_{4}^{4}}{\kappa_{5}^{4}}\hspace{5cm}\nonumber\\
\pm\frac{2\kappa_{4}^{2}}{\kappa_{5}^{2}}\,\sqrt{\frac{\kappa_{4}^{4}}{\kappa_{5}^{4}}+
\frac{\kappa_{4}^{2}}{3}V(\varphi)-\frac{\kappa_{4}^{2}}{3}f'^{2}R+\frac{2\kappa_{4}^{2}}{3}f'V'+
\frac{\kappa_{4}^{2}}{3}\lambda-\frac{\kappa_{5}^{4}}{36}\lambda^{2}
-\frac{\mathcal{C}}{a^{4}}}\,\bigg)\Bigg]d\varphi\,\,.
\end{eqnarray}
\end{widetext}

After presentation of the main equations of the setup in Jordan
frame, in the next section we consider the scalar perturbation of
the metric since the key test of any inflation model is the spectrum
of perturbations produced due to quantum fluctuations of the fields
about their homogeneous background values.

\section{Perturbations in Jordan frame}

In a warped DGP braneworld model, the effective covariant equations
on the brane for an arbitrary brane metric and matter distribution
is given by \cite{Maa10}
\begin{equation}
G_{\mu\nu}=\kappa_{5}^{4}\Pi_{\mu\nu}-E_{\mu\nu}\,,
\end{equation}
where
\begin{equation}
\Pi_{\mu\nu}=-\frac{1}{4}{\tau}_{\mu\sigma}{\tau}_{\nu}^{\sigma}+
\frac{1}{12}{\tau}{\tau}_{\mu\nu}+
\frac{1}{8}q_{\mu\nu}\Big(\tau_{\rho\sigma}{\tau}^{\rho\sigma}
-\frac{1}{3}{\tau}^{2}\Big)\,.
\end{equation}
$\tau_{\mu\nu}$ is the total stress-tensor on the brane and is
defined as
\begin{equation}
\tau_{\,\,\nu}^{\mu}=-\kappa_{4}^{2}G_{\,\,\nu}^{\mu}
-\lambda\delta_{\,\,\nu}^{\mu}+T_{\,\,\nu}^{\mu}\,,
\end{equation}
where $T_{\mu\nu}$, the energy-momentum tensor of a scalar field
non-minimally coupled to induced gravity on the brane is given by
\begin{eqnarray}
T_{\mu\nu}=g_{\mu\nu}\Big(\frac{1}{2}fR-\frac{1}{2}g^{\alpha\beta}\partial_{\alpha}\varphi\,
\partial_{\beta}\varphi-V(\varphi)\Big)+\partial_{\mu}\varphi\,\partial_{\nu}\varphi\nonumber\\
-fR_{\mu\nu}-
\Big(g_{\mu\nu}\Box-\nabla_{\mu}\nabla_{\nu}\Big)f\,.\hspace{2.5cm}
\end{eqnarray}
Also we have
\begin{equation}
E_{\mu\nu}=C_{MRS}^{\quad\quad N}\,\, n^{M}\,\,n_{N}\,\,
{q^{R}}_{\mu}\,\,{q^{S}}_{\nu},
\end{equation}
where $C_{MRS}^{\quad\quad N}$ is the five dimensional Weyl tensor
and $n_{A}$ is the spacelike unit vector normal to the brane.

Depending on the choice of gauge (coordinates), there are many
different ways of characterizing cosmological perturbations. In
longitudinal gauge, the scalar metric perturbations of the FRW
background are given by \cite{Bar80,Muk92,Ber95}
\begin{equation}
ds^{2}=-\big(1+2\Phi\big)dt^{2}+a^{2}(t)\big(1-2\Psi\big)\delta_{i\,j}\,dx^{i}dx^{j},
\end{equation}
where $a(t)$ is the scale factor on the brane, $\Phi=\Phi(t,x)$ and
$\Psi=\Psi(t,x)$ are the metric perturbations. For the above
perturbed metric, one can obtain the perturbed field equations as
follows

\begin{equation}
-3H(H\Phi+\dot{\Psi})-\frac{k^{2}}{a^{2}}=\frac{\kappa_{4}^{2}}{2}\delta
\rho_{eff}\,,
\end{equation}
\begin{equation}
\ddot{\Psi}+3H(H\Phi+\dot{\Psi})+H\dot{\Phi}+2\dot{H}\Phi+\frac{1}{3a^{2}}k^{2}(\Phi-\Psi)=
\frac{\kappa_{4}^{2}}{2}\delta p_{eff}\,,
\end{equation}
\begin{eqnarray}
\dot{\Psi}+H\Phi=\frac{\kappa_{4}^{2}}{2}\Big[
\frac{\kappa_{5}^{2}}{6\kappa_{4}^{2}}\rho_{\varphi}\dot{\varphi}\delta\varphi
-\frac{\kappa_{5}^{2}}{6\kappa_{4}^{2}}\rho_{\varphi}\int(\delta
T_{i}^{0})\,d x^{i}\Big]\nonumber\\
+\frac{1}{2}\int(\delta E_{i}^{0})\,d x^{i}\,,\hspace{2.5cm}.
\end{eqnarray}
\begin{equation}
\Psi-\Phi=8\pi
G\frac{\kappa_{4}^{2}H}{\kappa_{5}^{2}(\dot{H}+2H^{2})-H}
a^{2}\delta \pi_{E}\,.
\end{equation}
The anisotropic stress perturbation is defined as $\delta
\pi_{ij}=[\partial_{i}\partial_{j}+(k^{2}/3)\delta_{ij}]\delta\pi$,
\, where $\pi$ is the trace of $\pi_{ij}$\,. So, $\delta{\pi_{E}}$
is the anisotropic stress perturbation. In the Eqs. (28) and (29),
$\rho_{eff}$ and $p_{eff}$ can be obtained from the standard
Friedmann equation $H^{2}=\frac{\kappa_{4}^{2}}{3}\rho_{eff}$\, as
follows
\begin{eqnarray}
\rho_{eff}=\rho_{\varphi}+\lambda+\frac{6\kappa_{4}^{2}}{\kappa_{5}^{4}}\hspace{5cm}\nonumber\\
\pm\frac{6}{\kappa_{5}^{2}}\,\sqrt{\frac{\kappa_{4}^{4}}{\kappa_{5}^{4}}+
\frac{\kappa_{4}^{2}}{3}\rho_{\varphi}+
\frac{\kappa_{4}^{2}}{3}\lambda-\frac{\kappa_{5}^{4}}{36}\lambda^{2}-\frac{\mathcal{C}}{a^{4}}}\,\,.\hspace{1cm}
\end{eqnarray}

By using the continuity equation,
$\dot{\rho}_{eff}+3H(\rho_{eff}+p_{eff})=0$, one can deduce
\begin{eqnarray}
p_{eff}=p_{\varphi}\pm\frac{\kappa_{4}^{2}}{\kappa_{5}^{2}}\frac{\rho_{\varphi}+p_{\varphi}
-\frac{4}{\kappa_{4}^{2}}\frac{\mathcal{C}}{a^4}}{\sqrt{\frac{\kappa_{4}^{4}}{\kappa_{5}^{4}}+
\frac{\kappa_{4}^{2}}{3}\rho_{\varphi}+
\frac{\kappa_{4}^{2}}{3}\lambda-\frac{\kappa_{5}^{4}}{36}\lambda^{2}-\frac{\mathcal{C}}{a^{4}}}}\hspace{1cm}\nonumber\\
-\lambda-\frac{6\kappa_{4}^{2}}{\kappa_{5}^{4}}\mp\frac{6}{\kappa_{5}^{2}}\,
\sqrt{\frac{\kappa_{4}^{4}}{\kappa_{5}^{4}}+
\frac{\kappa_{4}^{2}}{3}\rho_{\varphi}+
\frac{\kappa_{4}^{2}}{3}\lambda-\frac{\kappa_{5}^{4}}{36}\lambda^{2}-\frac{\mathcal{C}}{a^{4}}}\,.\hspace{0.65cm}
\end{eqnarray}
So, the perturbed effective density and pressure can be written as
\begin{equation}
\delta\rho_{eff}=\delta\rho_{\varphi}\pm\frac{\kappa_{4}^{2}}{\kappa_{5}^{2}}\,\frac{\delta\rho_{\varphi}
-\frac{3}{\kappa_{4}^{2}}\delta
E_{0}^{0}}{\sqrt{\frac{\kappa_{4}^{4}}{\kappa_{5}^{4}}+
\frac{\kappa_{4}^{2}}{3}\rho_{\varphi}+
\frac{\kappa_{4}^{2}}{3}\lambda-\frac{\kappa_{5}^{4}}{36}\lambda^{2}-E_{0}^{0}}}\,,
\end{equation}
where $E_{0}^{0}=\frac{\mathcal{C}}{a^{4}}$ and
\begin{eqnarray}
\delta p_{eff}=\delta p_{\varphi}
\pm\frac{\kappa_{4}^{2}}{\kappa_{5}^{2}}\,\frac{\delta p_{\varphi}
-\frac{1}{\kappa_{4}^{2}}\delta
E_{0}^{0}}{\sqrt{\frac{\kappa_{4}^{4}}{\kappa_{5}^{4}}+
\frac{\kappa_{4}^{2}}{3}\rho_{\varphi}+
\frac{\kappa_{4}^{2}}{3}\lambda-\frac{\kappa_{5}^{4}}{36}\lambda^{2}-E_{0}^{0}}}\hspace{0.8cm}\nonumber\\
-\lambda-\frac{6\kappa_{4}^{2}}{\kappa_{5}^{4}}\mp\frac{6}{\kappa_{5}^{2}}\,
\sqrt{\frac{\kappa_{4}^{4}}{\kappa_{5}^{4}}+
\frac{\kappa_{4}^{2}}{3}\rho_{\varphi}+
\frac{\kappa_{4}^{2}}{3}\lambda-\frac{\kappa_{5}^{4}}{36}\lambda^{2}-\frac{\mathcal{C}}{a^{4}}}\,.\hspace{0.68cm}
\end{eqnarray}
$\delta E_{0}^{0}$ can be calculated from the general definition of
$\delta E_{\nu}^{\mu}$ as
\begin{equation}
\delta E_{\,\nu}^{\mu}=-\kappa_{4}^{2}\left(%
\begin{array}{cc}
  -\delta\rho_{E} & a \delta q_{E}\\
  a^{-1} \delta q_{E} & \frac{1}{3}\delta\rho_{E}\delta^{i}_{\,j}
  +(\delta\pi_{\,E})^{i}_{j}
  \\
\end{array}%
\right).
\end{equation}
The (gauge-invariant) scalar perturbations of $E_{\,\nu}^{\mu}$ can
be parameterized as an effective fluid with density perturbation
$\delta\rho_{E}$, isotropic pressure perturbation
$\frac{1}{3}\,\delta\rho_{E}$, anisotropic stress perturbation
$\delta\pi_{E}$ and energy flux perturbation $\delta q_{E}$ (see
\cite{Koy06,Lan01}). Also $\delta\rho_{\varphi}$ and $\delta
p_{\varphi}$ take the following forms
\begin{equation}
\delta\rho_{\varphi}=\dot{\varphi}\delta
\dot{\varphi}-\dot{\varphi}^{2}\Phi+V'\delta\varphi+\delta\rho_{nmc},
\end{equation}
where
\begin{eqnarray}
\delta\rho_{nmc}=-2\Bigg[3H^{2}+\Big(\Box+\nabla_{0}\nabla^{0}\Big)f\Bigg]\Phi
-6H^{2}f'+f\delta R_{0}^{0}\nonumber\\
-2\Bigg[\Box f-\frac{1}{2}f R\Bigg]\Phi
-\Bigg[\Box+\nabla_{0}\nabla^{0}\Bigg]f'\delta\varphi+\Bigg[\Big(\frac{k^{2}}{a^{2}}-3\dot{H}\Big)\Phi\nonumber\\
-\frac{2k^{2}}{a^{2}}\Psi-3\Big(\ddot{\Psi}+4H\dot{\Psi}+H\dot{\Phi}+\dot{H}
+4H^{2}\Phi\Big)\Bigg]f\,,\hspace{0.9cm}
\end{eqnarray}
and
\begin{equation}
\delta p_{\varphi}=\dot{\varphi}\delta
\dot{\varphi}-\dot{\varphi}^{2}\Phi-V'\delta\varphi+\delta
p_{nmc}\,,
\end{equation}\\
where
\begin{eqnarray}
\delta p_{nmc}=-\frac{\delta^{i}_{j}}{3}\Bigg\{g^{jk}\Bigg[f R_{ki}-\big(g_{ki}\,\Box
-\nabla_{k}\nabla_{i}\big)f\hspace{1.8cm}\nonumber\\
-6g_{ki}\,f(\dot{H}+2H^{2})\Bigg]\Phi
-g^{jk}\Bigg[f'\Big(R_{ki}+6g_{ki}(\dot{H}+H^{2})\Big)\hspace{0.3cm}\nonumber\\
-g_{ki}\,f
\bigg(\Big(\frac{k^{2}}{a^{2}}-3\dot{H}\Big)\Phi-\frac{2k^{2}}{a^{2}}\Psi
-3\Big(\ddot{\Psi}+4H\dot{\Psi}+H\dot{\Phi}\hspace{0.7cm}\nonumber\\
+\dot{H}+4H^{2}\Phi\Big)\bigg)+f\,\delta R_{ki}+\Big(g_{ki}\Box -\nabla_{k}\nabla_{i}\Big)f'\delta\varphi\hspace{1cm}\nonumber\\
-\Big(\Box f-\frac{1}{2}f\Big)\delta
g_{ki}\Bigg]\Bigg\}.\hspace{1cm}
\end{eqnarray}

Equations (37) and (39) in the minimal case and within the slow-roll
conditions reduce to $\delta\rho_{\varphi}=\frac{dV}{d\varphi}\delta
\varphi$ and $\delta p_{\varphi}=- \frac{dV}{d\varphi}\delta
\varphi$\, respectively. By perturbing the equation of motion of the
scalar field (11), one obtains
\begin{eqnarray}
\delta\ddot{\varphi}+3H\delta\dot{\varphi}+\Big(V''+\frac{k^{2}}{a^{2}}-\frac{1}{2}Rf''\Big)\delta\varphi
=\dot{\varphi}\Big(3\dot{\Psi}+\dot{\Phi}\Big)\nonumber\\
+\Phi\Big(Rf'-2V'\Big)+f'\bigg[\Big(\frac{k^{2}}{a^{2}}-3\dot{H}\Big)\Phi-\frac{2k^{2}}{a^{2}}\Psi\hspace{1cm}\nonumber\\
-3\Big(\ddot{\Psi}+4H\dot{\Psi}
+H\dot{\Phi}+\dot{H}\Phi+4H^{2}\Phi\Big)\bigg]\,.\hspace{0.5cm}
\end{eqnarray}

Now the scalar perturbations can be decomposed to an entropy or
isocurvature perturbation (the projection orthogonal to the
trajectory), and adiabatic or curvature perturbations (projection
parallel to the trajectory). The isocurvature perturbations are
generated if inflation is driven by more than one scalar field
\cite{Lan00,Lan07,Bas99,Gor01} or it interacts with other fields
such as the induced gravity on the brane \cite{Lop04,Kal05}. The
adiabatic perturbations are generated if the inflaton field is the
only field in inflation period \cite{Bas99,Gor01,Lop04,Kal05,Maa00}.
Here, since the inflaton field is non-minimally coupled to the
induced gravity on the brane, the entropy perturbations are
presented in this setup \cite{Maa00,Sea10}. A gauge-invariant
primordial curvature perturbation $\zeta$, can be defined as follows
\cite{Bar83}
\begin{equation}
\zeta=\Psi-\frac{H}{\dot{\rho}}\delta\rho\,\,.
\end{equation}
This definition is valid to first order in the cosmological
perturbations on scales outside the horizon. On uniform density
hypersurfaces where $\delta\rho=0$, the above quantity reduces to
the curvature perturbation, $\Psi$. In the warped DGP model and
within the Jordan frame, we should redefine Eq. (42) as
\begin{equation}
\zeta=\Psi-\frac{H}{\dot{\rho}_{eff}}\delta\rho_{eff}\,\,.
\end{equation}
Now, by using the energy conservation equation for linear
perturbations (in an arbitrary gauge)
\begin{equation}
\dot{\delta\rho}_{eff}+3H(\delta\rho_{eff}+\delta
p_{eff})+3(\rho_{eff}+p_{eff})\dot{\Psi}=0,
\end{equation}
we can find the variation of $\zeta$ with respect to the conformal
time as
\begin{equation}
\dot{\zeta}=\dot{\Psi}+\frac{\dot{\delta\rho}_{eff}}{3(\rho_{eff}+p_{eff})}-\frac{\dot{\rho}_{eff}
+\dot{p}_{eff}}{3(\rho_{eff}+p_{eff})^{2}}\delta\rho_{eff}\,\,,
\end{equation}
where $\dot{\rho}_{eff}$ and $\dot{p}_{eff}$ are given by time
derivatives of equations (32) and (33) respectively.

One can split the pressure perturbation (in any gauge) into
adiabatic and entropic (non-adiabatic) parts (see for instance Ref.
\cite{Wan00})
\begin{equation}
\delta p_{eff}=c_{s}^{2}\delta\rho_{eff}+\dot{p}_{eff}\Gamma\,\,,
\end{equation}
where $c_{s}^{2}=\frac{\dot{p}_{eff}}{\dot{\rho}_{eff}}$ is the
sound effective velocity. The non-adiabatic part is $\delta
p_{nad}=\dot{p}_{eff}\Gamma$\,, where $\Gamma$ represents the
displacement between hypersurfaces of uniform pressure and density.
From equations (34)-(40) we can deduce
\begin{eqnarray}
\delta
p_{nad}=(1-c_{s}^{2})\delta\rho_{eff}-\Bigg(2V'\delta\varphi+\delta\rho_{nmc}-\delta
p_{nmc}\Bigg)\hspace{0.5cm}\nonumber\\
\Bigg(1\pm\frac{\kappa_{4}^{2}}{\kappa_{5}^{2}}
\frac{1}{\sqrt{\frac{\kappa_{4}^{4}}{\kappa_{5}^{4}}+
\frac{\kappa_{4}^{2}}{3}\rho_{\varphi}+
\frac{\kappa_{4}^{2}}{3}\lambda-\frac{\kappa_{5}^{4}}{36}\lambda^{2}-E_{0}^{0}}}\Bigg)\hspace{1.8cm}\nonumber\\
\mp\frac{2\kappa_{4}^{2}}{\kappa_{5}^{2}}
\frac{\frac{1}{\kappa_{4}^{2}}\delta
E_{0}^{0}}{\sqrt{\frac{\kappa_{4}^{4}}{\kappa_{5}^{4}}+
\frac{\kappa_{4}^{2}}{3}\rho_{\varphi}+
\frac{\kappa_{4}^{2}}{3}\lambda-\frac{\kappa_{5}^{4}}{36}\lambda^{2}-E_{0}^{0}}}\hspace{2.4cm}\nonumber\\
\mp \frac{\kappa_{4}^{4}}{6\kappa_{5}^{2}}\frac{\Big(\delta
\rho_{\varphi}-\frac{1}{\kappa_{4}^{2}}\delta
E_{0}^{0}\Big)\Big(\rho_{\varphi}+p_{\varphi}-\frac{4}{\kappa_{4}^{2}}E_{0}^{0}\Big)}{\Big[\frac{\kappa_{4}^{4}}{\kappa_{5}^{4}}+
\frac{\kappa_{4}^{2}}{3}\rho_{\varphi}+
\frac{\kappa_{4}^{2}}{3}\lambda-\frac{\kappa_{5}^{4}}{36}\lambda^{2}
-E_{0}^{0}\Big]^{3/2}}\,.\hspace{1cm}
\end{eqnarray}

Using the equations (28)-(30) we can rewrite this relation as
\begin{eqnarray}
\delta
p_{nad}=-\frac{6}{\kappa_{4}^{2}}\bigg(1-c_{s}^{2}-{\cal{J}}\bigg)\frac{k}{a^{2}}\Psi\hspace{3.8cm}\nonumber\\
-\frac{6}{\kappa_{4}^{2}}{\cal{K}}\bigg(H\Phi+\dot{\Psi}\bigg)
+\frac{3}{\kappa_{4}^{2}}{\cal{J}}\delta
E_{0}^{0}+\frac{2}{\kappa_{4}^{2}}\delta
E_{0}^{0}\bigg({\cal{I}}-1\bigg)\hspace{1.3cm}\nonumber\\
+\Bigg(\delta p_{nmc}-\delta\rho_{nmc}
-2\frac{V'}{\dot{\varphi}}\int(\delta
T_{i}^{0})_{nmc}\,dx^{i}\hspace{2.4cm}\nonumber\\
+\frac{6\kappa_{4}^{2}}{\kappa_{5}^{2}}\frac{V'}{\rho_{\varphi}\dot{\varphi}}\int\delta
E_{i}^{0}\,dx^{i}\Bigg){\cal{I}}\,,\hspace{1.8cm}
\end{eqnarray}\\

where ${\cal{K}}$,\, ${\cal{J}}$ and ${\cal{I}}$ are defined as\\\\
\begin{widetext}
\begin{eqnarray}
{\cal{K}}=\frac{6\kappa_{4}^{2}}{\kappa_{5}^{2}}\frac{V'}{\rho_{\varphi}\dot{\varphi}}{\cal{I}}-3H{\cal{J}}
\hspace{12cm}\nonumber\\
+\frac{-3\bigg(2V'\dot{\varphi}+\dot{\rho}_{nmc}-\dot{p}_{nmc}\bigg){\cal{I}}
+\frac{24}{\kappa_{4}^{2}}E_{0}^{0}H\bigg({\cal{I}}-1\bigg)-3{\cal{J}}{\cal{I}}\bigg(\dot{\varphi}\ddot{\varphi}+V'\dot{\varphi}
+\dot{\rho}_{nmc}+\frac{12}{\kappa_{4}^{2}}E_{0}^{0}H\bigg)}{3\dot{\varphi}^{2}-\frac{1}{H}\bigg(\frac{1}{2}f'\dot{\varphi}
+\dot{\rho}_{nmc}\bigg) {\cal{I}}
+\bigg(3\dot{\varphi}^{2}-\frac{12}{\kappa_{4}^{2}}E_{0}^{0}\bigg)\bigg({\cal{I}}-1\bigg)}\,,
\end{eqnarray}
\end{widetext}
\begin{equation}
{\cal{J}}=\frac{\kappa_{4}^{4}}{6\kappa_{5}^{4}}
\frac{\Big(\rho_{\varphi}+p_{\varphi}-\frac{4}{\kappa_{4}^{2}}E_{0}^{0}\Big)}{\bigg(\frac{\kappa_{4}^{4}}{\kappa_{5}^{4}}+
\frac{\kappa_{4}^{2}}{3}\rho_{\varphi}+
\frac{\kappa_{4}^{2}}{3}\lambda-\frac{\kappa_{5}^{4}}{36}\lambda^{2}-E_{0}^{0}\bigg)^{3/2}\,{\cal{I}}}\,,
\end{equation}
and
\begin{equation}
{\cal{I}}=\Bigg(1\pm\frac{\kappa_{4}^{2}}{\kappa_{5}^{2}}
\frac{1}{\sqrt{\frac{\kappa_{4}^{4}}{\kappa_{5}^{4}}+
\frac{\kappa_{4}^{2}}{3}\rho_{\varphi}+
\frac{\kappa_{4}^{2}}{3}\lambda-\frac{\kappa_{5}^{4}}{36}\lambda^{2}-E_{0}^{0}}}\Bigg)\,,
\end{equation}
respectively.  Now we can rewrite the equation governing on the
variation of $\zeta$ versus the time in terms of the model's
parameters. From equations (44)-(48) we find
\begin{eqnarray}
\dot{\zeta}=\frac{\kappa_{4}^{2}\rho_{eff}}{9H(\rho_{eff}+p_{eff})}\Bigg(\delta
p_{nmc}-\delta\rho_{nmc}\hspace{3.4cm}
\nonumber\\
-\frac{2V'}{\dot{\varphi}}\int(\delta
T_{i}^{0})_{nmc}\,dx^{i}
+\frac{6\kappa_{4}^{2}}{\kappa_{5}^{2}}\frac{V'}{\rho_{\varphi}\dot{\varphi}}\int\delta
E_{i}^{0}\,dx^{i}\Bigg){\cal{I}}\hspace{2cm}\nonumber\\
+\frac{\rho_{eff}\delta
E_{0}^{0}}{3H(\rho_{eff}+p_{eff})}\Bigg({\cal{J}}+\frac{2}{3}({\cal{I}}-1)\Bigg)\hspace{3.5cm}\nonumber\\
-\frac{2(H\Phi+\dot{\Psi})}{3H(\rho_{eff}+p_{eff})}\rho_{eff}{\cal{K}}\hspace{3.4cm}
\end{eqnarray}
In the minimal case and within the standard model, the entropy
perturbation vanishes for long wavelength; we have $\dot{\zeta}=0$
and the primordial spectrum of perturbation is due to adiabatic
perturbations. But, it is obvious from equation (48) that in a
DGP-inspired non-minimal setup, there is a non-vanishing
contribution of the non-adiabatic perturbations, leading to
non-vanishing $\dot{\zeta}$, which affects the primordial spectrum
of perturbation. We note that isocurvature perturbations are free to
evolve on superhorizon scales, and the amplitude at the present day
depends on the details of the entire cosmological evolution from the
time that they are formed. On the other hand, because all
super-Hubble radius perturbations evolve in the same way, the shape
of the isocurvature perturbation spectrum is preserved during this
evolution \cite{Lid00b,Cid07}.

Here we are going to obtain scalar and tensorial perturbations in
our model. We take into account the slow-roll approximation at the
large scales, $k \ll aH$, where we need to describe the
non-decreasing modes. Then by using the relation between Ricci
scalar and $H$ and $\dot{H}$, we find from equation (41)
\begin{equation}
3H\delta\dot{\varphi}+\Big(V''-\frac{1}{2}f''R\Big)\delta\varphi\simeq
\Phi\Big(2 f'R-2V'\Big)\,.
\end{equation}
We note that the reason for large scale assumption is that the
scales of cosmological interest (e.g. for large-scale CMB
anisotropies) have spent most of their time far outside the Hubble
radius and have re-entered only relatively recently in the Universe
history. In this respect, in the large scale the condition $k \ll
aH$ is an acceptable assumption. As has been shown in Refs.
\cite{Amn06,Amn07}, when this condition is satisfied, $\dot{\Phi}$,
$\dot{\Psi}$ and $\ddot{\Phi}$ can be neglected. In fact, for the
longitudinal post-Newtonian limit to be satisfied, we require that
$\Delta\Psi\gg a^{2}H^{2}\times(\Psi,\, \dot{\Psi},\, \ddot{\Psi})$,
and similarly for other gradient terms \cite{Amn06,Amn07}. For a
plane wave perturbation with wavelength $\lambda$, we see that
$H^{2}\Psi$ is much smaller than $\Delta \Psi$ when $\lambda\ll
\frac{1}{H}$. The requirement that $\dot{\Psi}$ be also negligible
implies the condition $\frac{d \log \Psi}{d\zeta}\ll
\frac{1}{(\lambda H^{2})^{2}}$ (with $\zeta=\log a$), which holds if
condition $\lambda\ll \frac{1}{H}$ is satisfied for perturbation
growth. This argument can be applied for $\ddot{\Psi}$ and the other
metric potential, $\Phi$ too. By adopting a similar reasoning, form
Eq. (30) we have
\begin{equation}
\Phi\simeq
\frac{\frac{\kappa_{5}^{2}}{6}\rho_{\varphi}+\frac{1}{\dot{\varphi}\delta\varphi}\int\delta
E_{i}^{0}\,dx^{i}}{6H\bigg(1+\frac{\kappa_{5}^{2}}{18}\rho_{\varphi}f\bigg)}\dot{\varphi}\delta\varphi\,.
\end{equation}
In writing the above equation we used the relation $\int
(T_{i}^{0})_{nmc}\,dx^{i}=2f\Big(H\Phi+\dot{\Psi}\Big)$. By using
equation (53) and (54), we can deduce

\begin{eqnarray}
3H\delta\dot{\varphi}+\bigg(V''-\frac{1}{2}f''R\bigg)\delta\varphi\hspace{3.5cm}\nonumber\\
\simeq \bigg(2f'R-2V'\bigg)\hspace{3.5cm}\nonumber\\
\times\frac{\frac{\kappa_{5}^{2}}{6}\rho_{\varphi}
+\frac{1}{\dot{\varphi}\delta\varphi}\int\delta
E_{i}^{0}\,dx^{i}}{6H\left(1+\frac{\kappa_{5}^{2}}{18}\rho_{\varphi}f\right)}\dot{\varphi}\delta\varphi\,.\hspace{1cm}
\end{eqnarray}

By defining a function ${\cal{F}}$ as
\begin{equation}
{\cal{F}}\equiv\frac{\delta\varphi}{V'}\,,
\end{equation}
equation (55) can be rewritten as
\begin{widetext}
\begin{eqnarray}
\frac{{\cal{F}}'}{{\cal{F}}}=
-\frac{\bigg(V'-f'R\bigg)\bigg(\frac{\kappa_{5}^{2}}{6}\rho_{\varphi}+\frac{1}{\dot{\varphi}\delta\varphi}\int\delta
E_{i}^{0}\,dx^{i}\bigg)}{6\kappa_{4}^{2}\bigg(1+\frac{\kappa_{5}^{2}}{18}\rho_{\varphi}f\bigg)
\bigg(\rho_{\varphi}+\lambda+\frac{6\kappa_{4}^{2}}{\kappa_{5}^{4}}
\pm\frac{6}{\kappa_{5}^{2}}\,\sqrt{\frac{\kappa_{4}^{4}}{\kappa_{5}^{4}}+
\frac{\kappa_{4}^{2}}{3}\rho_{\varphi}+
\frac{\kappa_{4}^{2}}{3}\lambda-\frac{\kappa_{5}^{4}}{36}\lambda^{2}-\frac{\mathcal{C}}{a^{4}}}\bigg)}
+\frac{\frac{1}{2}f''R-V''}{\frac{1}{2}f'R-V'}-\frac{V''}{V'}\,.\hspace{2cm}
\end{eqnarray}
\end{widetext}
A solution of this equation is
${\cal{F}}={\cal{C}}\exp(\int\frac{{\cal{F}}'}{{\cal{F}}}d\varphi)$,
where ${\cal{C}}$ is an integration constant. So, from equation (56)
we find
\begin{widetext}
\begin{eqnarray}
\delta\varphi={\cal{C}}\,V'\exp\Bigg[-\int\Bigg(\frac{\Big(V'-f'R\Big)
\Big(\frac{\kappa_{5}^{2}}{6}\rho_{\varphi}+\frac{1}{\dot{\varphi}\delta\varphi}\int\delta
E_{i}^{0}\,dx^{i}\Big)}{6\kappa_{4}^{2}\Big(1+\frac{\kappa_{5}^{2}}{18}\rho_{\varphi}f\Big)
\Big(\rho_{\varphi}+\lambda+\frac{6\kappa_{4}^{2}}{\kappa_{5}^{4}}
\pm\frac{6}{\kappa_{5}^{2}}\,\sqrt{\frac{\kappa_{4}^{4}}{\kappa_{5}^{4}}+
\frac{\kappa_{4}^{2}}{3}\rho_{\varphi}+
\frac{\kappa_{4}^{2}}{3}\lambda-\frac{\kappa_{5}^{4}}{36}\lambda^{2}-\frac{\mathcal{C}}{a^{4}}}\Big)}\hspace{3cm}\nonumber\\
+\frac{\frac{1}{2}f''R-V''}{\frac{1}{2}f'R-V'}-\frac{V''}{V'}\Bigg)d\varphi
\Bigg]\,.\hspace{1cm}
\end{eqnarray}
\end{widetext}
For simplicity we define the following quantity\\\\

\begin{widetext}
\begin{equation}
{\cal{G}}=\frac{-2\Big(V'-\frac{1}{2}f'R\Big)
\Big(\frac{\kappa_{5}^{2}}{6}\rho_{\varphi}+\frac{1}{\dot{\varphi}\delta\varphi}\int\delta
E_{i}^{0}\,dx^{i}\Big)}{6\Big(1+\frac{\kappa_{5}^{2}}{18}\rho_{\varphi}f\Big)
\Big(\rho_{\varphi}+\lambda+\frac{6\kappa_{4}^{2}}{\kappa_{5}^{4}}
\pm\frac{6}{\kappa_{5}^{2}}\,\sqrt{\frac{\kappa_{4}^{4}}{\kappa_{5}^{4}}+
\frac{\kappa_{4}^{2}}{3}\rho_{\varphi}+
\frac{\kappa_{4}^{2}}{3}\lambda-\frac{\kappa_{5}^{4}}{36}\lambda^{2}-\frac{\mathcal{C}}{a^{4}}}
\Big)}-\frac{f''R-2V''}{\frac{1}{2}f'R-V'}+\frac{2V''}{V'}.
\end{equation}
\end{widetext}

As we have stated, brane parameters cannot be determined freely and
are influenced by bulk physics through boundary conditions (see for
instance \cite{Sht07} for details). In our case, the term
$\frac{1}{\dot{\varphi}\delta\varphi}\int\delta E_{i}^{0}\,dx^{i}$
in Eqs. (58) and (59) which is a non-trivial contribution of the
bulk on the brane is neglected in our forthcoming arguments. This
means that we assume backreaction due to metric perturbations in the
bulk can be neglected (we refer the reader to
\cite{Maa10,Hir06,Koy07a,Koy08,Ste93} for details and justification
of this assumption). Based on the arguments provided in
\cite{Maa10}, our assumption of neglecting the bulk-brane
interactions in this study is viable. Now with definition (59), Eq.
(58) can be rewritten as
\begin{equation}
\delta\varphi={\cal{C}}\,V'\exp\bigg(\int {\cal{G}}
d\varphi\bigg)\,.
\end{equation}
So, the density perturbation is given by
\begin{equation}
A_{s}^{2}=\frac{k^{3}}{2\pi^{2}}\,\exp\bigg(2\int {\cal{G}}
d\varphi\bigg)\,,
\end{equation}
where the effects of the non-minimal coupling of the scalar field
and induced gravity on the brane are hidden in the definition of
${\cal{G}}$. The scale-dependence of the perturbations is described
by the spectral index as
\begin{equation}
n_{s}-1=\frac{d \ln A_{S}^{2}}{d \ln k}\,.
\end{equation}
The interval in wave number is related to the number of e-folds by
the relation $$d \ln k(\varphi)=d N(\varphi)\,.$$ So we obtain
\begin{widetext}
\begin{eqnarray}
n_{s}=1-3\epsilon+\frac{2}{3}\eta\hspace{15cm}\nonumber\\
+\Bigg[\frac{-2\Big(V'-\frac{1}{2}f'R\Big)
\Big(\frac{\kappa_{5}^{2}}{6}\rho_{\varphi}\Big)}{6\Big(1+\frac{\kappa_{5}^{2}}{18}\rho_{\varphi}f\Big)
\Big(\rho_{\varphi}+\lambda+\frac{6\kappa_{4}^{2}}{\kappa_{5}^{4}}
\pm\frac{6}{\kappa_{5}^{2}}\,\sqrt{\frac{\kappa_{4}^{4}}{\kappa_{5}^{4}}+
\frac{\kappa_{4}^{2}}{3}\rho_{\varphi}+
\frac{\kappa_{4}^{2}}{3}\lambda-\frac{\kappa_{5}^{4}}{36}\lambda^{2}-\frac{\mathcal{C}}{a^{4}}}
\Big)}\frac{9H^{2}}{\frac{1}{2}f'R-V'}+\frac{2V''}{V'}\Bigg]\hspace{4cm}\nonumber\\
\times\Bigg[\frac{\kappa_{4}^{2}}{3}
+\frac{1}{V}\bigg(\frac{\kappa_{4}^{2}}{3}\lambda-\frac{\kappa_{4}^{2}f'^{2}R}{3}+\frac{2\kappa_{4}^{2}f'V'}{3}
+\frac{2\kappa_{4}^{4}}{\kappa_{5}^{4}}\hspace{10cm}\nonumber\\
\pm\frac{2\kappa_{4}^{2}}{\kappa_{5}^{2}}\,\sqrt{\frac{\kappa_{4}^{4}}{\kappa_{5}^{4}}+
\frac{\kappa_{4}^{2}}{3}V(\varphi)-\frac{\kappa_{4}^{2}}{3}f'^{2}R+\frac{2\kappa_{4}^{2}}{3}f'V'+
\frac{\kappa_{4}^{2}}{3}\lambda-\frac{\kappa_{5}^{4}}{36}\lambda^{2}-\frac{\mathcal{C}}{a^{4}}}\,\bigg)\Bigg]^{-1}
\Bigg(\frac{V'-\frac{1}{2}f'R}{3V}\Bigg).\hspace{1.1cm}
\end{eqnarray}
\end{widetext}
The running of the spectral index in our setup is given by\\\\

\begin{widetext}
\begin{eqnarray}
\alpha=\frac{d n_{s}}{d \ln k}\hspace{16.4cm}\nonumber\\
=6\epsilon^{2}+2\epsilon\eta-\left[\frac{\frac{1}{2}f''R-V''}{H^{4}}\right]\left[V'''-\frac{1}{2}f'''R\right]
+\frac{\frac{1}{2}\big(f''R-2V''\big)^{2}}{\big(\frac{1}{2}f'R-V'\big)^{2}}+\Bigg[\dot{H}+\frac{V''}{V'}\bigg(\Big(V''
-\frac{1}{2}f''R\Big)\Big(1+\frac{3H^{4}V'}{2V''}\Big)+\dot{H}
\bigg)\Bigg]\hspace{0.7cm}\nonumber\\
\times\Bigg[\frac{-4\Big(V'-\frac{1}{2}f'R\Big)
\Big(\frac{\kappa_{5}^{2}}{6}\rho_{\varphi}\Big)}{6\Big(1+\frac{\kappa_{5}^{2}}{18}\rho_{\varphi}f\Big)
\Big(\rho_{\varphi}+\lambda+\frac{6\kappa_{4}^{2}}{\kappa_{5}^{4}}
\pm\frac{6}{\kappa_{5}^{2}}\,\sqrt{\frac{\kappa_{4}^{4}}{\kappa_{5}^{4}}+
\frac{\kappa_{4}^{2}}{3}\rho_{\varphi}+
\frac{\kappa_{4}^{2}}{3}\lambda-\frac{\kappa_{5}^{4}}{36}\lambda^{2}-\frac{\mathcal{C}}{a^{4}}}
\Big)}\Bigg]\left[\frac{V'-\frac{1}{2}f'R}{3H^{4}}\right]+{\cal{G}}'
+\frac{f'''R-2V'''}{\frac{1}{2}f'R-V'}-\frac{3\ddot{H}}{H^{2}}\hspace{0.5cm}\nonumber\\.
\end{eqnarray}
\end{widetext}

The tensor perturbations amplitude of a given mode when leaving the
Hubble radius are given by
\begin{equation}
A_{T}^{2}=\frac{4\kappa_{4}^{2}}{25\pi}H^{2}\Bigg|_{k=aH}\,.
\end{equation}
In our setup and within the slow-roll approximation, we find
\begin{widetext}
\begin{eqnarray}
A_{T}^{2}=\frac{4\kappa_{4}^{2}}{25\pi}V\Bigg[\frac{\kappa_{4}^{2}}{3}
+\frac{1}{V}\bigg(\frac{\kappa_{4}^{2}}{3}\lambda-\frac{\kappa_{4}^{2}f'^{2}R}{3}+\frac{2\kappa_{4}^{2}f'V'}{3}
+\frac{2\kappa_{4}^{4}}{\kappa_{5}^{4}}\hspace{9cm}\nonumber\\
\hspace{4cm}\pm\frac{2\kappa_{4}^{2}}{\kappa_{5}^{2}}\,\sqrt{\frac{\kappa_{4}^{4}}{\kappa_{5}^{4}}+
\frac{\kappa_{4}^{2}}{3}V(\varphi)-\frac{\kappa_{4}^{2}}{3}f'^{2}R+\frac{2\kappa_{4}^{2}}{3}f'V'+
\frac{\kappa_{4}^{2}}{3}\lambda-\frac{\kappa_{5}^{4}}{36}\lambda^{2}}\,\bigg)\Bigg]\,.\hspace{1cm}
\end{eqnarray}
\end{widetext}
The tensor spectral index is given by
\begin{equation}
n_{T}=\frac{d \ln A_{T}^{2}}{d \ln k}\,,
\end{equation}
that in our model it takes the following form
\begin{widetext}
\begin{eqnarray}
n_{T}=
\Bigg(\frac{V'}{3V}\Bigg)\Bigg(\frac{f'R-V'}{V'}\Bigg)\Bigg[\frac{\kappa_{4}^{2}}{3}
+\frac{1}{V}\bigg(\frac{\kappa_{4}^{2}}{3}\lambda-\frac{\kappa_{4}^{2}f'^{2}R}{3}+\frac{2\kappa_{4}^{2}f'V'}{3}
+\frac{2\kappa_{4}^{4}}{\kappa_{5}^{4}}\hspace{6cm}\nonumber\\
\hspace{3.4cm}\pm\frac{2\kappa_{4}^{2}}{\kappa_{5}^{2}}\,\sqrt{\frac{\kappa_{4}^{4}}{\kappa_{5}^{4}}+
\frac{\kappa_{4}^{2}}{3}V(\varphi)-\frac{\kappa_{4}^{2}}{3}f'^{2}R+\frac{2\kappa_{4}^{2}}{3}f'V'+
\frac{\kappa_{4}^{2}}{3}\lambda-\frac{\kappa_{5}^{4}}{36}\lambda^{2}}\,\bigg)\Bigg]^{-1}\Sigma\,,\hspace{1cm}
\end{eqnarray}
\end{widetext}
where $\Sigma$ is defined as
\begin{widetext}
\begin{eqnarray}
\Sigma\equiv\frac{\kappa_{4}^{2}}{3}\bigg(\frac{V'-2f''f'R+2f''V'
+2f'V''}{H^{2}}\bigg)\hspace{10cm}\nonumber\\
\hspace{4cm}\times\bigg(1\pm\frac
{{\kappa_{{4}}}^{2}}{{\kappa_{{5}}}^{2}} \frac {1}{\sqrt {{\frac
{{\kappa_{{4}}}^{4}}{{\kappa_{{5}}}^{4}}}+\frac{\kappa_{4}^{2}}{3}V(\varphi)
-\frac{\kappa_{4}^{2}}{3}f'^{2}R+\frac{2\kappa_{4}^{2}}{3}f'V'+\frac{{\kappa_{{4}}}^
{2}}{3}\lambda-\frac{\kappa_{5}^{4}}{36}\lambda^{2}}}\bigg)\,.\hspace{1cm}
\end{eqnarray}
\end{widetext}
In terms of the slow-roll parameters, the tensor (gravitational
wave) spectral index can be expressed as
\begin{equation}
n_{T}=-2\epsilon\,.
\end{equation}

The ratio between the amplitudes of tensor and scalar perturbations
(tensor-to-scalar ratio) is given by
\begin{equation}
r\equiv\frac{A_{T}^{2}}{A_{S}^{2}}\simeq\frac{8\pi\kappa_{4}^{2}}{25}\frac{\exp
\Bigg(\int - {\cal{G}}d\varphi\Bigg)}{{\cal{C}}^{2} V'^{2}k^{3}}\,.
\end{equation}

After a detailed calculation of the perturbations in Jordan frame,
now we present an explicit example to see how previous equations
work.

\section{An explicit example: Monomial case with $f\sim\varphi^{2}$ and
$V\sim\varphi^{2m}$}

In this part, we take a monomial form of $f(\varphi)$ as
\begin{equation}
f(\varphi)=\xi\varphi^{2}\,,
\end{equation}
where $\xi$ is a constant parameter. Also we choose the following
form of the original scalar field potential in Jordan frame
\begin{equation}
V=\frac{b}{2m}\varphi^{2m}\,,
\end{equation}
with constant $b$. In which follows, we intend to study two types of
potentials: quadratic potential with $m=1$ and quartic potential
with $m=2$. Further, we shall compare the outcomes of these two
cases. By using equations (72) and (73) we rewrite the slow-roll
parameters (Eqs.(13) and (14)) as
\begin{widetext}
\begin{equation}
\epsilon=\left\{\begin{array}{ll}
\frac{2}{\kappa_{4}^{2}}\left(\frac{b-\xi
R}{b^{2}\varphi^{3}}\right)\left(b\varphi-8\xi^{2}\varphi R+8\xi
b\varphi\right)\\\times\frac{1\pm\frac{\kappa_{4}^{2}}{\kappa_{5}^{2}}
\left(\sqrt{\frac{\kappa_{4}^{4}}{\kappa_{5}^{4}}+\frac{\kappa_{4}^{2}}{6}b\varphi^{2}
+\frac{\kappa_{4}^{2}}{6}b\varphi^{2}-\frac{4\kappa_{4}^{2}}{3}\xi^{2}\varphi^{2}R
+\frac{4\kappa_{4}^{2}}{3}\xi
b\varphi^{2}+\frac{\kappa_{4}^{2}}{3}\lambda
-\frac{\kappa_{5}^{4}}{36}\lambda^{2}}\right)^{-\frac{1}{2}}}{\left[1+\frac{2\lambda}{b\varphi^{2}}
-\frac{8\xi^{2}R}{b}+8\xi+\frac{12\kappa_{4}^{2}}{\kappa_{5}^{4}b\varphi^{2}}
\pm\frac{12}{\kappa_{5}^{2}b\varphi^{2}}\sqrt{\frac{\kappa_{4}^{4}}{\kappa_{5}^{4}}+\frac{\kappa_{4}^{2}}{6}b\varphi^{2}
-+\frac{\kappa_{4}^{2}}{6}b\varphi^{2}-\frac{4\kappa_{4}^{2}}{3}\xi^{2}\varphi^{2}R
+\frac{4\kappa_{4}^{2}}{3}\xi b
\varphi^{2}+\frac{\kappa_{4}^{2}}{3}\lambda
-\frac{\kappa_{5}^{4}}{36}\lambda^{2}}\right]^{2}}
\quad\quad\quad m=1\vspace{0.5cm} \\\\
\frac{8}{\kappa_{4}^{2}}\left(\frac{1}{b\varphi^{5}}-\frac{\xi
R}{b^{2}\varphi^{7}}\right)\left(b\varphi^{3}-8\xi^{2}\varphi
R+16\xi b\varphi^{3}\right)\\
\times\frac{1\pm\frac{\kappa_{4}^{2}}{\kappa_{5}^{2}}
\left(\sqrt{\frac{\kappa_{4}^{4}}{\kappa_{5}^{4}}+\frac{\kappa_{4}^{2}}{12}b\varphi^{4}
-\frac{4\kappa_{4}^{2}}{3}\xi^{2}\varphi^{2}R+\frac{4\kappa_{4}^{2}}{3}\xi
b\varphi^{4}+\frac{\kappa_{4}^{2}}{3}\lambda
-\frac{\kappa_{5}^{4}}{36}\lambda^{2}}\right)^{-\frac{1}{2}}}{\left[1+\frac{4\lambda}{b\varphi^{4}}
-\frac{16\xi^{2}R}{b\varphi^{2}}+16\xi+\frac{24\kappa_{4}^{2}}{\kappa_{5}^{4}b\varphi^{4}}
\pm\frac{24}{\kappa_{5}^{2}b\varphi^{4}}\sqrt{\frac{\kappa_{4}^{4}}{\kappa_{5}^{4}}+\frac{\kappa_{4}^{2}}{12}b\varphi^{4}
-\frac{4\kappa_{4}^{2}}{3}\xi^{2}\varphi^{2}R+\frac{4\kappa_{4}^{2}}{3}\xi
b\varphi^{4}+\frac{\kappa_{4}^{2}}{3}\lambda
-\frac{\kappa_{5}^{4}}{36}\lambda^{2}}\right]^{2}}, \quad \quad
\quad\quad\quad\, m=2\vspace{0.5cm} \\
\end{array}\right.
\end{equation}
\end{widetext}
and
\begin{widetext}
\begin{equation}
\eta=\left\{\begin{array}{ll}
\frac{2}{\kappa_{4}^{2}}\left(\frac{b-\xi
R}{b\varphi^{2}}\right)\\\times\frac{1}{\left[1
+\frac{2\lambda}{b\varphi^{2}}
-\frac{8\xi^{2}R}{b}+8\xi+\frac{12\kappa_{4}^{2}}{\kappa_{5}^{4}b\varphi^{2}}
\pm\frac{12}{\kappa_{5}^{2}b\varphi^{2}}\sqrt{\frac{\kappa_{4}^{4}}{\kappa_{5}^{4}}+\frac{\kappa_{4}^{2}}{6}b\varphi^{2}
-+\frac{\kappa_{4}^{2}}{6}b\varphi^{2}-\frac{4\kappa_{4}^{2}}{3}\xi^{2}\varphi^{2}R
+\frac{4\kappa_{4}^{2}}{3}\xi
b\varphi^{2}+\frac{\kappa_{4}^{2}}{3}\lambda
-\frac{\kappa_{5}^{4}}{36}\lambda^{2}}\right]}
\quad\,\,\quad  m=1\vspace{0.5cm} \\\\
\frac{12}{\kappa_{4}^{2}}\left(\frac{1}{\varphi^{2}}-\frac{\xi
R}{3b\varphi^{4}}\right)\\\times\frac{1}{\left[1+\frac{4\lambda}{b\varphi^{4}}
-\frac{16\xi^{2}R}{b\varphi^{2}}+16\xi+\frac{24\kappa_{4}^{2}}{\kappa_{5}^{4}b\varphi^{4}}
\pm\frac{24}{\kappa_{5}^{2}b\varphi^{4}}\sqrt{\frac{\kappa_{4}^{4}}{\kappa_{5}^{4}}+\frac{\kappa_{4}^{2}}{12}b\varphi^{4}
-\frac{4\kappa_{4}^{2}}{3}\xi^{2}\varphi^{2}R+\frac{4\kappa_{4}^{2}}{3}\xi
b\varphi^{4}+\frac{\kappa_{4}^{2}}{3}\lambda
-\frac{\kappa_{5}^{4}}{36}\lambda^{2}}\right]}
\quad \quad \quad \quad\quad  m=2\vspace{0.5cm}\\
\end{array}\right.
\end{equation}
\end{widetext}
Other inflation parameters such as $n_{S}$,\, $n_{T}$ and $r$ can be
expressed in terms of $\epsilon$ and $\eta$. We neglect presentation
of these quantities here due to very lengthy structure of these
equations. In which follows we perform an analysis on these
parameters space.

\begin{figure*}
\flushleft\leftskip-9em{\includegraphics[width=2.5in]{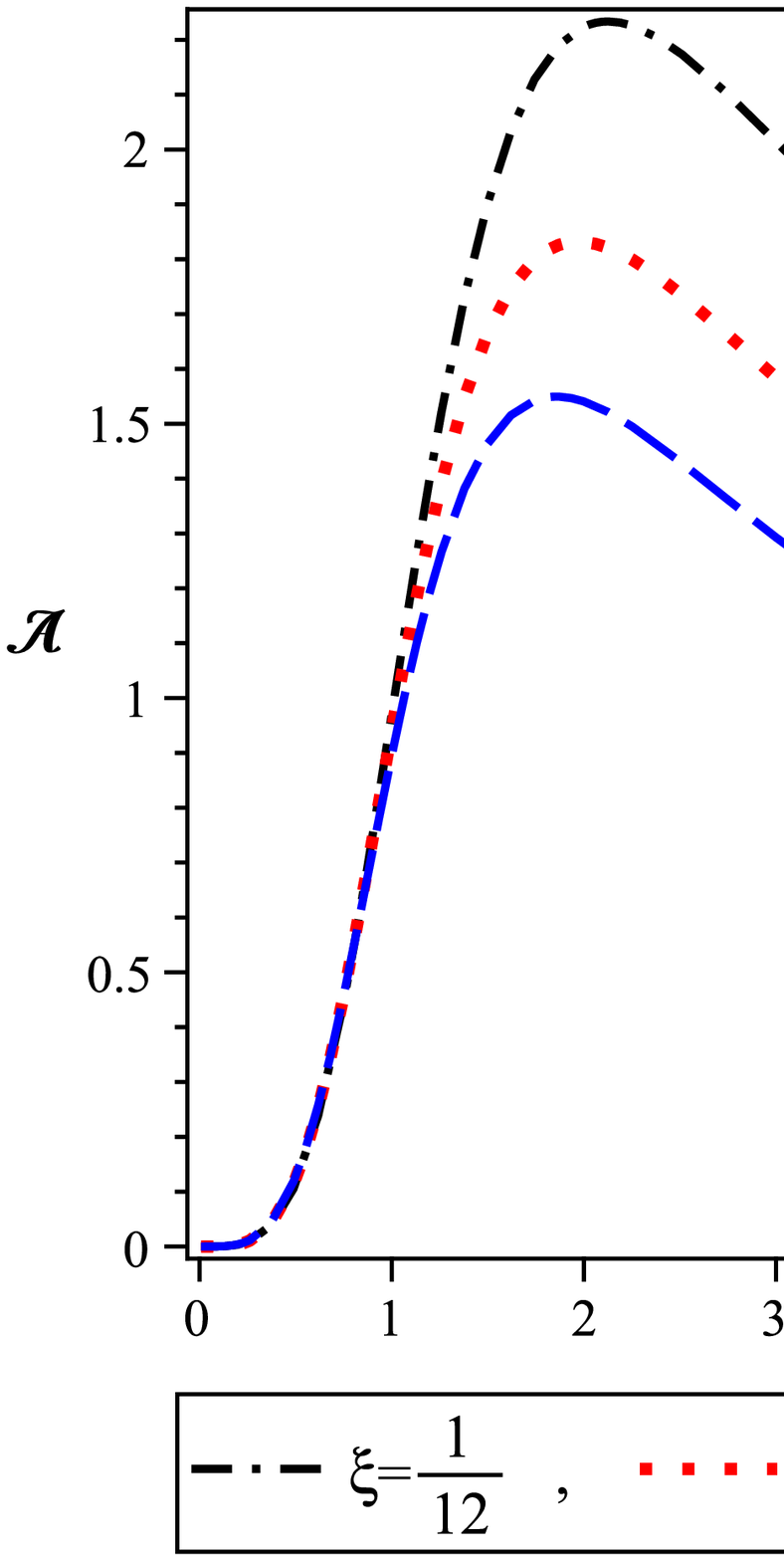}}\hspace{3.6cm}
{\includegraphics[width=2.5in]{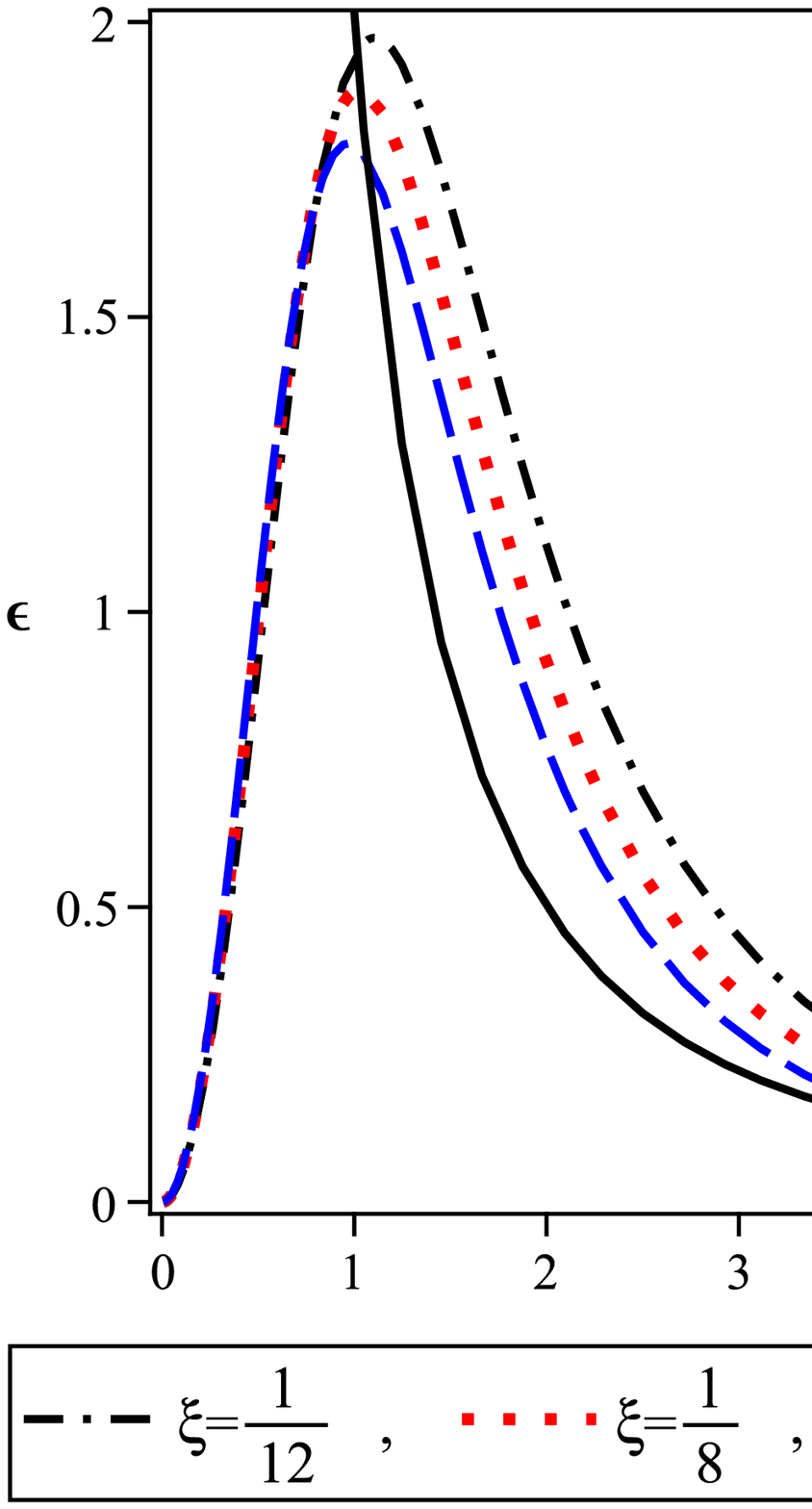}} \caption{\label{fig:1}The
evolution of the correctional factor ${\cal{A}}$ (left panel) and
the first slow-roll parameter $\epsilon$ (right panel) versus the
scalar field with a quadratic potential. The presence of the
correctional factor, ${\cal{A}}$, causes the $\epsilon$ to behave as
the standard 4D case in the large field regime. In the small field
regime, the behavior of $\epsilon$ deviates from the standard 4D
behavior. }
\end{figure*}

\begin{figure*}
\flushleft\leftskip-6em{\includegraphics[width=2.5in]{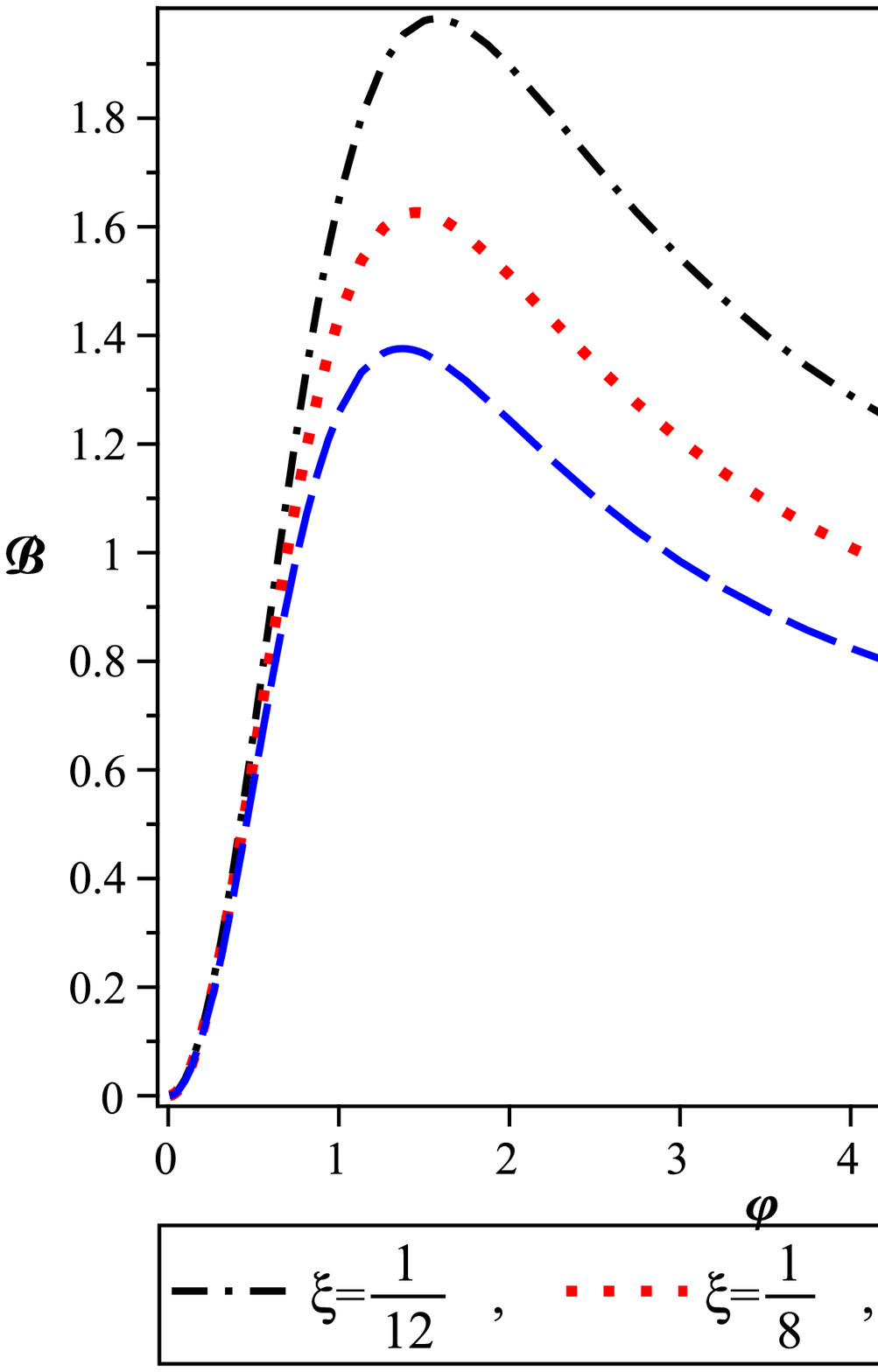}}\hspace{3cm}
{\includegraphics[width=2.5in]{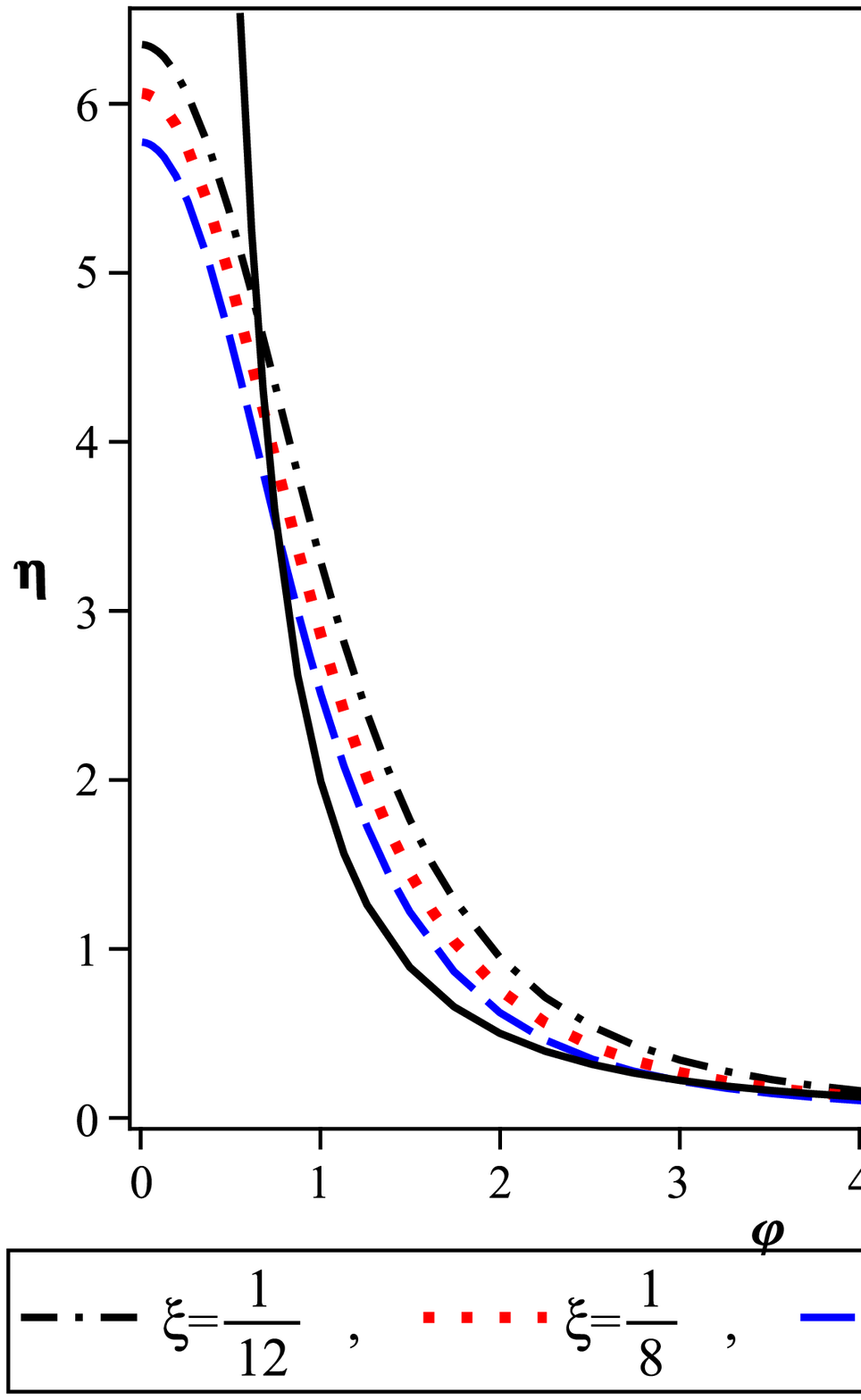}} \caption{\label{fig:2}The
evolution of the correctional factor ${\cal{B}}$ (left panel) and
the second slow-roll parameter $\eta$ (right panel) versus the
scalar field with a quadratic potential. The effect of the
correctional factor causes the $\eta$ to follow a behavior which
deviates from the standard 4D behavior in the small field regime.
There is a maximum value of $\eta$ at $\varphi=0$.}
\end{figure*}

\begin{figure*}
\flushleft\leftskip-6em{\includegraphics[width=2.5in]{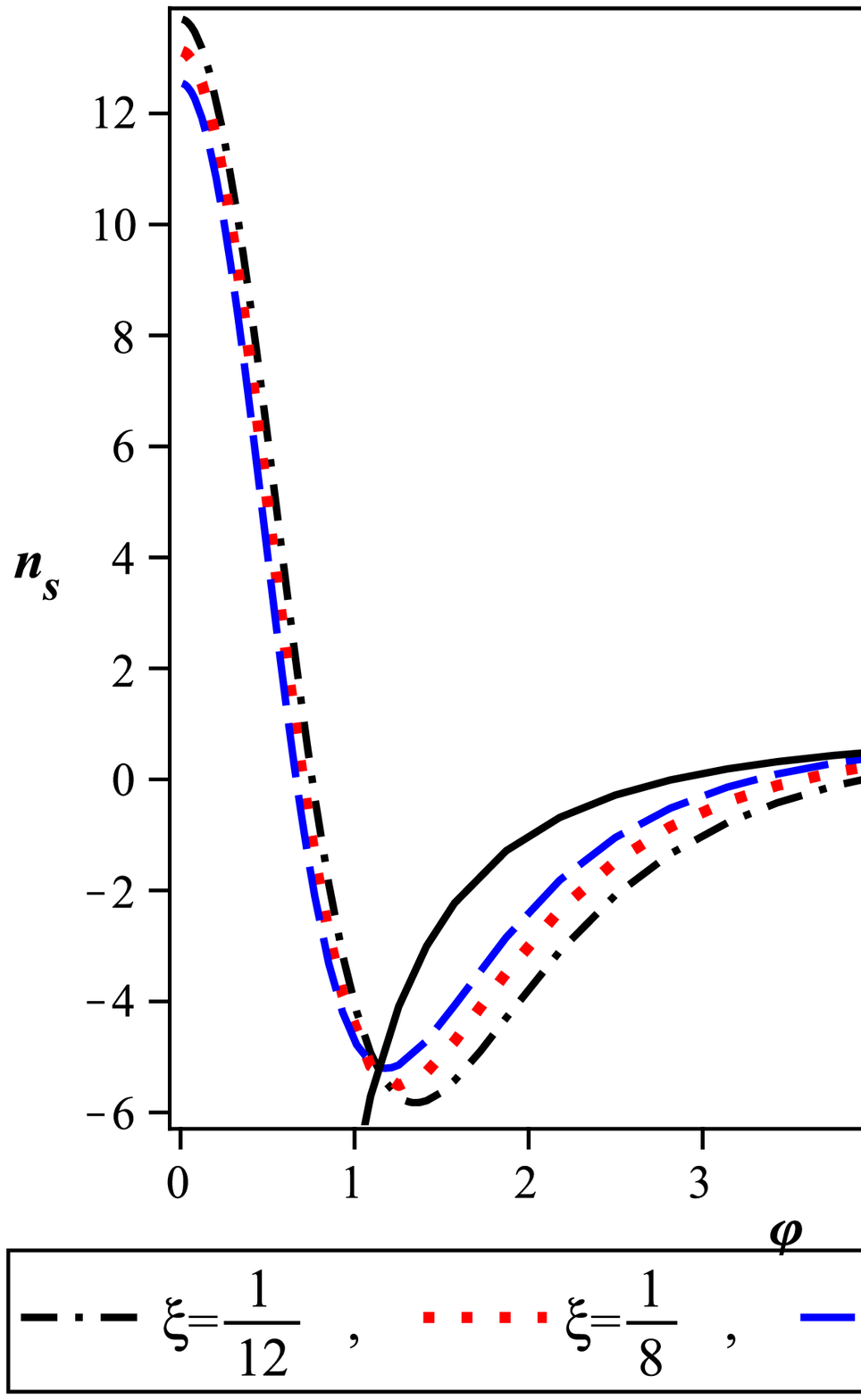}}\hspace{3cm}
{\includegraphics[width=2.5in]{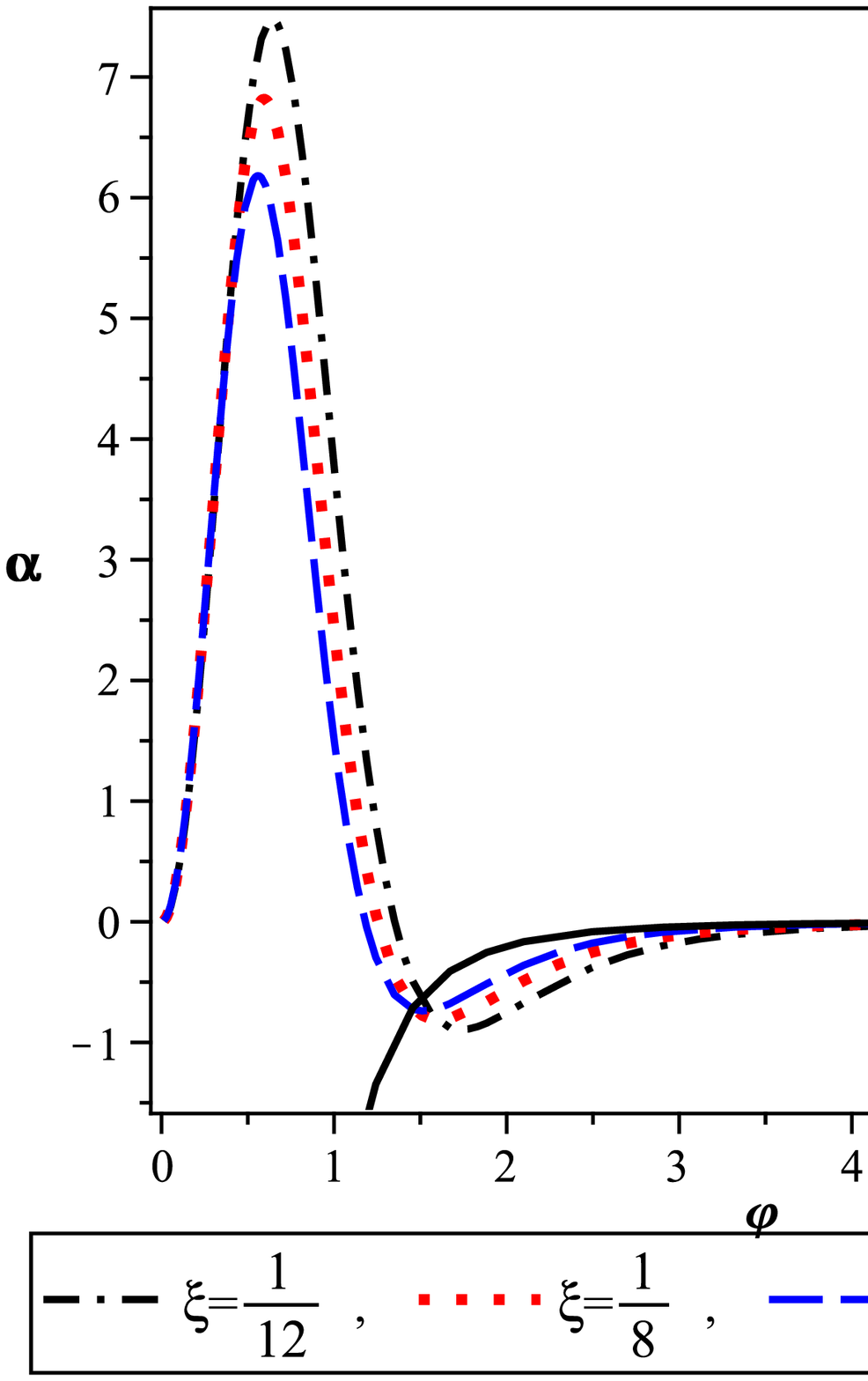}} \caption{\label{fig:3}The
evolution of the scalar spectral index (left panel) and running of
the spectral index (right panel) versus the scalar field with a
quadratic potential. In the large scalar field regime, the behavior
of $n_{s}$ and $\alpha$ are similar to the standard 4D one.}
\end{figure*}

\begin{figure}
\flushleft\leftskip-7em{\includegraphics[width=2.5in]{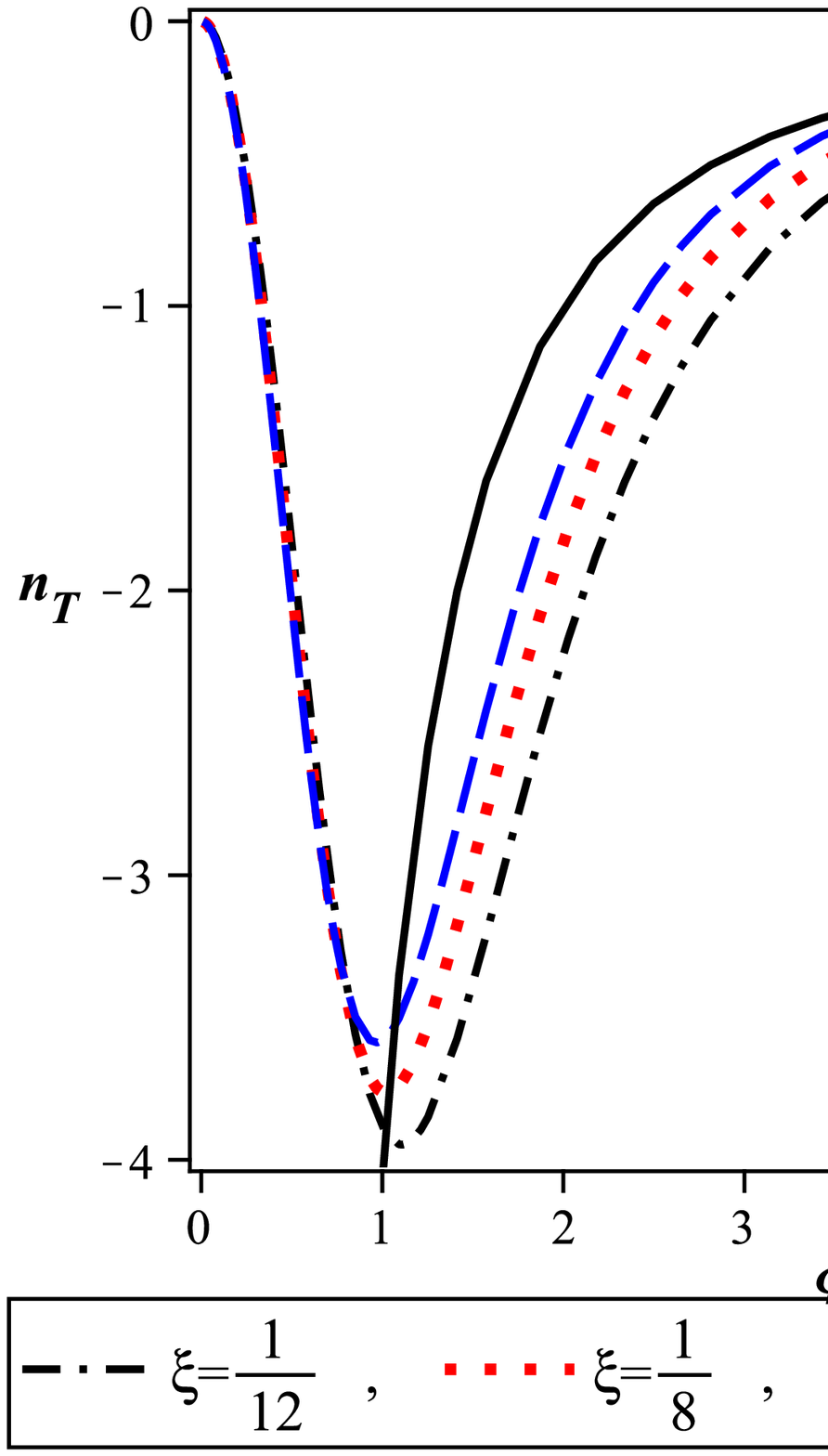}}\hspace{3cm}
\caption{\label{fig:4}The evolution of the tensor to scalar spectral
indices ratio versus the scalar field with a quadratic potential.
The behavior of $r$ in the large field regime is similar to the
standard 4D one.}
\end{figure}

\begin{figure}
\vspace{0.5cm}\flushleft\leftskip-3em{\includegraphics[width=2.5in]{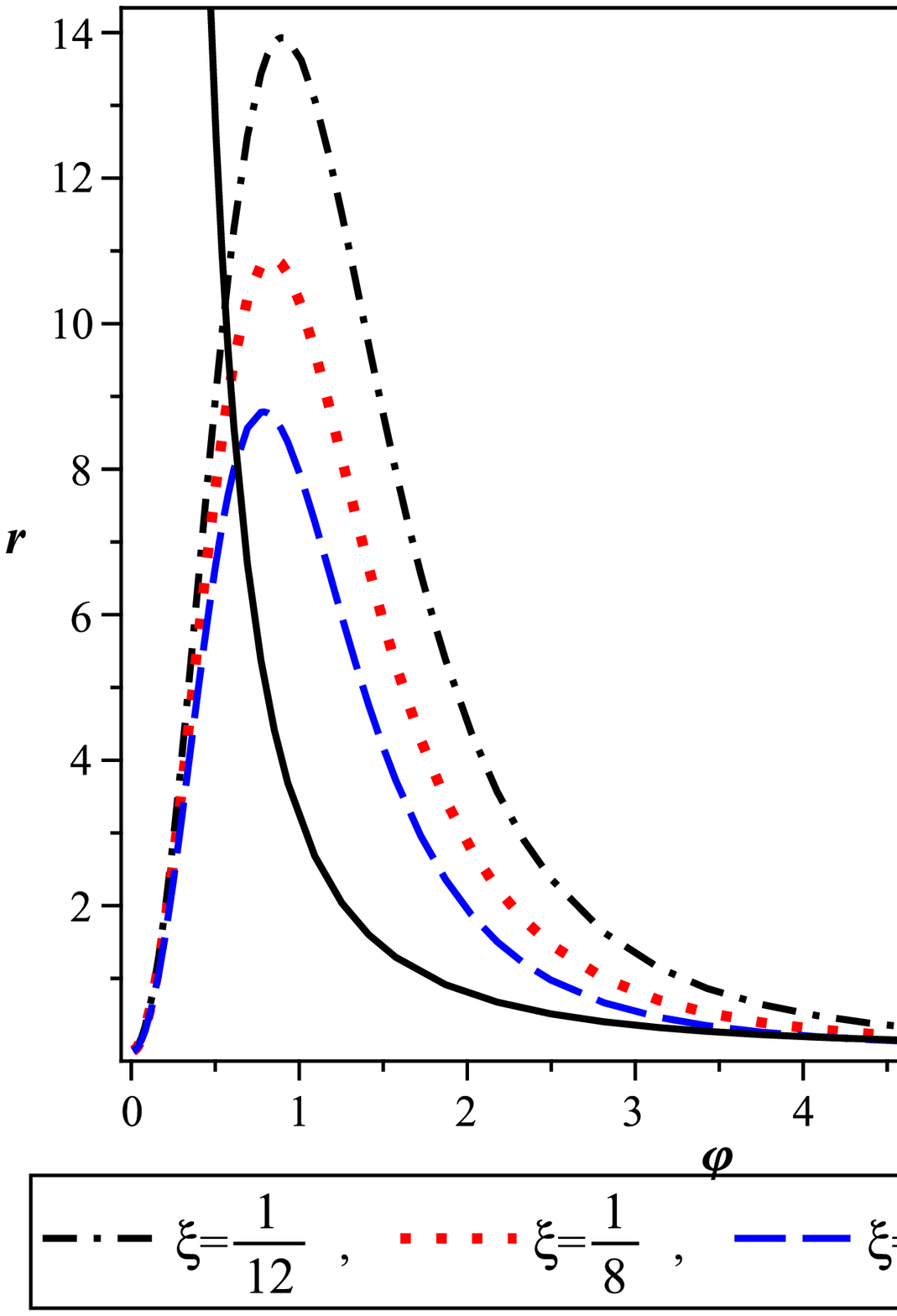}}\hspace{3cm}
\caption{\label{fig:5}The evolution of the tensor to scalar spectral
indices ratio versus the scalar field with a quadratic potential.
The behavior of $r$ in the large field regime is similar to the
standard 4D one.}
\end{figure}

\subsection{\label{sec:level2}Quadratic Potential: $V(\varphi)=\frac{b}{2}\varphi^{2}$}

The first inflaton potential we analyze is the quadratic potential,
the case with $m=1$ in equation (73). The following figures are
created as the outcome of our analysis of the model parameter space
(we note that in all figures we have set
$\kappa_{4}=\kappa_{5}=b=1$). Since $R=6(\dot{H}+2H^{2})$ and $H$ is
nearly constant in the inflation epoch, we can consider in our
numerical analysis $R$ to be approximately a constant and we set it
to unity for simplicity. Nevertheless, we will consider a more
general case by ignoring this assumption and adopting some reliable
ansatz in our numerical analysis. We note also that all of our
numerical analysis in this paper are performed for \emph{normal
branch} of this DGP-inspired model since this branch is ghost-free
\cite{Koy05,Izu07,Koy07b,Rha06}. In the left panel of
figure~\ref{fig:1}, behavior of ${\cal{A}}(\varphi)$ as a correction
factor to the standard result is depicted versus the scalar field.
In this figure (and almost in all figures of this paper) we consider
three values for $\xi$: $\frac{1}{12}$, $\frac{1}{8}$ and
$\frac{1}{6}$. We note that $\xi=\frac{1}{6}$ is the conformal
coupling of the standard general relativity
\cite{Far96,Far00,Far99,Fuj03}.

The left panel of figure~\ref{fig:1} shows that as the scalar field
decreases from the initial large values, ${\cal{A}}(\varphi)$
increases toward a maximum and then decreases. This maximum has
different values for different $\xi$. As $\xi$ increases, the value
of the maximum decreases and occurs in smaller values of the scalar
field. Also, for each value of $\varphi$, the value of
${\cal{A}}(\varphi)$ decreases as $\xi$ increases. The behavior of
${\cal{A}}(\varphi)$ affects the behavior of the first slow-roll
parameter $\epsilon$. This can be seen in the right panel of
figure~\ref{fig:1}. At large scalar field regime, the value of
$\epsilon$ in warped DGP model is smaller than the corresponding
value in the standard four dimensional model (here we note that in
all of our figures the solid, black line curve represents the
evolution of corresponding parameter in the standard 4D model). As
the scalar field decreases, $\epsilon$ increases. For some value of
the scalar field, $\epsilon$ takes the same value in both warped DGP
and the standard four-dimensional model. For this value of the
scalar field, ${\cal{A}}(\varphi)=1$. But, at some value of scalar
field, $\epsilon$ reaches its maximum and then decreases. During
this evolution, the behavior of $\epsilon$ in the warped DGP model
is similar to the standard 4D case. With more reduction of the
scalar field, $\epsilon$ deviates from 4D behavior and as the scalar
field decreases, $\epsilon$ decreases similar to the correctional
factor ${\cal{A}}(\varphi)$. This deviation from the 4D behavior is
due to the presence of the brane tension. If there is no brane
tension (also, with the zero effective cosmological constant), we
attain the pure DGP model and the slow roll parameter always behaves
as what it does in 4D model. In high energy regime, the effect of
scalar field dominates the brane tension, but in low energy regime,
where the scalar field becomes small, the brane tension's effect
becomes dominant in the dynamics of the model and so we can see the
deviation of the standard 4D model. During the reduction of
$\epsilon$, in some value of scalar field where ${\cal{A}}(\varphi)$
reaches to unity, the value of $\epsilon$ becomes equal to the 4D
one again. We note that as for ${\cal{A}}(\varphi)$, the maximum
value of $\epsilon$ depends on the value of $\xi$ too. As $\xi$
increases, the maximum becomes smaller and take places in smaller
value of the scalar field. It means that for larger $\xi$, the 4D
behavior lasts in wider domain of the scalar field values. For all
values of $\xi$, it is possible for $\epsilon$ to reach unity and so
the inflation has a graceful exit in this setup without need to any
additional mechanism. In our setup, the slow-roll parameter reaches
to unity twice. But, we know that the inflation occurs when
${\epsilon,\eta}\ll 1$. So, the first reaching of $\epsilon$ to
unity, which take places in larger scalar field value, is the end of
inflation since it reaches to unity from values
smaller than $1$.\\
The behavior of the second correctional factor,
${\cal{B}}(\varphi)$, is more or less similar to
${\cal{A}}(\varphi)$. While the scalar field decreases, ${\cal{B}}$
increases to a maximum and then decreases (see the left panel of
figure~\ref{fig:2}). From the right panel of figure~\ref{fig:2}, we
can see the effect of the evolution of ${\cal{B}}$ on the second
slow-roll parameter, $\eta$. $\eta$ in the warped DGP model always
increases by reduction of the scalar field. This is similar to the
behavior of $\eta$ in the standard four-dimensional case. However,
due to the presence of the correctional factor ${\cal{B}}$, $\eta$
in the warped DGP model does not increase strictly as it does in 4D
model (see the right panel of figure~\ref{fig:2}). There is a
maximum value for $\eta$ at $\varphi=0$. This maximum, for smaller
$\xi$, has larger value. Since $\eta$ can attain the unit value too,
the graceful exit from the inflationary phase in this model is
guaranteed. We notify that in non-minimal inflation on the warped
DGP brane within Jordan frame with a quadratic potential, both
$\epsilon$ and $\eta$ are always positive. The next parameters that
we consider are the scalar and tensorial spectral indices (shown as
$n_{s}$ and $n_{T}$ respectively). In figures~\ref{fig:3} (the left
panel) and ~\ref{fig:4}, we have shown the behavior of the scalar
and tensorial spectral indices versus the scalar field. One can
realize the effect of first and second slow-roll parameters in the
behavior of spectral indices. In the large values of the scalar
field, both parameters behave similar to the corresponding
parameters in the standard 4-dimensional model. It means that both
scalar and spectral indices decrease by reduction of the scalar
field strength. However, at some values of the scalar field, $n_{s}$
and $n_{T}$ reach a minimum and after that they increase, in
contrast with the standard 4D case. The minimum value of these
parameters decreases by reduction of $\xi$ and take places in larger
values of the scalar field. So, for larger values of $\xi$, the
standard behavior of $n_{s}$ and $n_{T}$ last in larger domain of
$\varphi$ values. The general behavior of $n_{s}$ and $n_{T}$ is
very similar to $\epsilon$ and $\eta$: similarity with the standard
four-dimensional case in the large scalar field regime and deviation
from it in the small scalar field regime.

\begin{table*}
\caption{\label{tab:table1}The values of some inflation parameters
with a quadratic potential in Jordan frame at the time that physical
scales crossed the horizon.}
\begin{ruledtabular}
\begin{tabular}{ccccc}
$\xi$&$n_{s}$&$r$&$\alpha$ \\ \hline\\
$0$& $1.000000000$ &$2.022875231\times 10^{-13}$&-3.029161336$\times
10^{-101}$
\\\\
$\frac{1}{12}$& $0.9667051476$ &$0.1575567389$&-1.670874757$\times 10^{-38}$\\\\
$\frac{1}{8}$& $0.9653928105$ &$0.2094435340$&-2.537874469$\times 10^{-38}$\\\\
$\frac{1}{6}$& $0.9665654212$ &$0.3137965930$&-3.269176612$\times 10^{-38}$\\\\
$observation$& $0.968\pm 0.012$ &$ < 0.24(95\% CL) $&$-0.022 \pm 0.020$\\
\end{tabular}
\end{ruledtabular}
\end{table*}

In the right panel of figure ~\ref{fig:3} we see the evolution of
the running of the scalar spectral index, $\alpha$, versus the
scalar field. In the large scalar field regime, the behavior of
$\alpha$ is similar to the corresponding parameter in the standard
4D case and decreases by decreasing the scalar field value. But, at
some value of the scalar field, $\alpha$ reaches its minimum value
and then increases toward a maximum and after that, it decreases
again. The minimum value of $\alpha$ take places in smaller scalar
field values by increasing $\xi$. So, as $\xi$ increases, the 4D
behavior of $\alpha$ lasts in larger domain of the scalar field. The
last parameter that we are going to consider, is the ratio between
the amplitudes of the tensor and scalar perturbations ($r$). We have
shown the behavior of this ratio versus the scalar field in
figure~\ref{fig:5}. Its behavior is similar to the behavior of
$\epsilon$ in general. As the scalar field decreases, $r$ increases
toward a maximum in some values of the scalar field. Then, it begins
to decrease. In other words, its evolution in the large scalar field
region obeys the standard 4D behavior and in the small scalar field
region, it evolves differently. Similar to other parameters, the
extremum value of $r$ depends on the value of $\xi$. For larger
$\xi$, the extremum value of $r$ becomes smaller and take places in
smaller value of $\varphi$. So, for larger value of $\xi$, the ratio
between the amplitudes of the tensor and scalar perturbations in
warped DGP model, in larger domain of large $\varphi$, behaves as 4D
model one.\\
Now we proceed to calculate some inflation parameters with a
quadratic potential at the time that physical scales crossed the
horizon. To find the value of the scalar field at the end of
inflation, we set one of the slow-roll parameters, $\epsilon$ or
$\eta$, equal to unity to get $\varphi_{f}$. To find the value of
the scalar field at the time of horizon crossing, we have to adopt
another strategy: the horizon crossing occurred about $60$ e-folds
before the end of the inflation. So the definition of the number of
e-folds helps us to find the value of the scalar field at the
horizon crossing time, $\varphi_{hc}$. Now we rewrite the Friedmann
equation (12) in the high energy limit ($\rho\gg\lambda$) as follows
\begin{eqnarray}
H^{2}\simeq
\Bigg(\frac{\kappa_{4}^{2}}{3}V-\frac{\kappa_{4}^{2}}{3}f'^{2}R+\frac{2\kappa_{4}^{2}}{3}f'V'\Bigg)\hspace{2.7cm}\nonumber\\
\times\Bigg[1-\frac{2\kappa_{4}^{2}}{\kappa_{5}^{2}}\Bigg(\frac{\kappa_{4}^{2}}{3}V-\frac{\kappa_{4}^{2}}{3}f'^{2}R
+\frac{2\kappa_{4}^{2}}{3}f'V'\Bigg)^{-\frac{1}{2}}\,\Bigg].\hspace{0.6cm}
\end{eqnarray}
So, the number of e-folds by using equation (21) can be expressed as
\begin{eqnarray}
N=3\int_{\varphi_{hc}}^{\varphi_{f}}
\frac{\Big(\frac{\kappa_{4}^{2}}{3}V-\frac{\kappa_{4}^{2}}{3}f'^{2}R+\frac{2\kappa_{4}^{2}}{3}f'V'\Big)}{\frac{1}{2}f'R-V'}
\hspace{2cm}\nonumber\\
\times\Bigg[1-\frac{2\kappa_{4}^{2}}{\kappa_{5}^{2}}\Bigg(\frac{\kappa_{4}^{2}}{3}V-\frac{\kappa_{4}^{2}}{3}f'^{2}R
+\frac{2\kappa_{4}^{2}}{3}f'V'\Bigg)^{-\frac{1}{2}}\,\Bigg]d\varphi\,.\nonumber\\
\end{eqnarray}

\begin{figure*}
\flushleft\leftskip-9em{\includegraphics[width=2.5in]{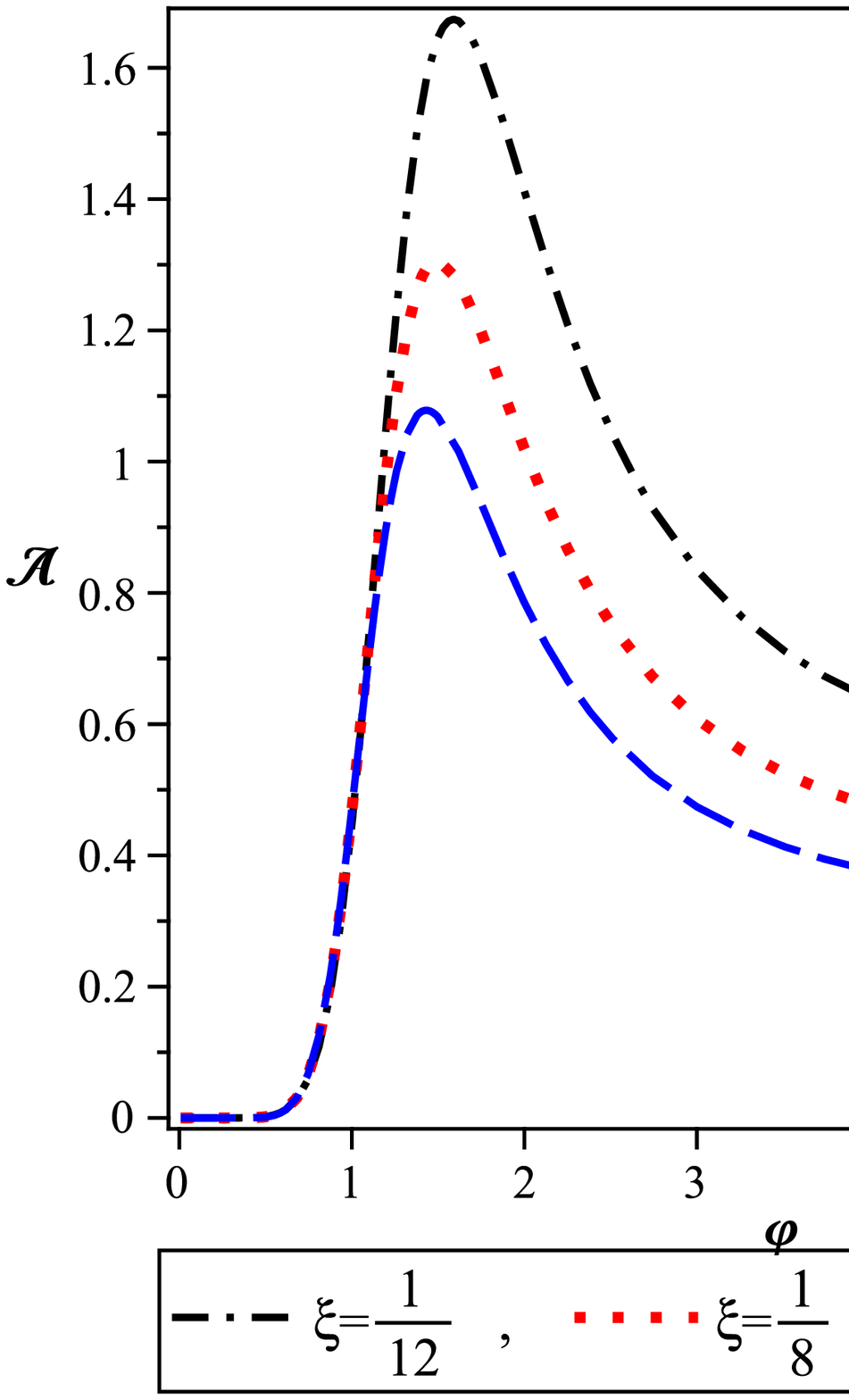}}\hspace{3.6cm}
{\includegraphics[width=2.5in]{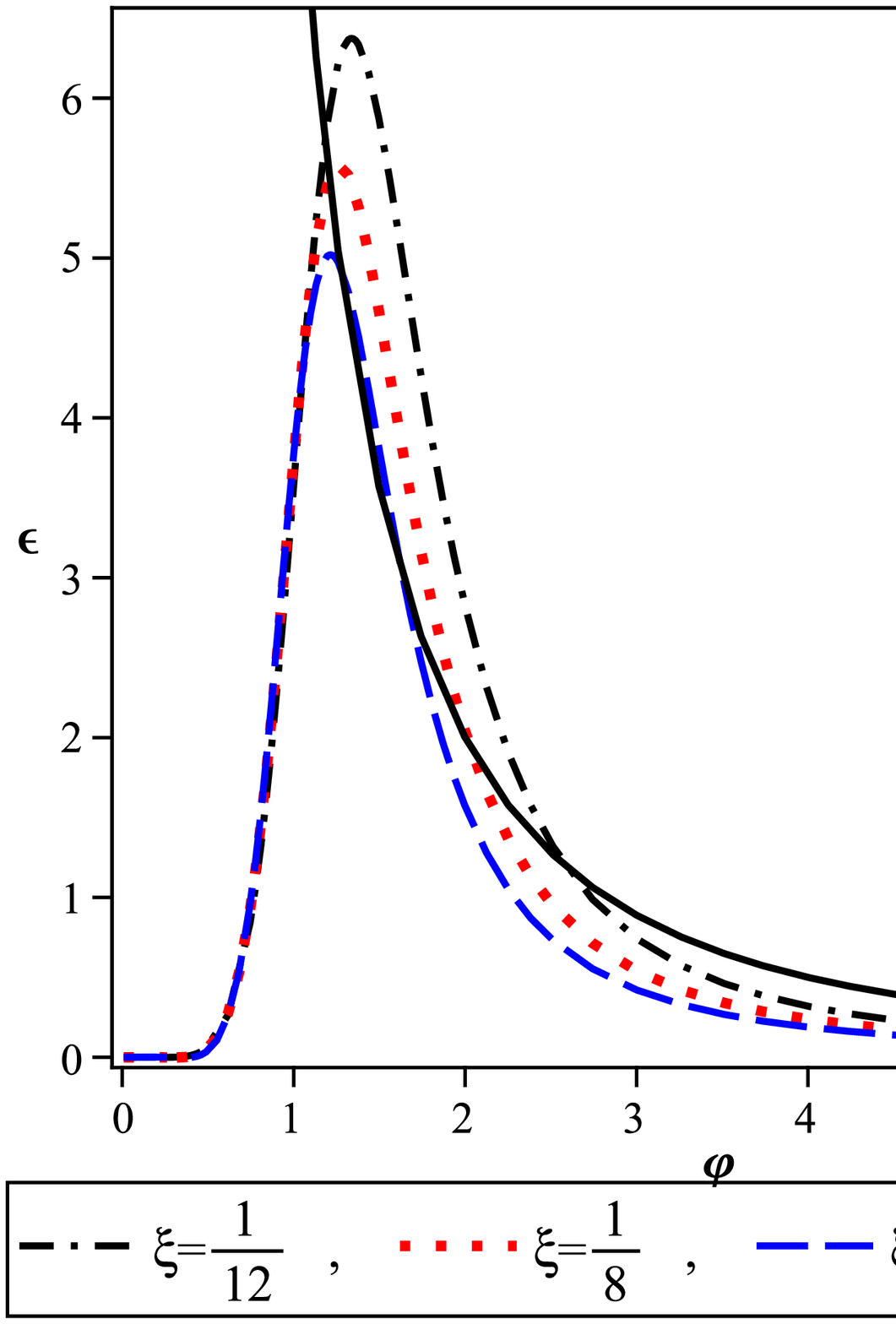}} \caption{\label{fig:6}The
evolution of the correctional factor ${\cal{A}}$ (left panel) and
the first slow-roll parameter $\epsilon$ (right panel) versus the
scalar field with a quartic potential. The braneworld and
non-minimal nature of the model through the existence of ${\cal{A}}$
causes the $\epsilon$ to behave as the standard 4D case just in the
large field regime. In the small field regime, the behavior of
$\epsilon$ deviates from the standard 4D behavior considerably.}
\end{figure*}

\begin{figure*}
\flushleft\leftskip-7em{\includegraphics[width=2.5in]{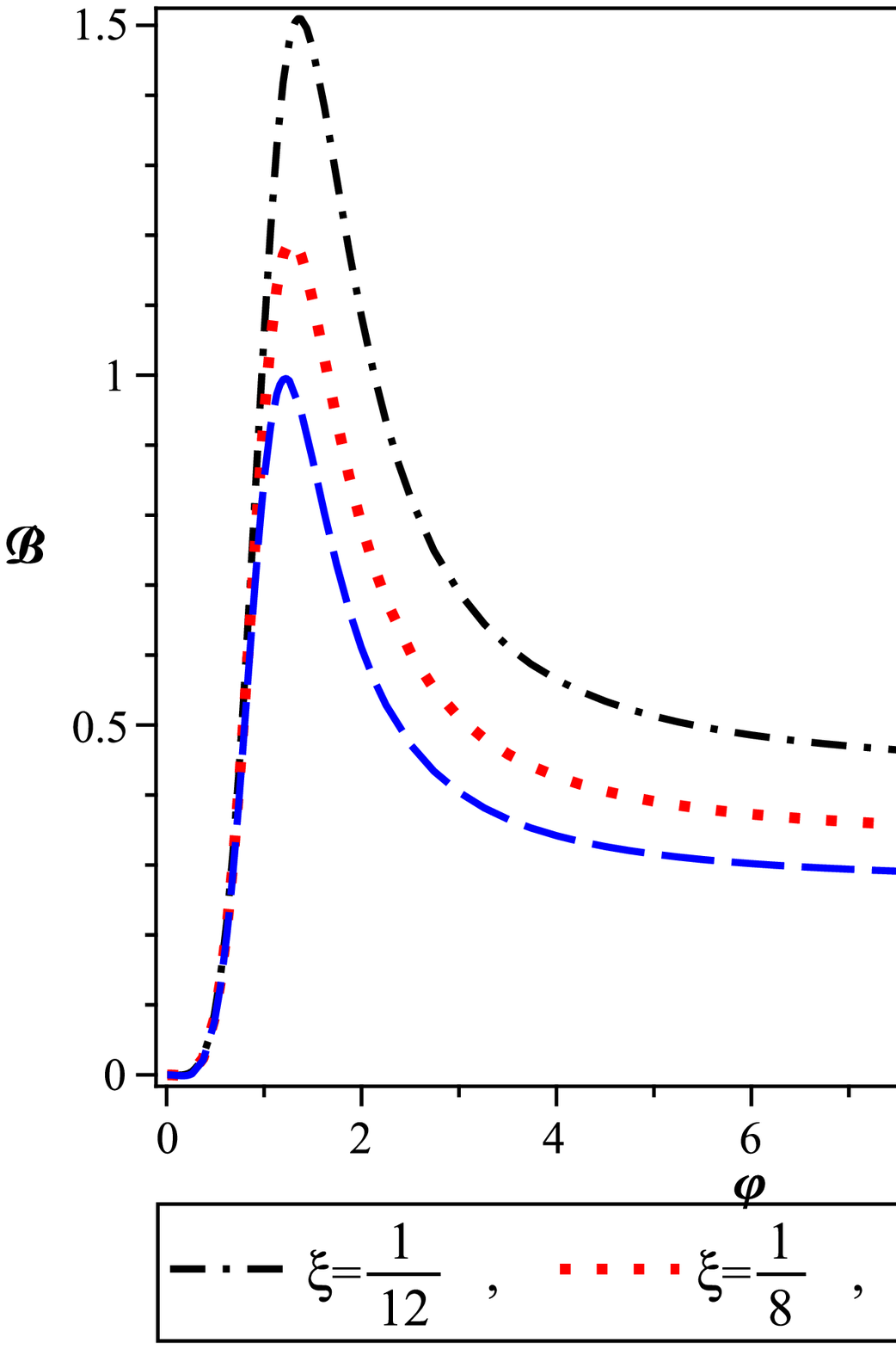}}\hspace{2.7cm}
{\includegraphics[width=2.5in]{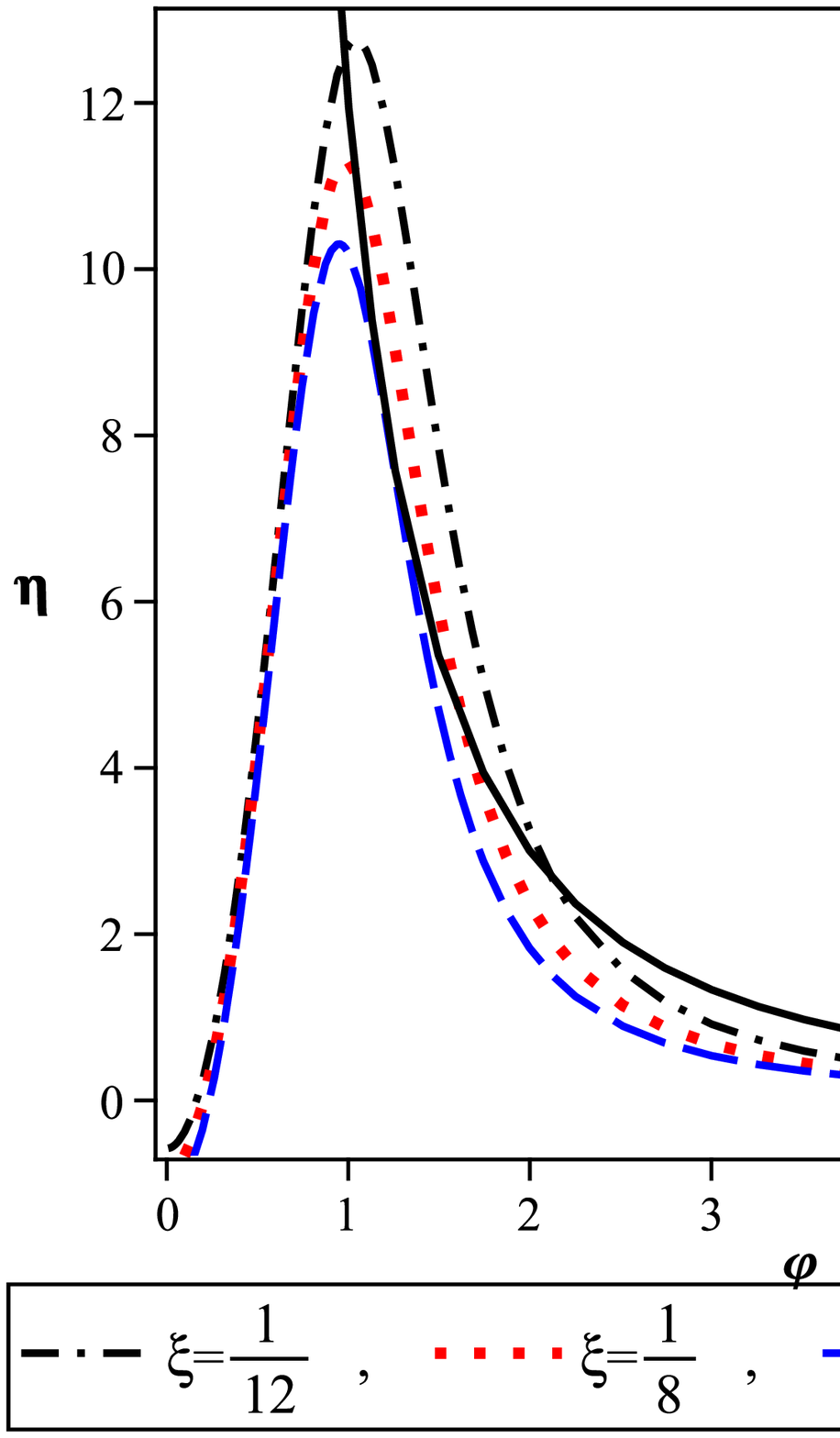}} \caption{\label{fig:7}The
evolution of the correctional factor ${\cal{B}}$ (left panel) and
the second slow-roll parameter $\eta$ (right panel) versus the
scalar field with a quartic potential. The effect of the
correctional factor causes the $\eta$ to follow a behavior which
deviates from the standard 4D behavior in the small field regime.}
\end{figure*}

\begin{figure*}
\flushleft\leftskip-7em{\includegraphics[width=2.5in]{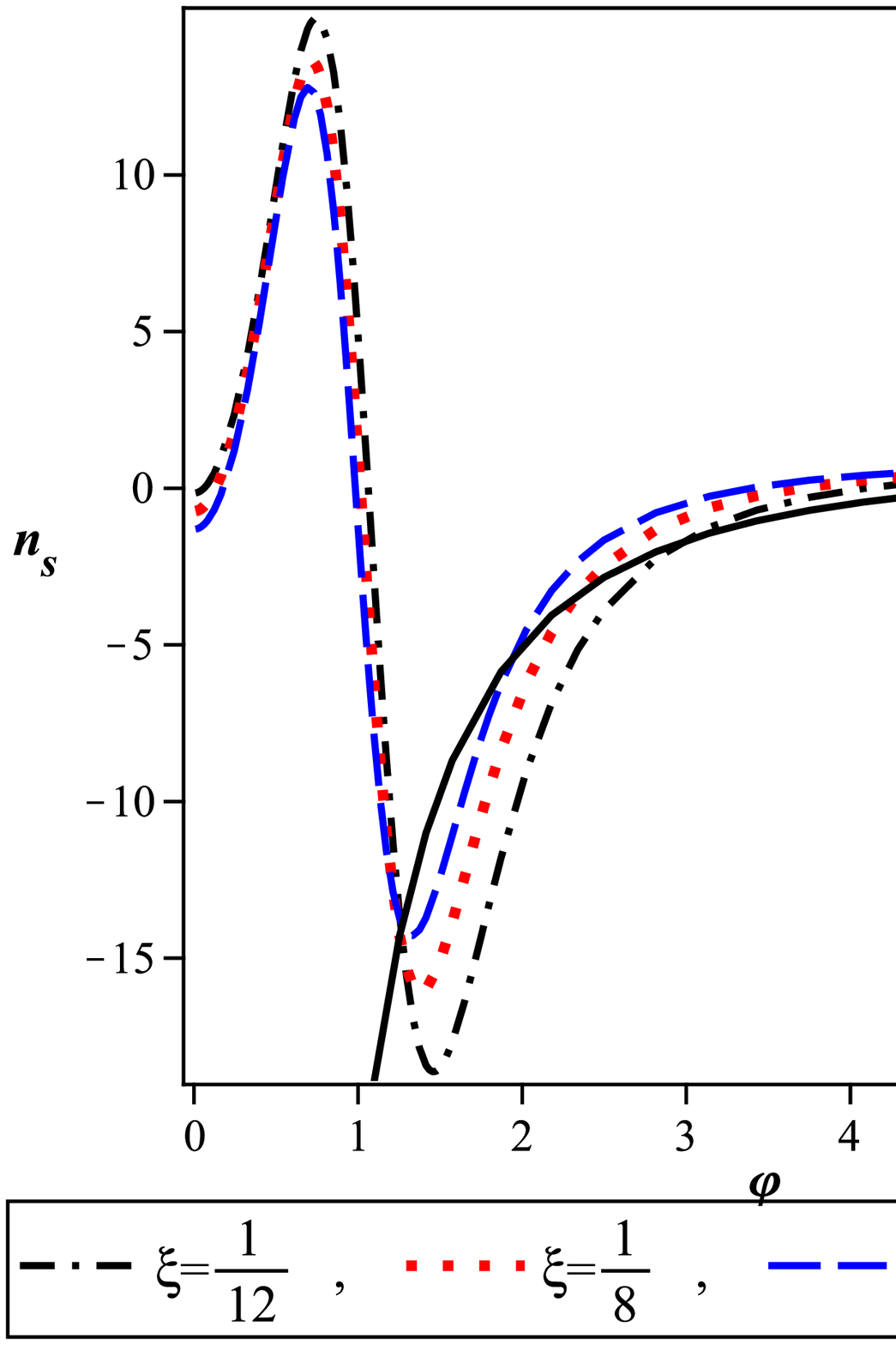}}\hspace{3cm}
{\includegraphics[width=2.5in]{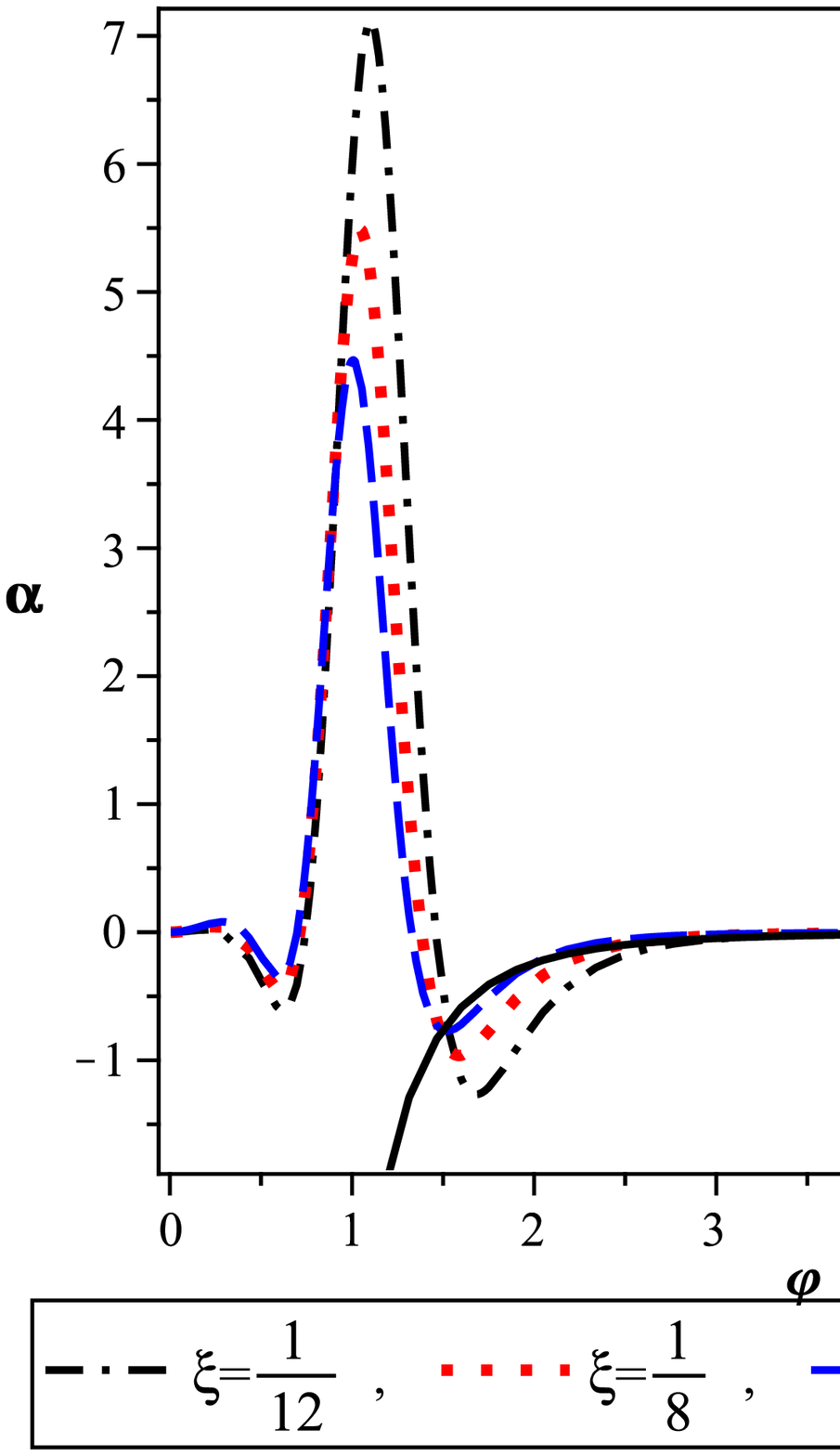}} \caption{\label{fig:8}The
evolution of the scalar spectral index (left panel) and the running
of the spectral index (right panel) versus the scalar field with a
quartic potential. In the large and small scalar field regime, the
scalar spectral index and its running decrease by reduction of the
scalar field (as the 4D case).}
\end{figure*}

\begin{figure*}
\flushleft\leftskip-5em{\includegraphics[width=2.5in]{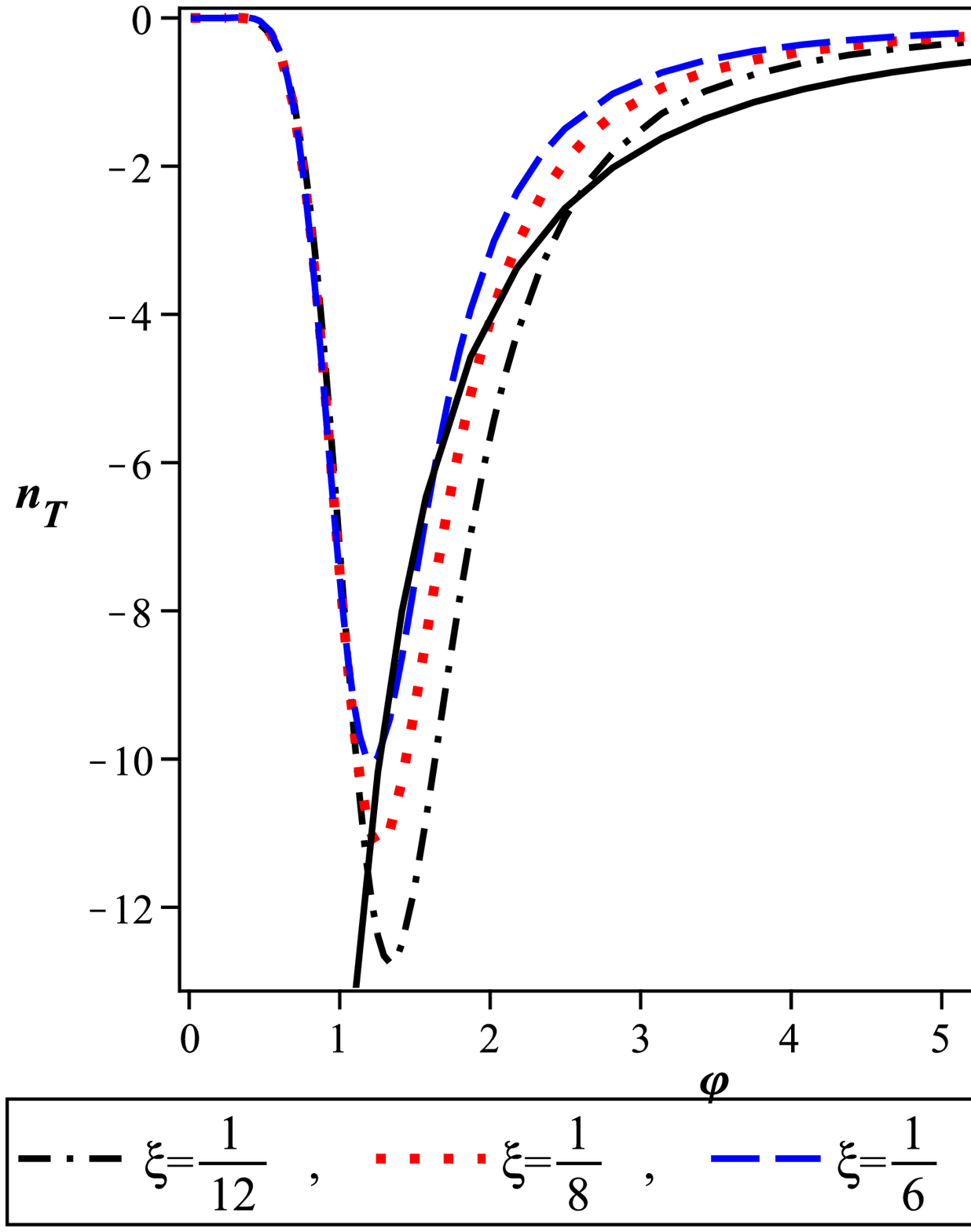}}\hspace{2.6cm}
{\includegraphics[width=2.5in]{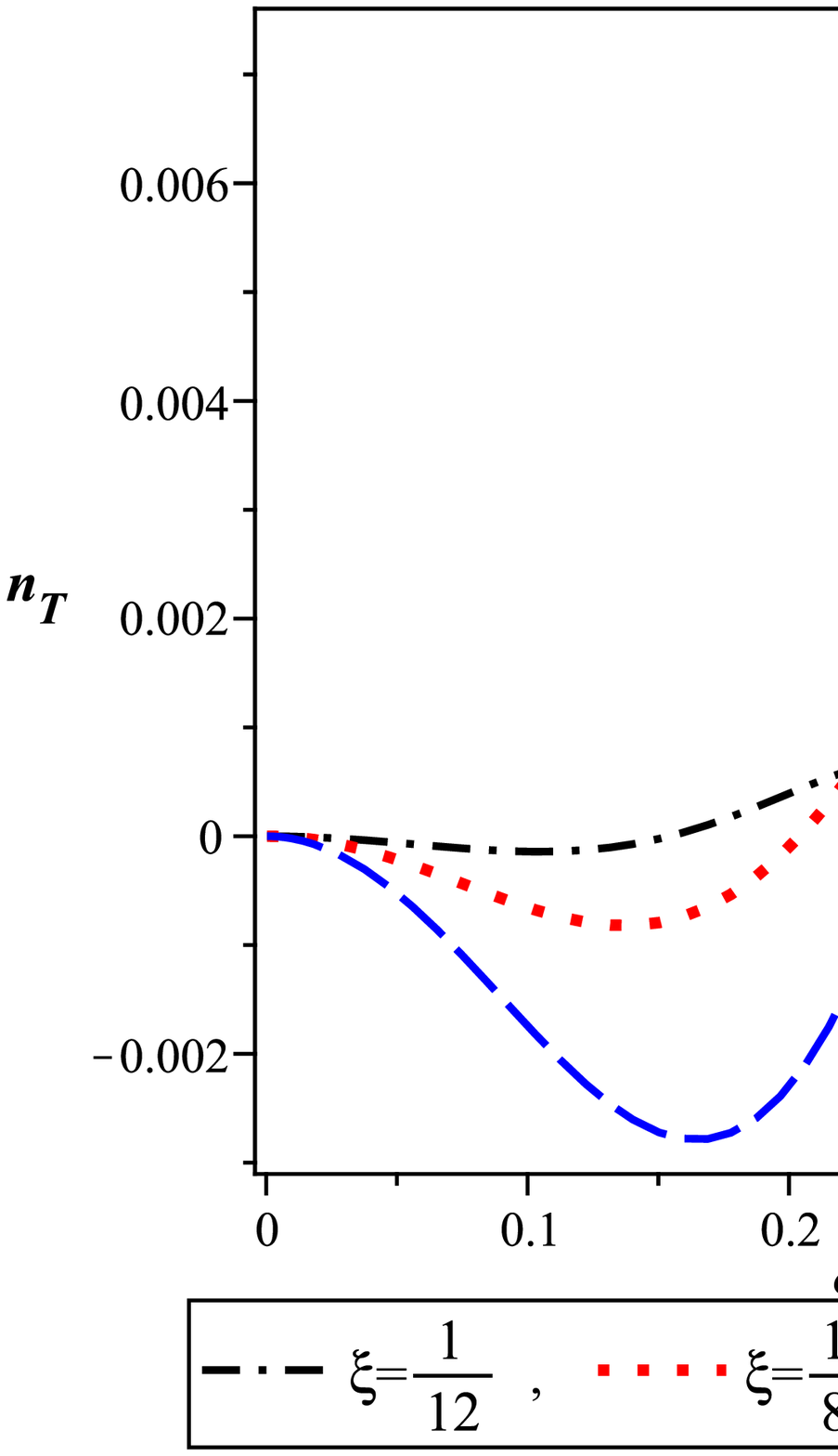}} \caption{\label{fig:9}The
evolution of the tensor spectral index versus the scalar field with
a quartic potential. In two extremal region of the scalar field, the
tensor spectral index decreases by reduction of the scalar field (as
the 4D case). The right panel shows the behavior of $n_{T}$ in very
small values of the scalar field as a special feature of the model
with quartic potential.}
\end{figure*}

\begin{figure}
\flushleft\leftskip-9em{\includegraphics[width=2.5in]{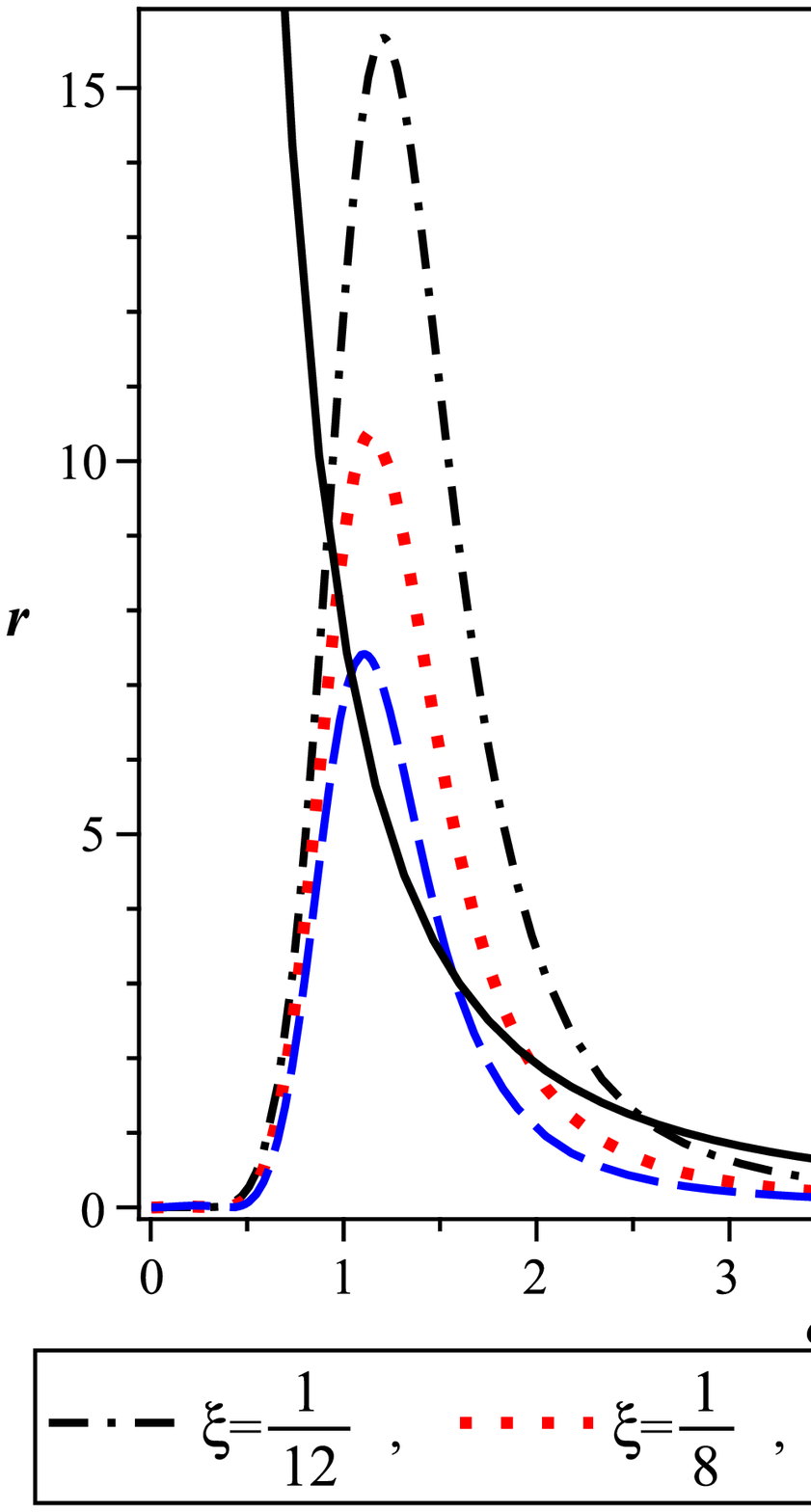}}\hspace{3cm}
\caption{\label{fig:10}The evolution of the tensor to scalar ratio
versus the scalar field with a quartic potential. The behavior of
$r$ in the large scalar field regime is similar to the 4D behavior.}
\end{figure}

\begin{figure*}
\flushleft\leftskip3em{\includegraphics[width=2.5in]{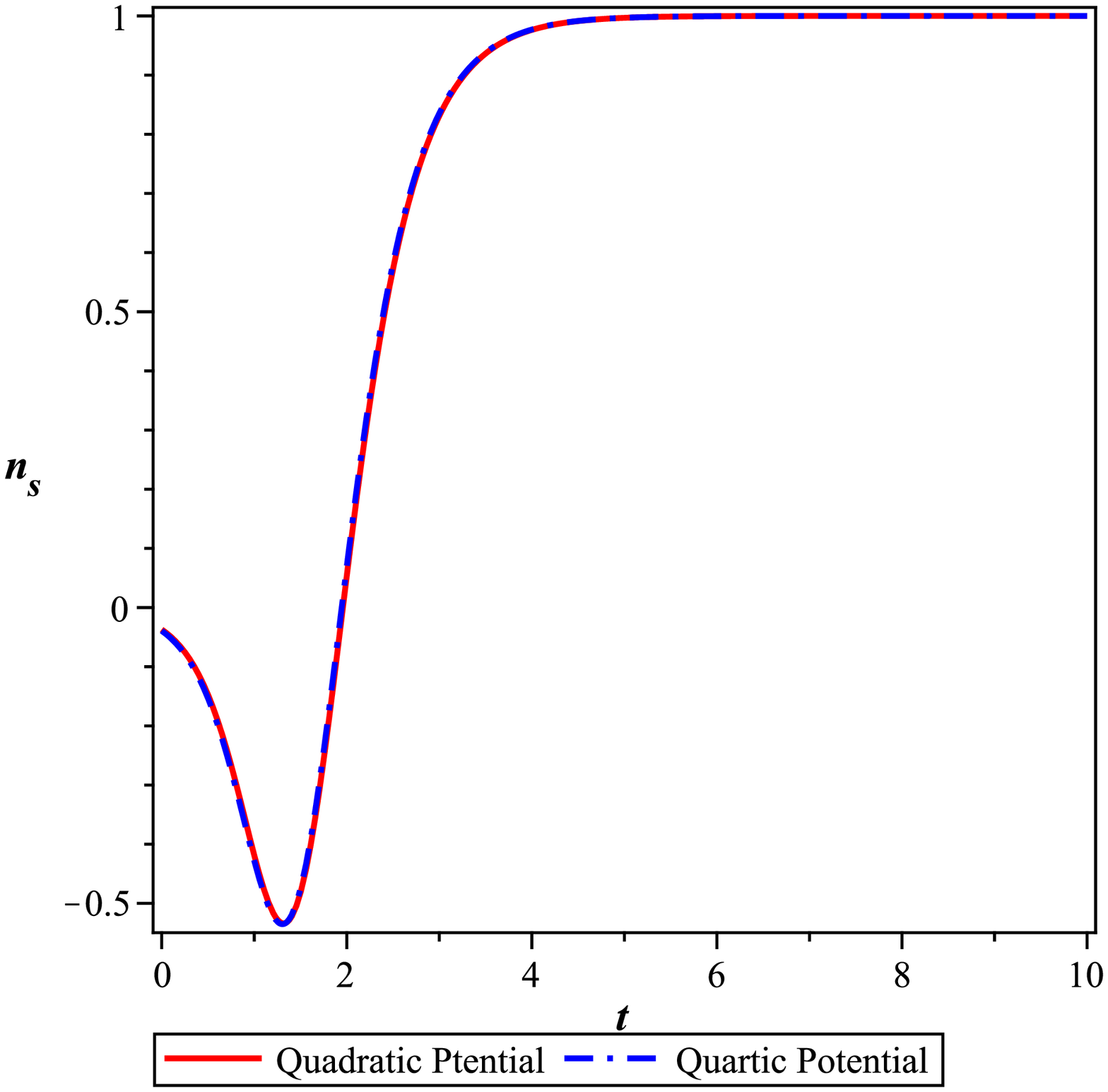}}\hspace{2cm}
{\includegraphics[width=2.5in]{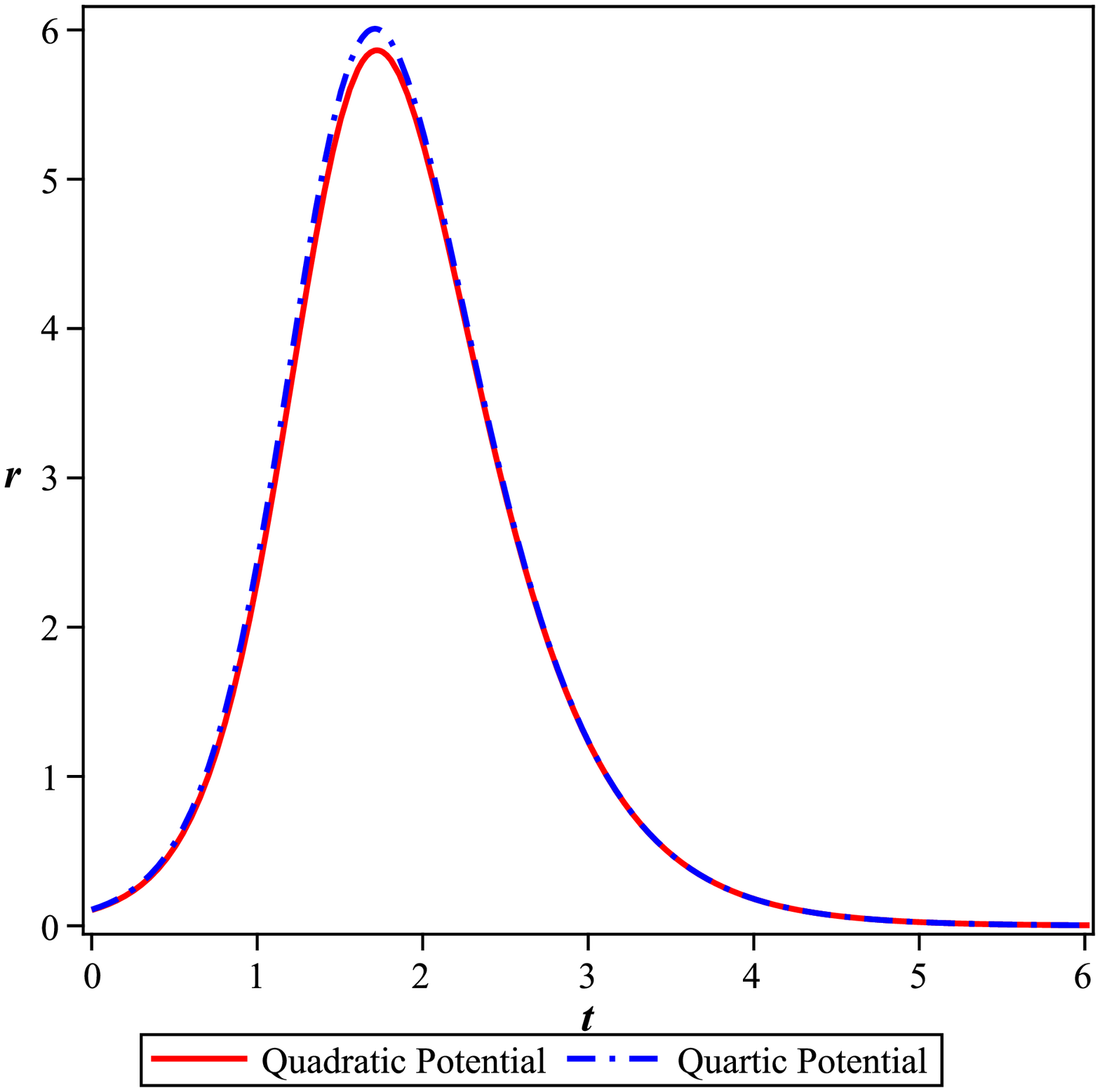}} \caption{\label{fig:5ab} The
evolution of the scalar spectral index (left panel) and
tensor-to-scalar ratio (right panel) for quadratic and quartic
potentials with adopted exponential ansatz and $\xi=\frac{1}{6}$.}
\end{figure*}

We must solve the above integral in order to find $\varphi_{hc}$. In
appendix {\bf A}, we have presented the solution of the integral
(77), where we assumed $\varphi_{hc}\gg\varphi_{f}$. Then we found
$\varphi_{hc}$ from that solution and substitute it in the equations
(63), (64) and (71) in order to find the values of these parameters
at the time of the horizon crossing. Our analysis shows that
although for all values of $0\leq\xi\leq\frac{1}{6}$, in the warped
DGP model with a quadratic potential in Jordan frame we have $0.966
\leq n_{s}\leq 1$ (so, the spectrum of the scalar perturbation is
nearly scale invariant and red-tilted), but just for $\frac{1}{8}$
we arrive at $r\approx0.22$\, which is observationally more reliable
\cite{Kom11}. In this case the value of $r$ at the time of horizon
crossing, decreases by decreasing $\xi$. Table~\ref{tab:table1}
shows the value of the $n_{s}$, $r$ and $\alpha$ when the physical
scales crossed the horizon for three different values of $\xi$. For
comparison we have listed also the corresponding recently realized
observational data. Note that the observational parameters are
defined at $k_{0}=0.002$ Mpc$^{-1}$ where $k_{0}$ denotes the value
of $k$ when universe scale crosses the Hubble horizon during
inflation. Also these parameters are obtained via WMAP+BAO+H$_{0}$
Mean data, where "Mean" refers to the mean of the posterior
distribution of each parameter. The quoted errors for $n_{s}$ show
the $68\%$ confidence levels (CL) (see \cite{Kom11} for details). As
the table shows, there is relatively good agreement between our
results and recent observation. But note that the running of the
spectral index in our setup is extra-ordinary close to zero. It is
negative and in this respect viable. $n_{s}$ and $r$ are in good
agreement with observation.

\subsection{Quartic Potential:
$V(\varphi)=\frac{b}{4}\varphi^{4}$}

The second potential we consider is the quartic potential i.e. the
case with $m=2$ in equation (73). The left panel of
figure~\ref{fig:6} shows the behavior of the correctional factor,
${\cal{A}}$ versus the scalar field in this case. In the large
scalar field regime, ${\cal{A}}$ increases by reduction of the
scalar field. So, in this situation $\epsilon$ increases and its
behavior mimics the behavior of $\epsilon$ in the standard 4D case
(see the right panel of figure~\ref{fig:6}). However, the growth of
${\cal{A}}$ by reduction of the scalar field stopes at some value of
the scalar field (which attains larger values for smaller $\xi$) and
then it decreases. Similarly, by reduction of the scalar field
$\epsilon$ reaches a maximum and its growth stopes. This maximum has
larger value for smaller $\xi$. By further reduction of the scalar
field, it deviates from the 4D behavior and decreases by reduction
of the scalar field strength. In contrast with the quadratic
potential where for small values of the scalar field the minimum of
both ${\cal{A}}$ and $\epsilon$ were located at $\varphi_{min}=0$,
here both ${\cal{A}}$ and $\epsilon$ have minimums located at some
non-vanishing values of the scalar field. In fact, for quartic
potential in this setup, $\epsilon$ has relatively more complicated
structure than the quadratic case in the small scalar field regime.
In the scale adopted in figure~\ref{fig:6}, this behavior is not so
evident, but it shall be more evident in figures of $n_{s}$ and
$n_{T}$ versus the scalar field (as we will see later). Note that as
$\xi$ increases, $\epsilon$ mimics the 4D behavior in a relatively
wider domain of $\varphi$ values.

In the next step, we consider the evolution of the correctional
factor ${\cal{B}}$ and the second slow-roll parameter $\eta$ versus
$\varphi$ as shown in figure~\ref{fig:7}. The general behavior of
${\cal{B}}$ and $\eta$ is similar to ${\cal{A}}$ and $\epsilon$.
But, since the minimum value of these parameters occurs at
$\varphi=0$, only in the large scalar field regime these parameters
evolve similar to the corresponding parameters in 4D case. It should
be noticed that for $\xi=\frac{1}{6}$ (the conformal coupling), the
correctional factor ${\cal{B}}$ is always less than unity. It means
that for this value of $\xi$, the value of $\eta$ in warped DGP
model is always smaller than the value of this parameter in 4D
model. We note also that in contrast with the quadratic potential
case, the slow-roll parameters can be negative in some values of the
scalar field.

In figures~\ref{fig:8} (the left panel) and~\ref{fig:9}, we have
shown the behavior of the scalar and tensor spectral index versus
the scalar field. As we expected from the evolution of $\epsilon$,
$n_{s}$ and $n_{T}$ at two extremal regimes of the scalar field
evolve as they do in the standard four-dimensional model. At these
two extremal regimes, $n_{s}$ and $n_{T}$ evolve from larger values
to the smaller values by reduction of the scalar field. For other
(intermediate) values of the scalar field, these parameters increase
as the scalar field decreases. The behavior of the running of the
scalar spectral index is shown in the right panel of
figure~\ref{fig:8}. In the large scalar field regime, $\alpha$
behaves as it does in 4D and decreases by reduction of the scalar
field. This 4D behavior lasts in a wider domain of the scalar field
for the larger values of $\xi$. But, at some value of the scalar
field, $\alpha$ reaches its minimum and then increases to a maximum.
After that, as scalar field decreases, there are other minimum and
maximum values for $\alpha$, providing a relatively complicated
structure relative to the quadratic potential case. This feature is
shown in the right panel of the figure~\ref{fig:9} by adopting a
smaller scale than the left panel one. Evidently there is a
different structure of $n_{T}$ relative to the quadratic potential
case where there was no minimum other than $\varphi_{min}=0$ in the
small field regime.

Next we consider the tensor-to-scalar ratio, $r$. The result of this
consideration is shown in figure~\ref{fig:10}. For quartic
potential, $r$ has more complicated behavior relative to the
quadratic potential case in the small scalar field regime similar to
the behavior of $\epsilon$, $n_{s}$ and $n_{T}$ in this regime. In
two extremal regimes of the scalar field (large and small scalar
field regimes), $r$ in the warped DGP model increases as the scalar
field decreases. This is the same as the behavior of $r$ in the
standard 4D case. For other (intermediate) values of the scalar
field, it decreases by reduction of the scalar field.

\begin{table*}
\caption{\label{tab:table2}The values of some inflation parameters
with a quartic potential in Jordan frame at the time that physical
scales crossed the horizon.}
\begin{ruledtabular}
\begin{tabular}{ccccc}
$\xi$&$n_{s}$&$r$&$\alpha$ \\ \hline\\
$0$& $0.9999999525$ &$3.979559439\times10^{-7}$&-6.534716905 $\times
10^{-58}$\\\\
$\frac{1}{12}$& $1.000000000$ &$1.745217481\times10^{-21}$&-1.238633640 $\times 10^{-51}$\\\\
$\frac{1}{8}$& $0.9999999999$ &$1.137780227\times10^{-21}$& -1.647520071 $\times 10^{-51}$\\\\
$\frac{1}{6}$& $0.9999999996$ &$9.156848134\times10^{-22}$& -8.485475908 $\times 10^{-51}$\\
\end{tabular}
\end{ruledtabular}
\end{table*}

Some inflation parameters calculated for quartic potential at the
time that physical scales have crossed the horizon are shown in
table~\ref{tab:table2}. Similar to the quadratic potential case, the
Friedmann equation and the number of e-folds are given via equations
(76) and (77) but now with quartic potential. The solution of
integral (77) with a quartic potential is presented in appendix {\bf
B}. By using that result and finding $\varphi_{hc}$ for this case,
we obtain the values of the scalar spectral index, its running and
the tensor to scalar ratio at the time of horizon crossing. The
results for three values of $\xi$ is shown in
table~\ref{tab:table2}. We see that although for different values of
$\xi$ the scalar spectral index is nearly scale invariant and
red-tilted, the running of the spectral index and the tensor to
scalar ratio increase by reduction of $\xi$. We note that the
corresponding observational data and the conditions for calculations
of these quantities are the same as what we have done for production
of table~\ref{tab:table2}.

Before presenting our analysis in the Einstein frame, we note that
in our previous numerical analysis we argued that since $H$ is
nearly constant in inflation epoch, the Ricci scalar
$R=6(\dot{H}+2H^{2})$ is also nearly constant in this epoch. With
this assumption, we have set $R=1$ in our numerical analysis. Now we
consider a more general case to have more generic results: we
consider the following ansatz for scale factor and scalar field
$$a(t)=a_{0}e^{\nu t},\,\quad \varphi=\varphi_{0}e^{-\vartheta t}$$
where $\nu$ and $\vartheta$ are positive constants. Note that these
ansatz are chosen by taking into account the inflationary nature of
the solutions for scale factor and a decreasing nature of the scalar
field. Applying these ansatz to equation (11) and performing our
numerical analysis for quadratic and quartic potentials with
$\nu=10$, $a_{0}=\varphi_{0}=1$ and $\vartheta=1$, we find for
$n_{s}$ the results that are shown in the left panel of figure
\ref{fig:5ab}. These results are more generic than the case that we
set the Ricci scalar to be a constant due to constancy of $H$ in
inflation era. Also, the right panel of figure ~\ref{fig:5ab} shows
the results of our numerical calculation of the tensor-to-scalar
ratio, $r$, for quadratic and quartic potentials adopting the above
ansatz. Comparison of these more general results with the
corresponding results obtained by a constant Ricci scalar shows that
the results obtained by assumption of a constant Ricci scalar are
actually reasonable in some sense. In fact, this comparison shows
that the assumption of a constant Ricci scalar due to constancy of
the Hubble parameter in inflation epoch is relatively a viable
assumption. We have checked also the situation with ansatz
$$a(t)=\bigg(t^{2}+\frac{t_{0}}{1-\nu}\bigg)^{\frac{1}{1-\nu}},\,\quad\quad
\varphi=\varphi_{0}t^{-\delta}$$ where we assume $\nu< 1$,\,
$t_{0}>0$ (see for instance \cite{Cai07}) and $\delta>0$. Although
this is not an exponentially solution of the scale factor, but the
previous argument is applicable more or less even with this ansatz
(for instance with $\nu=0.9$ and $\delta=3$). We note that the
general case without adopting ansatz is far more difficult to find
analytical or even numerical results.

\section{Inflation on the warped DGP brane in Einstein Frame}
Up to now, we have considered the situation in Jordan frame. We can
pass from Jordan to Einstein frame by making the following conformal
transformation [22,24,41]
\begin{equation}
\hat{q}_{\mu\nu}=\Omega^{2} q_{\mu\nu}\,,
\end{equation}
where the parameter $\Omega$ is defined as
\begin{equation}
\Omega^{2}=1+\kappa_{4}^{2}f(\varphi).
\end{equation}
Under this transformation, action (1) in Einstein frame becomes
\begin{widetext}
\begin{eqnarray}
S=\frac{1}{2\kappa_{5}^{2}}\int
d^{5}x\sqrt{-g^{(5)}}\Bigg[R^{(5)}-2\Lambda_{5}\Bigg]\hspace{12cm}\nonumber\\
-\int d^{4}x\sqrt{-\hat{q}}
\Bigg[\frac{1}{2\kappa_{4}^{2}}\,\hat{R}-\frac{3}{4}\left(\frac{\kappa_{4}^{2}f'(\varphi)}
{1+\kappa_{4}^{2}f(\varphi)}\right)^{2}\hat{q}^{\mu\nu}\partial_{\mu}\varphi\partial_{\nu}\varphi
-\frac{1}{2}\Omega^{-4}\hat{q}^{\mu\nu}\partial_{\mu}\varphi\partial_{\nu}\varphi-\Omega^{-4}\lambda
-\Omega^{-4}V(\varphi)\Bigg].\hspace{1cm}
\end{eqnarray}
\end{widetext}
Now, we define a new scalar field $\hat{\varphi}$ in
Einstein frame as follows
\begin{equation}
\frac{d\hat{\varphi}}{d\varphi}=\kappa_{4}^{-1}\sqrt{\frac{2(1+\kappa_{4}^{2}f(\varphi))+3\kappa_{4}^{2}f'^{2}(\varphi)}
{2(1+\kappa_{4}^{2}f(\varphi))^{2}}},
\end{equation}
and the corresponding potential $\hat{V}$ defined in Einstein frame
is
\begin{equation}
\hat{V}(\hat{\varphi})=\left[1+\kappa_{4}^{2}f(\varphi(\hat{\varphi}))\right]^{-2}V(\varphi(\hat{\varphi})).
\end{equation}
The general condition for flatness of the potential at the large
field limit is
\begin{equation}
\lim_{\varphi\rightarrow\infty}\frac{V}{f^{2}}=Const.>0.
\end{equation}
The condition $f(\varphi)\gg \kappa_{4}^{-2}$ for $\varphi \gg
\kappa_{4}^{-1}$ is required for the potential to be bounded from
below and the location of the global minimum is well localized
around the small field value. Even though the condition (83)
actually determines the flatness of the potential at the large field
limit, it is not necessarily required in generic inflation models.
Depending on the shape of the potential, it might still be possible
to have sufficient time of exponential expansion for some finite
region of field value, $\varphi$ \cite{Par08}.

The generalized cosmological dynamics of this setup in Einstein
frame is given by the following Friedmann equation
\begin{eqnarray}
\hat{H}^{2}=\frac{\kappa_{4}^{2}}{3}\rho_{\hat{\varphi}}+\frac{\kappa_{4}^{2}}{3}\hat{\lambda}
+\frac{2\kappa_{4}^{4}}{\kappa_{5}^{4}}\hspace{4cm}\nonumber\\
\pm\frac{2\kappa_{4}^{2}}{\kappa_{5}^{2}}\,\sqrt{\frac{\kappa_{4}^{4}}{\kappa_{5}^{4}}+
\frac{\kappa_{4}^{2}}{3}\rho_{\hat{\varphi}}+
\frac{\kappa_{4}^{2}}{3}\hat{\lambda}-\frac{\Lambda_{5}}{6}-\frac{\mathcal{C}}{\hat{a}^{4}}}.\hspace{0.5cm}
\end{eqnarray}
where\,  $\hat{}$ \, refers to parameters written in Einstein frame.
In Friedmann equation (84) we defined
$\hat{\lambda}=\frac{1}{(1+\kappa_{4}^{2}f(\varphi))^{2}}\lambda$
and $\hat{a}=(1+\kappa_{4}^{2}f(\varphi))^{1/2}a$. Also,
$\rho_{\hat{\varphi}}$ the energy-density corresponding to the now
minimally coupled scalar field in Einstein frame is defined as
follows
\begin{equation}
\rho_{\hat{\varphi}}=\frac{1}{2}\left(\frac{d
\hat{\varphi}}{d\hat{t}}\right)^{2}+\hat{V}(\hat{\varphi}),
\end{equation}
and the corresponding pressure is given by
\begin{equation}
p_{\hat{\varphi}}=\frac{1}{2}\left(\frac{d
\hat{\varphi}}{d\hat{t}}\right)^{2}-\hat{V}(\hat{\varphi}).
\end{equation}
where $\hat{t}=(1+\kappa_{4}^{2}f(\varphi))^{1/2}t$ .

In this step, similar to the Jordan frame case, we introduce the
effective cosmological constant on the brane in Einstein frame as
follows
\begin{equation}
\hat{\Lambda}_{eff}=\kappa_{4}^{2}\hat{\lambda}+\frac{6\kappa_{4}^{4}}{\kappa_{5}^{4}}
\pm\frac{\sqrt{6}\kappa_{4}^{4}}{\kappa_{5}^{4}}
\sqrt{(2\kappa_{4}^{2}\hat{\lambda}-\Lambda_{5})\frac{\kappa_{5}^{4}}{\kappa_{4}^{4}}+6}\,.
\end{equation}
By putting the effective cosmological constant equal to zero, we
find
\begin{equation}
\Lambda_{5}=-\frac{\kappa_{5}^{4}}{6}\hat{\lambda}^{2}.
\end{equation}
So, we can rewrite Friedmann equation (84) as follows
\begin{eqnarray}
\hat{H}^{2}=\frac{\kappa_{4}^{2}}{3}\rho_{\hat{\varphi}}+\frac{\kappa_{4}^{2}}{3}\hat{\lambda}
+\frac{2\kappa_{4}^{4}}{\kappa_{5}^{4}}\hspace{4cm}\nonumber\\
\pm\frac{2\kappa_{4}^{2}}{\kappa_{5}^{2}}\,\sqrt{\frac{\kappa_{4}^{4}}{\kappa_{5}^{4}}+
\frac{\kappa_{4}^{2}}{3}\rho_{\hat{\varphi}}+
\frac{\kappa_{4}^{2}}{3}\hat{\lambda}-\frac{\kappa_{5}^{4}}{36}\hat{\lambda}^{2}-\frac{\mathcal{C}}{\hat{a}^{4}}}\,,\hspace{0.5cm}
\end{eqnarray}
and the second Friedmann equation can be expressed as
\begin{equation}
\frac{d\hat{H}}{d\hat{t}}=\frac{\kappa_{4}^{2}}{6\hat{H}}\frac{d\rho_{\hat{\varphi}}}{d\hat{t}}
\pm\frac{\kappa_{4}^{2}}{\kappa_{5}^{2}}\frac{\frac{\kappa_{4}^{2}}{6\hat{H}}\frac{d\rho_{\hat{\varphi}}}{d\hat{t}}
+\frac{\mathcal{C}}{\hat{a}^{4}}}{\sqrt{\frac{\kappa_{4}^{4}}{\kappa_{5}^{4}}+
\frac{\kappa_{4}^{2}}{3}\rho_{\hat{\varphi}}+
\frac{\kappa_{4}^{2}}{3}\hat{\lambda}-\frac{\kappa_{5}^{4}}{36}\hat{\lambda}^{2}-\frac{\mathcal{C}}{\hat{a}^{4}}}}\,\,.
\end{equation}
The equation of motion of the scalar field in Einstein frame now is
given by
\begin{equation}
\frac{d^{2}\hat{\varphi}}{d
\hat{t}^{2}}+3\hat{H}\frac{d\hat{\varphi}}{d\hat{t}}+\frac{d\hat{V}}{d\hat{\varphi}}=0\,.
\end{equation}
In the slow-roll approximation where
$\left(\frac{d\hat{\varphi}}{d\hat{t}}\right)^{2}\ll
\hat{V}(\hat{\varphi})$ and
$\frac{d^{2}\hat{\varphi}}{d\hat{t}^{2}}\ll|3\hat{H}\frac{d\hat{\varphi}}{d\hat{t}}|$,
energy density and equation of motion for the scalar field take the
following forms respectively
\begin{equation}
\hat{\rho}\approx \hat{V}(\hat{\varphi})\,,
\end{equation}
\begin{equation}
3\hat{H}\frac{d\hat{\varphi}}{d\hat{t}}+\frac{d\hat{V}}{d\hat{\varphi}}=0\,.
\end{equation}
Now the Friedmann equation can be expressed as follows
\begin{eqnarray}
\hat{H}^{2}=\frac{\kappa_{4}^{2}}{3}\hat{V}+\frac{\kappa_{4}^{2}}{3}\hat{\lambda}
+\frac{2\kappa_{4}^{4}}{\kappa_{5}^{4}}\hspace{4cm}\nonumber\\
\pm\frac{2\kappa_{4}^{2}}{\kappa_{5}^{2}}\,\sqrt{\frac{\kappa_{4}^{4}}{\kappa_{5}^{4}}+
\frac{\kappa_{4}^{2}}{3}\hat{V}+
\frac{\kappa_{4}^{2}}{3}\hat{\lambda}-\frac{\kappa_{5}^{4}}{36}\hat{\lambda}^{2}
-\frac{\mathcal{C}}{\hat{a}^{4}}}\,\,.\hspace{0.5cm}
\end{eqnarray}
We define the slow-roll parameters in Einstein frame as
\begin{equation}
\hat{\epsilon}\equiv
-\frac{1}{\hat{H}^{2}}\frac{d\hat{H}}{d\hat{t}},
\end{equation}
\begin{equation}
\hat{\eta}\equiv
-\frac{1}{\hat{H}}\frac{(d^{2}\hat{H})/(d\hat{t}^{2})}{(d\hat{H})/(d\hat{t})}\,.
\end{equation}
In the slow-roll approximation, from Eq. (94) we find\\\\

\begin{widetext}
\begin{equation}
\hat{\epsilon}=\frac{1}{2\kappa_{4}^{2}}\Bigg(\frac{(d\hat{V})/(d\hat{\varphi})}{\hat{V}}\Bigg)^{2}\left[\frac{1
\pm\frac{\kappa_{4}^{2}}{\kappa_{5}^{2}}\bigg(1-\frac{36\hat{H}^{2}}{\kappa_{4}^{2}\left(\frac{d\hat{V}}{d\hat{\varphi}}\right)^{2}}
\frac{\mathcal{C}}{\hat{a}^{4}}\bigg)\bigg(\sqrt
{\frac{\kappa_{4}^{4}}{\kappa_{5}^{4}}+
\frac{\kappa_{4}^{2}}{3}\hat{V}+
\frac{\kappa_{4}^{2}}{3}\hat{\lambda}-\frac{\kappa_{5}^{4}}{36}\hat{\lambda}^{2}
-\frac{\mathcal{C}}{\hat{a}^{4}}}\,\,\bigg)^{-\frac{1}{2}} }{\Bigg(1
+\frac{1}{\hat{V}}\bigg(\hat{\lambda}+\frac{6\kappa_{4}^{2}}{\kappa_{5}^{4}}
\pm\frac{6}{\kappa_{5}^{2}}\,\sqrt{\frac{\kappa_{4}^{4}}{\kappa_{5}^{4}}+
\frac{\kappa_{4}^{2}}{3}\hat{V}+
\frac{\kappa_{4}^{2}}{3}\hat{\lambda}-\frac{\kappa_{5}^{4}}{36}\hat{\lambda}^{2}
-\frac{\mathcal{C}}{\hat{a}^{4}}}\bigg)\Bigg)^{2}}\right],
\end{equation}
\end{widetext}
and
\begin{widetext}
\begin{equation}
\hat{\eta}\simeq\frac{1}{\kappa_{4}^{2}}\Bigg(\frac{(d^{2}\hat{V})/(d\hat{\varphi}^{2})}{\hat{V}}\Bigg)\left[1
+\frac{1}{\hat{V}}\Bigg(\hat{\lambda}+\frac{6\kappa_{4}^{2}}{\kappa_{5}^{4}}
\pm\frac{6}{\kappa_{5}^{2}}\,\sqrt{\frac{\kappa_{4}^{4}}{\kappa_{5}^{4}}+
\frac{\kappa_{4}^{2}}{3}\hat{V}+
\frac{\kappa_{4}^{2}}{3}\hat{\lambda}-\frac{\kappa_{5}^{4}}{36}\hat{\lambda}^{2}
-\frac{\mathcal{C}}{\hat{a}^{4}}}\Bigg)\right]^{-1}.
\end{equation}
\end{widetext}

In equations (97) and (98), the terms in the brackets are
corrections to the standard 4-dimensional model. These corrections
are contributions originating from braneworld nature of the setup.

For warped DGP model with non-minimally coupled scalar field on the
brane, the number of e-folds in Einstein frame becomes
\begin{widetext}
\begin{eqnarray}
\hat{N}=-\int_{\hat{\varphi}_{hc}}^{\hat{\varphi}_{f}}\Bigg(\frac{d\hat{\varphi}}{d\varphi}\Bigg)^{2}
\Bigg(\frac{3\hat{V}}{d\hat{V}/d\varphi}\Bigg)\hspace{12cm}\nonumber\\
\times\Bigg[\frac{\kappa_{4}^{2}}{3}
+\frac{1}{\hat{V}}\bigg(\frac{\kappa_{4}^{2}}{3}\hat{\lambda}+\frac{2\kappa_{4}^{4}}{\kappa_{5}^{4}}
\pm\frac{2\kappa_{4}^{2}}{\kappa_{5}^{2}}\,\sqrt{\frac{\kappa_{4}^{4}}{\kappa_{5}^{4}}+
\frac{\kappa_{4}^{2}}{3}\hat{V}+
\frac{\kappa_{4}^{2}}{3}\hat{\lambda}-\frac{\kappa_{5}^{4}}{36}\hat{\lambda}^{2}
-\frac{\mathcal{C}}{\hat{a}^{4}}}\,\,\bigg)\Bigg]d\varphi\,\,.\hspace{1cm}
\end{eqnarray}
\end{widetext}

In the next section we consider the scalar perturbation of the
metric in Einstein frame.

\section{Perturbations in Einstein Frame}

The effective covariant equations on the brane in a warped DGP
braneworld scenario and in Einstein frame are given by
\begin{equation}
\hat{G}_{\mu\nu}=\kappa_{5}^{4}\hat{\Pi}_{\mu\nu}-\hat{E}_{\mu\nu},
\end{equation}
where
\begin{equation}
\hat{\Pi}_{\mu\nu}=-\frac{1}{4}{\hat{\tau}}_{\mu\sigma}{\hat{\tau}}_{\nu}^{\sigma}+
\frac{1}{12}{\hat{\tau}}{\hat{\tau}}_{\mu\nu}+
\frac{1}{8}\hat{q}_{\mu\nu}\Big({\hat{\tau}}_{\rho\sigma}{\hat{\tau}}^{\rho\sigma}
-\frac{1}{3}{\hat{\tau}}^{2}\Big),
\end{equation}
and $\hat{\tau}_{\mu\nu}$ is the total stress-tensor on the brane
and is defined as
\begin{equation}
\hat{\tau}_{\,\,\nu}^{\mu}=-\kappa_{4}^{2}\hat{G}_{\,\,\nu}^{\mu}
-\hat{\lambda}\delta_{\,\,\nu}^{\mu}+\hat{T}_{\,\,\nu}^{\mu}\,.
\end{equation}
$\hat{T}_{\mu\nu}$, the energy-momentum tensor of the scalar field
in Einstein frame which now is minimally coupled to the induced
gravity on the brane, is given by (compare this result with
corresponding equation in Jordan frame, Eq. (25))
\begin{equation}
\hat{T}_{\mu\nu}=\partial_\mu \hat{\varphi}\,\partial_\nu
\hat{\varphi}-\hat{q}_{\mu \nu}\Big(\frac{1}{2}\hat{q}^{\alpha\beta}
\partial_{\alpha}\hat{\varphi}\,\partial_{\beta}\hat{\varphi}+\hat{V}(\hat{\varphi})\Big).
\end{equation}
Also we have
\begin{equation}
\hat{E}_{\mu\nu}=C_{MRS}^{\quad\quad N}\,\, n^{M}\,\,n_{N}\,\,
{\hat{q}^{R}}_{\mu}\,\,{\hat{q}^{S}}_{\nu}=\Omega^{4}E_{\mu\nu}
\end{equation}

Since in Einstein frame $d\hat{s}^{2}=\Omega^{2}ds^{2}$, the scalar
metric perturbations of the FRW background (Eq. (27)) is translated
to
\begin{equation}
d\hat{s}^{2}=-\big(1+2\hat{\Phi}\big)d\hat{t}^{2}+\hat{a}^{2}(\hat{t})\big(1-2\hat{\Psi}\big)\delta_{i\,j}\,dx^{i}dx^{j}.
\end{equation}
where $\hat{a}(\hat{t})$ is the scale factor on the brane in
Einstein frame, $\hat{\Phi}=\hat{\Phi}(\hat{t},x)$ and
$\hat{\Psi}=\hat{\Psi}(\hat{t},x)$ are the metric perturbations. For
the above perturbed metric one can obtain the temporal part of the
perturbed field equations in Einstein frame:
\begin{equation}
-3\hat{H}(\hat{H}\hat{\Phi}+\frac{d\hat{\Psi}}{d\hat{t}})-\frac{\hat{k}^{2}}{\hat{a}^{2}}=\frac{\kappa_{4}^{2}}{2}\delta
\hat{\rho}_{eff}
\end{equation}
\begin{eqnarray}
\frac{d^{2}\hat{\Psi}}{d\hat{t}^{2}}+3\hat{H}(\hat{H}\hat{\Phi}+\frac{d\hat{\Psi}}{d\hat{t}})+\hat{H}\frac{d\Phi}{d\hat{t}}
+2\frac{d\hat{H}}{d\hat{t}}\hat{\Phi}\hspace{2cm}\nonumber\\
+\frac{1}{3\hat{a}^{2}}\hat{k}^{2}(\hat{\Phi}-\hat{\Psi})=
\frac{\kappa_{4}^{2}}{2}\delta \hat{p}_{eff}\,,\hspace{0.8cm}
\end{eqnarray}
\begin{equation}
\frac{d\hat{\Psi}}{d\hat{t}}+\hat{H}\hat{\Phi}=
\frac{\kappa_{4}^{2}}{2}\Big(\frac{\kappa_{5}^{4}\rho_{\hat{\varphi}}}{6\kappa_{4}^{2}}
\frac{d\hat{\varphi}}{d\hat{t}}\delta \hat{\varphi}
\Big)+\frac{1}{2}\int(\delta \hat{E}_{i}^{0})\,d x_{i}\,,
\end{equation}
\begin{equation}
\hat{\Psi}-\hat{\Phi}=8\pi
G\frac{\kappa_{4}^{2}\hat{H}}{\kappa_{5}^{2}(\frac{d\hat{H}}{d\hat{t}}+2\hat{H}^{2})-\hat{H}}
\hat{a}^{2}\delta{\hat{\pi}_{\hat{E}}}\,.
\end{equation}
In the last equation, $\delta{\hat{\pi}_{\hat{E}}}$ is anisotropic
stress perturbation in the Einstein frame. In Eqs. (106) and (107),
$\hat{\rho}_{eff}$ and $\hat{p}_{eff}$ can be obtained from the
standard Friedmann equation
$\hat{H}^{2}=\frac{\kappa_{4}^{2}}{3}\hat{\rho}_{eff}$, as follows
\begin{eqnarray}
\hat{\rho}_{eff}=\rho_{\hat{\varphi}}+\hat{\lambda}+\frac{6\kappa_{4}^{2}}{\kappa_{5}^{4}}\hspace{4.9cm}\nonumber\\
\pm\frac{6}{\kappa_{5}^{2}}\,\sqrt{\frac{\kappa_{4}^{4}}{\kappa_{5}^{4}}+
\frac{\kappa_{4}^{2}}{3}\rho_{\hat{\varphi}}+
\frac{\kappa_{4}^{2}}{3}\hat{\lambda}-\frac{\kappa_{5}^{4}}{36}\hat{\lambda}^{2}-\frac{\mathcal{C}}{\hat{a}^{4}}}.\hspace{0.8cm}
\end{eqnarray}
By using the continuity equation,
$\frac{d}{d\hat{t}}\hat{\rho}_{eff}+3\hat{H}(\hat{\rho}_{eff}+\hat{p}_{eff})=0$,
one can deduce
\begin{eqnarray}
\hat{p}_{eff}=p_{\hat{\varphi}}\pm\frac{\kappa_{4}^{2}}{\kappa_{5}^{2}}\frac{\rho_{\hat{\varphi}}+p_{\hat{\varphi}}
-\frac{4}{\kappa_{4}^{2}}\frac{\mathcal{C}}{\hat{a}^4}}{\sqrt{\frac{\kappa_{4}^{4}}{\kappa_{5}^{4}}+
\frac{\kappa_{4}^{2}}{3}\rho_{\hat{\varphi}}+
\frac{\kappa_{4}^{2}}{3}\hat{\lambda}-\frac{\kappa_{5}^{4}}{36}\hat{\lambda}^{2}
-\frac{\mathcal{C}}{\hat{a}^{4}}}}-\hat{\lambda}\hspace{0.2cm}\nonumber\\
-\frac{6\kappa_{4}^{2}}{\kappa_{5}^{4}}
\mp\frac{6}{\kappa_{5}^{2}}\,\sqrt{\frac{\kappa_{4}^{4}}{\kappa_{5}^{4}}+
\frac{\kappa_{4}^{2}}{3}\rho_{\hat{\varphi}}+
\frac{\kappa_{4}^{2}}{3}\hat{\lambda}-\frac{\kappa_{5}^{4}}{36}\hat{\lambda}^{2}-\frac{\mathcal{C}}{\hat{a}^{4}}}\,.\hspace{0.7cm}
\end{eqnarray}
So, the perturbed effective density and pressure in Einstein frame
can be written as
\begin{equation}
\delta\hat{\rho}_{eff}=\delta\rho_{\hat{\varphi}}\pm\frac{\kappa_{4}^{2}}{\kappa_{5}^{2}}\,\frac{\delta\rho_{\hat{\varphi}}
-\frac{3}{\kappa_{4}^{2}}\delta
\hat{E}_{0}^{0}}{\sqrt{\frac{\kappa_{4}^{4}}{\kappa_{5}^{4}}+
\frac{\kappa_{4}^{2}}{3}\rho_{\hat{\varphi}}+
\frac{\kappa_{4}^{2}}{3}\hat{\lambda}-\frac{\kappa_{5}^{4}}{36}\hat{\lambda}^{2}-\hat{E}_{0}^{0}}}.
\end{equation}
and
\begin{eqnarray}
\delta \hat{p}_{eff}=\delta p_{\hat{\varphi}}
\pm\frac{6}{\kappa_{5}^{2}}\,\frac{\delta p_{\hat{\varphi}}
-\frac{1}{\kappa_{4}^{2}}\delta
\hat{E}_{0}^{0}}{\sqrt{\frac{\kappa_{4}^{4}}{\kappa_{5}^{4}}+
\frac{\kappa_{4}^{2}}{3}\rho_{\hat{\varphi}}+
\frac{\kappa_{4}^{2}}{3}\hat{\lambda}-\frac{\kappa_{5}^{4}}{36}\hat{\lambda}^{2}-\hat{E}_{0}^{0}}}
\hspace{0.2cm}\nonumber\\
\mp \frac{\kappa_{4}^{4}}{6\kappa_{5}^{2}}\frac{\Big(\delta
\rho_{\hat{\varphi}}-\frac{1}{\kappa_{4}^{2}}\delta
\hat{E}_{0}^{0}\Big)\Big(\rho_{\hat{\varphi}}+\hat{p}-\frac{4}{\kappa_{4}^{2}}\hat{E}_{0}^{0}\Big)}
{\Big[\frac{\kappa_{4}^{4}}{\kappa_{5}^{4}}+
\frac{\kappa_{4}^{2}}{3}\rho_{\hat{\varphi}}+
\frac{\kappa_{4}^{2}}{3}\hat{\lambda}-\frac{\kappa_{5}^{4}}{36}\hat{\lambda}^{2}
-\hat{E}_{0}^{0}\Big]^{3/2}}\,,\hspace{0.7cm}
\end{eqnarray}
where $\delta\hat{E}_{0}^{0}$ can be calculated from the following
relation
\begin{equation}
\delta \hat{E}_{\,\nu}^{\mu}=-\kappa_{4}^{2}\left(%
\begin{array}{cc}
  -\delta\hat{\rho}_{\hat{E}} & a \delta \hat{q}_{\hat{E}}\\
  \hat{a}^{-1} \delta \hat{q}_{\hat{E}} & \frac{1}{3}\delta\hat{\rho}_{\hat{E}}\delta^{i}_{\,j}
  +(\delta\hat{\pi}_{\,\hat{E}})^{i}_{j}
  \\
\end{array}%
\right)\,,
\end{equation}
that is written in Einstein frame. Also $\delta\rho_{\hat{\varphi}}$
and $\delta p_{\hat{\varphi}}$ take the following forms
\begin{equation}
\delta\rho_{\hat{\varphi}}=\frac{d\hat{\varphi}}{d\hat{t}}\,\,\delta\left(\frac{d\hat{\varphi}}{d\hat{t}}\right)
-\left(\frac{d\hat{\varphi}}{d\hat{t}}\right)^{2}\hat{\Phi}+
\frac{d\hat{V}}{d\hat{\varphi}}\delta \hat{\varphi},
\end{equation}
\begin{equation}
\delta
p_{\hat{\varphi}}=\frac{d\hat{\varphi}}{d\hat{t}}\,\,\delta\left(\frac{d\hat{\varphi}}{d\hat{t}}\right)
-\left(\frac{d\hat{\varphi}}{d\hat{t}}\right)^{2}\hat{\Phi}-
\frac{d\hat{V}}{d\hat{\varphi}}\delta \hat{\varphi}.
\end{equation}
These equations in the slow-roll regime reduce to
$\delta\rho_{\hat{\varphi}}=\frac{d\hat{V}}{d\hat{\varphi}}\delta
\hat{\varphi}$ and $\delta p_{\hat{\varphi}}=-
\frac{d\hat{V}}{d\hat{\varphi}}\delta \hat{\varphi}$\, respectively.
By perturbing the equation of motion of the scalar field (91) one
can find
\begin{eqnarray}
\delta\frac{d^{2}\hat{\varphi}}{d\hat{t}^{2}}+3\hat{H}\delta\left(\frac{d\hat{\varphi}}{d\hat{t}}\right)
+\left(\frac{d^{2}\hat{V}}{d\hat{\varphi}^{2}}
+\frac{\hat{k}^{2}}{\hat{a}^{2}}\right)\delta \hat{\varphi}\hspace{2cm}\nonumber\\
=\frac{d\hat{\varphi}}{d\hat{t}}\left(3\frac{d\hat{\Psi}}{d\hat{t}}+\frac{d\hat{\Phi}}{d\hat{t}}\right)
+\hat{\Phi}\left(-2\frac{d\hat{V}}{d\hat{\varphi}}\right).\hspace{0.9cm}
\end{eqnarray}
In Einstein frame and within the warped DGP model, we should
redefine equation (43) as
\begin{equation}
\hat{\zeta}=\hat{\Psi}-\frac{\hat{H}}{(d\hat{\rho}_{eff})/(d\hat{t})}\delta\hat{\rho}_{eff}.
\end{equation}
where $\hat{\Psi}$ is an Einstein frame quantity. Now, by using the
energy-conservation equation for linear perturbations,
\begin{equation} \frac{d}{d\hat{t}}\delta
\hat{\rho}_{eff}+3\hat{H}(\delta \hat{\rho}_{eff}+\delta
\hat{p}_{eff})+3(\hat{\rho}_{eff}+\hat{p}_{eff})\frac{d\hat{\Psi}}{d\hat{t}}=0,
\end{equation}
we find the variation of $\hat{\zeta}$ with respect to the conformal
time as
\begin{equation}
\frac{d\hat{\zeta}}{d\hat{t}}=\frac{d\hat{\Psi}}{d\hat{t}}+\frac{\frac{d}{d\hat{t}}\delta
\hat{\rho}_{eff}}{3(\hat{\rho}_{eff}+\hat{p}_{eff})}
-\frac{\frac{d}{d\hat{t}}(\hat{\rho}_{eff}+\hat{p}_{eff})}{(\hat{\rho}_{eff}+\hat{p}_{eff})^{2}}\delta
\hat{\rho}_{eff},
\end{equation}
where $\frac{d\hat{\rho}_{eff}}{d\hat{t}}$ and $\frac{d
\hat{p}_{eff}}{d\hat{t}}$ are given by time derivatives of equations
(110) and (111) respectively.

Similar to the Jordan frame case, we split the pressure
perturbations into adiabatic and entropic parts as follows
\begin{equation}
\delta
\hat{p}_{eff}=c_{s}^{2}\delta\hat{\rho}_{eff}+\frac{d\hat{p}_{eff}}{d\hat{t}}\hat{\Gamma}\,.
\end{equation}
The non-adiabatic part is $\delta
p_{nad}=\frac{d\hat{p}_{eff}}{d\hat{t}}\hat{\Gamma}$\,, where
$\hat{\Gamma}$ is defined as
\begin{equation}
\hat{\Gamma}=\frac{\delta
\hat{p}_{eff}}{(d\hat{p}_{eff})/(d\hat{t})}-\frac{\delta\hat{\rho}_{eff}}{(d\hat{\rho}_{eff})/(d\hat{t})}.
\end{equation}
From Equations (112)-(116) we can deduce
\begin{widetext}
\begin{eqnarray}
\delta
\hat{p}_{nad}=\Bigg(1-\hat{c}_{s}^{2}\Bigg)\delta\hat{\rho}_{eff}-\Bigg(2\frac{d\hat{V}}{d\hat{\varphi}}\delta\hat{\varphi}
\Bigg)\Bigg(1\pm\frac{\kappa_{4}^{2}}{\kappa_{5}^{2}}
\frac{1}{\sqrt{\frac{\kappa_{4}^{4}}{\kappa_{5}^{4}}+
\frac{\kappa_{4}^{2}}{3}\rho_{\hat{\varphi}}+
\frac{\kappa_{4}^{2}}{3}\hat{\lambda}-\frac{\kappa_{5}^{4}}{36}\hat{\lambda}^{2}-\hat{E}_{0}^{0}}}\Bigg)\hspace{5cm}\nonumber\\
\mp\frac{2\kappa_{4}^{2}}{\kappa_{5}^{2}}
\frac{\frac{1}{\kappa_{4}^{2}}\delta
\hat{E}_{0}^{0}}{\sqrt{\frac{\kappa_{4}^{4}}{\kappa_{5}^{4}}+
\frac{\kappa_{4}^{2}}{3}\rho_{\hat{\varphi}}+
\frac{\kappa_{4}^{2}}{3}\hat{\lambda}-\frac{\kappa_{5}^{4}}{36}\hat{\lambda}^{2}-\hat{E}_{0}^{0}}}
\mp \frac{\kappa_{4}^{4}}{6\kappa_{5}^{2}}\frac{\Big(\delta
\rho_{\hat{\varphi}}-\frac{1}{\kappa_{4}^{2}}\delta
\hat{E}_{0}^{0}\Big)\Big(\rho_{\hat{\varphi}}+p_{\hat{\varphi}}-\frac{4}{\kappa_{4}^{2}}\hat{E}_{0}^{0}\Big)}
{\Big[\frac{\kappa_{4}^{4}}{\kappa_{5}^{4}}+
\frac{\kappa_{4}^{2}}{3}\rho_{\hat{\varphi}}+
\frac{\kappa_{4}^{2}}{3}\hat{\lambda}-\frac{\kappa_{5}^{4}}{36}\hat{\lambda}^{2}
-\hat{E}_{0}^{0}\Big]^{3/2}}\,.\hspace{1cm}
\end{eqnarray}
\end{widetext}
Using equations (106)-(108) we rewrite this relation
as follows
\begin{widetext}
\begin{eqnarray}
\delta
\hat{p}_{nad}=-\frac{6}{\kappa_{4}^{2}}\Bigg(1-\hat{c}_{s}^{2}-\hat{{\cal{J}}}\Bigg)\frac{\hat{k}}{\hat{a}^{2}}\hat{\Psi}
-\frac{6}{\kappa_{4}^{2}}\hat{{\cal{K}}}\Bigg(\hat{H}\hat{\Phi}+\frac{d\hat{\Psi}}{d\hat{t}}\Bigg)
+\frac{3}{\kappa_{4}^{2}}\hat{{\cal{J}}}\delta
\hat{E}_{0}^{0}+\frac{2}{\kappa_{4}^{2}}\delta
\hat{E}_{0}^{0}\Bigg(\hat{{\cal{I}}}-1\Bigg)+\Bigg(\frac{6\kappa_{4}^{2}}{\kappa_{5}^{2}}\frac{\frac{d\hat{V}}{d\varphi}}
{\rho_{\hat{\varphi}}\frac{d\hat{\varphi}}{d\hat{t}}}\int\delta
\hat{E}_{i}^{0}\,dx^{i}\Bigg)\hat{{\cal{I}}},\hspace{0.8cm}
\end{eqnarray}
\end{widetext}
where $\hat{{\cal{K}}}$\,,$\hat{{\cal{J}}}$ and $\hat{{\cal{I}}}$
are defined as
\begin{widetext}
\begin{eqnarray}
\hat{{\cal{K}}}=\frac{-3\bigg(2\frac{d\hat{V}}{d\hat{\varphi}}\frac{d\hat{\varphi}}{d\hat{t}}
\bigg)\hat{{\cal{I}}}
+\frac{24}{\kappa_{4}^{2}}\hat{E}_{0}^{0}\hat{H}\bigg(\hat{{\cal{I}}}-1\bigg)
-3\hat{{\cal{J}}}\hat{{\cal{I}}}\bigg(\frac{d\hat{\varphi}}{d\hat{t}}\frac{d^{2}\hat{\varphi}}{d\hat{t}^{2}}
+\frac{d\hat{V}}{d\varphi}\frac{d\hat{\varphi}}{d\hat{t}}+\frac{12}{\kappa_{4}^{2}}\hat{E}_{0}^{0}\hat{H}\bigg)}
{3\Big(\frac{d\hat{\varphi}}{d\hat{t}}\Big)^{2}+\bigg(3\Big(\frac{d\hat{\varphi}}{d\hat{t}}\Big)^{2}
-\frac{12}{\kappa_{4}^{2}}\hat{E}_{0}^{0}\bigg)\bigg(\hat{{\cal{I}}}-1\bigg)}
+\frac{6\kappa_{4}^{2}}{\kappa_{5}^{2}}\frac{\frac{d\hat{V}}{d\hat{\varphi}}}
{\rho_{\hat{\varphi}}\frac{d\hat{\varphi}}{d\hat{t}}}\hat{{\cal{I}}}-3\hat{H}\hat{{\cal{J}}}\,,
\end{eqnarray}
\end{widetext}
\begin{equation}
\hat{{\cal{J}}}=\frac{\kappa_{4}^{4}}{6\kappa_{5}^{4}}
\frac{\Big(\rho_{\hat{\varphi}}+p_{\hat{\varphi}}-\frac{4}{\kappa_{4}^{2}}\hat{E}_{0}^{0}\Big)}
{\bigg(\frac{\kappa_{4}^{4}}{\kappa_{5}^{4}}+
\frac{\kappa_{4}^{2}}{3}\rho_{\hat{\varphi}}+
\frac{\kappa_{4}^{2}}{3}\hat{\lambda}-\frac{\kappa_{5}^{4}}{36}\hat{\lambda}^{2}-\hat{E}_{0}^{0}\bigg)^{3/2}\,\hat{{\cal{I}}}}\,,
\end{equation}
and
\begin{equation}
\hat{{\cal{I}}}=\Bigg(1\pm\frac{\kappa_{4}^{2}}{\kappa_{5}^{2}}
\frac{1}{\sqrt{\frac{\kappa_{4}^{4}}{\kappa_{5}^{4}}+
\frac{\kappa_{4}^{2}}{3}\rho_{\hat{\varphi}}+
\frac{\kappa_{4}^{2}}{3}\hat{\lambda}-\frac{\kappa_{5}^{4}}{36}\hat{\lambda}^{2}-\hat{E}_{0}^{0}}}\Bigg)\,,
\end{equation}
respectively. Now we rewrite the equation of the variation of
$\hat{\zeta}$ versus time in terms of the model's parameters. From
equations (119)-(121) we find
\begin{eqnarray}
\frac{d\hat{\zeta}}{d\hat{t}}=\frac{\kappa_{4}^{2}\hat{\rho}_{eff}}{9\hat{H}(\hat{\rho}_{eff}+\hat{p}_{eff})}
\Bigg(\frac{6\kappa_{4}^{2}}{\kappa_{5}^{2}}\frac{\frac{d\hat{V}}{d\hat{\varphi}}}{\rho_{\hat{\varphi}}
\frac{d\hat{\varphi}}{d\hat{t}}}\int\delta\hat{E}_{i}^{0}\,dx^{i}\Bigg)\hat{{\cal{I}}}\hspace{0.9cm}\nonumber\\
+\frac{\hat{\rho}_{eff}\delta
\hat{E}_{0}^{0}}{3\hat{H}(\hat{\rho}_{eff}+\hat{p}_{eff})}\Bigg(\hat{{\cal{J}}}
+\frac{2}{3}(\hat{{\cal{I}}}-1)\Bigg)\hspace{1cm}\nonumber\\
-\frac{2(\hat{H}\hat{\Phi}+\frac{d\hat{\Psi}}{d\hat{t}})}
{3\hat{H}(\hat{\rho}_{eff}+\hat{p}_{eff})}\hat{\rho}_{eff}\hat{{\cal{K}}}.\hspace{1cm}
\end{eqnarray}

Here we are going to obtain scalar and tensorial perturbation in our
model. First let's rewrite equation (117) in the slow-roll
approximation at the large scales as follows
\begin{equation}
3\hat{H}\delta\Big(\frac{d\hat{\varphi}}{d\hat{t}}\Big)+\frac{d^{2}\hat{V}}{d\hat{\varphi}^{2}}\delta\hat{\varphi}\simeq
-2\hat{\Phi}\frac{d\hat{V}}{d\hat{\varphi}}.
\end{equation}
Also for equation (108) we have
\begin{equation}
\hat{\Phi}\simeq
\frac{\frac{\kappa_{5}^{2}}{6}\rho_{\hat{\varphi}}+\frac{1}{\frac{d\hat{\varphi}}{d\hat{t}}\delta\hat{\varphi}}\int\delta
\hat{E}_{i}^{0}\,dx^{i}}{6\hat{H}\bigg(1+\frac{\kappa_{5}^{2}}{18}\rho_{\hat{\varphi}}\bigg)}
\frac{d\hat{\varphi}}{d\hat{t}}\delta\hat{\varphi}.
\end{equation}
By using equations (129) and (130) we can deduce
\begin{eqnarray}
3H\delta\Big(\frac{d\hat{\varphi}}{d\hat{t}}\Big)
+\frac{d^{2}\hat{V}}{d\hat{t}^{2}}\delta\hat{\varphi}\simeq\hspace{5cm}\nonumber\\
-2\frac{d\hat{V}}{d\hat{t}}\frac{\frac{\kappa_{5}^{2}}{6}\rho_{\hat{\varphi}}
+\frac{1}{\frac{d\hat{\varphi}}{d\hat{t}}\delta\hat{\varphi}}\int\delta
\hat{E}_{i}^{0}\,dx^{i}}{6\hat{H}\bigg(1+\frac{\kappa_{5}^{2}}{18}\rho_{\hat{\varphi}}\bigg)}
\frac{d\hat{\varphi}}{d\hat{t}}\delta\hat{\varphi}.\hspace{1.3cm}
\end{eqnarray}
Now, similar to the Jordan frame case, by defining a function as
\begin{equation}
\hat{{\cal{F}}}=\delta\hat{\varphi}\bigg(\frac{d\hat{V}}{d\hat{t}}\bigg)^{-1},
\end{equation}
equation (131) can be written as follows\\\\

\begin{widetext}
\begin{eqnarray}
\frac{\hat{{\cal{F}}}'}{\hat{{\cal{F}}}}=-
\frac{\bigg(\frac{d\hat{V}}{d\hat{t}}\bigg)
\bigg(\frac{\kappa_{5}^{2}}{6}\rho_{\hat{\varphi}}+\frac{1}{\frac{d\hat{\varphi}}{d\hat{t}}\delta\hat{\varphi}}\int\delta
\hat{E}_{i}^{0}\,dx^{i}\bigg)}{3\kappa_{4}^{2}\bigg(1+\frac{\kappa_{5}^{2}}{18}\rho_{\hat{\varphi}}\bigg)
\bigg(\rho_{\hat{\varphi}}+\hat{\lambda}+\frac{6\kappa_{4}^{2}}{\kappa_{5}^{4}}
\pm\frac{6}{\kappa_{5}^{2}}\,\sqrt{\frac{\kappa_{4}^{4}}{\kappa_{5}^{4}}+
\frac{\kappa_{4}^{2}}{3}\rho_{\hat{\varphi}}+
\frac{\kappa_{4}^{2}}{3}\hat{\lambda}-\frac{\kappa_{5}^{4}}{36}\hat{\lambda}^{2}-\hat{E}_{0}^{0}}\bigg)}
-\frac{2\frac{d^{2}\hat{V}}{d\hat{\varphi}^{2}}}{\frac{d\hat{V}}{d\hat{\varphi}}}.
\end{eqnarray}
\end{widetext}

As said before, a solution of this equation is
$\hat{{\cal{F}}}={\cal{C}}\exp(\int\frac{\hat{{\cal{F}}}'}{\hat{{\cal{F}}}}d\hat{\varphi})$.
So, from equation (132) we find
\begin{widetext}
\begin{eqnarray}
\delta\hat{\varphi}={\cal{C}}\,\frac{d\hat{V}}{d\hat{\varphi}}
\times
\exp\Bigg[\int\Bigg(\frac{\Big(-\frac{d\hat{V}}{d\hat{t}}\Big)
\Big(\frac{\kappa_{5}^{2}}{6}\rho_{\hat{\varphi}}+\frac{1}{\frac{d\hat{\varphi}}{{d\hat{t}}}\delta\hat{\varphi}}\int\delta
\hat{E}_{i}^{0}\,dx^{i}\Big)}{\Big(6+\frac{\kappa_{5}^{2}}{3}\rho_{\hat{\varphi}}\Big)
\Big(\rho_{\hat{\varphi}}+\hat{\lambda}+\frac{6\kappa_{4}^{2}}{\kappa_{5}^{4}}
\pm\frac{6}{\kappa_{5}^{2}}\,\sqrt{\frac{\kappa_{4}^{4}}{\kappa_{5}^{4}}+
\frac{\kappa_{4}^{2}}{3}\rho_{\hat{\varphi}}+
\frac{\kappa_{4}^{2}}{3}\hat{\lambda}-\frac{\kappa_{5}^{4}}{36}\hat{\lambda}^{2}-\hat{E}_{0}^{0}}\Big)}
-\frac{2\frac{d^{2}\hat{V}}{d\hat{\varphi}^{2}}}{\frac{d\hat{V}}{d\hat{\varphi}}}\Bigg)d\hat{\varphi}\Bigg].\hspace{1cm}
\end{eqnarray}
\end{widetext}

Once again by the same reasons as have been presented after Eq. (59)
and for the sake of simplicity we neglect the nontrivial
contribution of the bulk manifold arising via the term
$\frac{1}{\frac{d\hat{\varphi}}{{d\hat{t}}}\delta\hat{\varphi}}\int\delta
\hat{E}_{i}^{0}\,dx^{i}$.  By defining the following quantity
\begin{widetext}
\begin{eqnarray}
\hat{{\cal{G}}}=-\frac{\Big(\frac{d\hat{V}}{d\hat{\varphi}}\Big)
\Big(\frac{\kappa_{5}^{2}}{6}\rho_{\hat{\varphi}}\Big)}{6\Big(1+\frac{\kappa_{5}^{2}}{18}\rho_{\hat{\varphi}}\Big)
\Big(\rho_{\hat{\varphi}}+\hat{\lambda}+\frac{6\kappa_{4}^{2}}{\kappa_{5}^{4}}
\pm\frac{6}{\kappa_{5}^{2}}\,\sqrt{\frac{\kappa_{4}^{4}}{\kappa_{5}^{4}}+
\frac{\kappa_{4}^{2}}{3}\rho_{\hat{\varphi}}+
\frac{\kappa_{4}^{2}}{3}\hat{\lambda}-\frac{\kappa_{5}^{4}}{36}\hat{\lambda}^{2}-\frac{\mathcal{C}}{\hat{a}^{4}}}\Big)},
\end{eqnarray}
\end{widetext}
equation (134) can be rewritten as
\begin{equation}
\delta\hat{\varphi}={\cal{C}}\,\frac{d\hat{V}}{d\hat{\varphi}}\exp
\int \hat{{\cal{G}}}d\hat{\varphi}.
\end{equation}\\\\\\
So, density perturbation is given by
\begin{equation}
\hat{A}_{s}^{2}=\frac{\hat{k}^{3}}{2\pi^{2}}\,{\cal{C}}^{2}\,\Bigg(\frac{d\hat{V}}{d\hat{\varphi}}\Bigg)^{2}
\exp \int
2\hat{{\cal{G}}}d\hat{\varphi}\Bigg|_{\hat{k}=\hat{a}\hat{H}}.
\end{equation}\\\\
The scale-dependence of the perturbations is described by the
spectral index as
\begin{equation}
\hat{n}_{s}-1=\frac{d \ln \hat{A}_{S}^{2}}{d \ln \hat{k}}\,.
\end{equation}
The interval in wave number is related to the number of e-folds by
the relation $d \ln \hat{k}(\hat{\varphi})=d
\hat{N}(\hat{\varphi})$, so we obtain
\begin{widetext}
\begin{eqnarray}
\hat{n}_{s}=1-3\hat{\epsilon}+\frac{2}{3}\hat{\eta}\hspace{14cm}\nonumber\\
+\Bigg[\frac{-2\Big(\frac{d\hat{V}}{d\hat{\varphi}}\Big)
\Big(\frac{\kappa_{5}^{2}}{6}\rho_{\hat{\varphi}}\Big)}{6\Big(1+\frac{\kappa_{5}^{2}}{18}\rho_{\hat{\varphi}}\Big)
\Big(\rho_{\hat{\varphi}}+\hat{\lambda}+\frac{6\kappa_{4}^{2}}{\kappa_{5}^{4}}
\pm\frac{6}{\kappa_{5}^{2}}\,\sqrt{\frac{\kappa_{4}^{4}}{\kappa_{5}^{4}}+
\frac{\kappa_{4}^{2}}{3}\rho_{\hat{\varphi}}+
\frac{\kappa_{4}^{2}}{3}\hat{\lambda}-\frac{\kappa_{5}^{4}}{36}\hat{\lambda}^{2}-\frac{\mathcal{C}}{\hat{a}^{4}}}
\Big)}-\frac{9\hat{H}^{2}}{\frac{d\hat{V}}{d\hat{\varphi}}}
+\frac{4\frac{d^{2}\hat{V}}{d\hat{\varphi}^{2}}}{\frac{d\hat{V}}{d\hat{\varphi}}}\Bigg]\hspace{3cm}\nonumber\\
\times\Bigg[\frac{\kappa_{4}^{2}}{3}
+\frac{1}{\hat{V}}\bigg(\frac{\kappa_{4}^{2}}{3}\hat{\lambda}
+\frac{2\kappa_{4}^{4}}{\kappa_{5}^{4}}-\frac{2\kappa_{4}^{2}}{\kappa_{5}^{2}}\,\sqrt{\frac{\kappa_{4}^{4}}{\kappa_{5}^{4}}+
\frac{\kappa_{4}^{2}}{3}\hat{V}+
\frac{\kappa_{4}^{2}}{3}\hat{\lambda}-\frac{\kappa_{5}^{4}}{36}\hat{\lambda}^{2}
-\frac{\mathcal{C}}{\hat{a}^{4}}}\,\bigg)\Bigg]^{-1}\Bigg(\frac{\frac{d\hat{V}}{d\hat{\varphi}}}{3\hat{V}}\Bigg).\hspace{1.1cm}
\end{eqnarray}
\end{widetext}

The running of the spectral index in our setup, in Einstein frame,
is given as follows\\
\begin{widetext}
\begin{eqnarray}
\hat{\alpha}=\frac{d \hat{n}_{s}}{d \ln \hat{k}}\hspace{16.7cm}\nonumber\\
=6\hat{\epsilon}^{2}+2\hat{\epsilon}\hat{\eta} +\hat{{\cal{G}}}'
+\frac{2\frac{d^{3}\hat{V}}{d\hat{\varphi}^{3}}}{\frac{d\hat{V}}{d\hat{\varphi}}}
-\frac{3\frac{d^{2}\hat{H}}{d\hat{t}^{2}}}{\hat{H}^{2}}+\frac{\Big(\frac{d^{2}\hat{V}}{d\hat{\varphi}^{2}}\Big)
\Big(\frac{d^{3}\hat{V}}{d\hat{\varphi}^{3}}\Big)}{\hat{H}^{4}}
+\frac{1}{2}\Bigg(\frac{2\frac{d^{2}\hat{V}}{d\hat{\varphi}^{2}}}{\frac{d\hat{V}}{d\hat{\varphi}}}\Bigg)^{2}
+\Bigg[\frac{d\hat{H}}{d\hat{t}}+\frac{\Big(\frac{d^{2}\hat{V}}{d\hat{\varphi}^{2}}\Big)^{2}}
{\frac{d\hat{V}}{d\hat{\varphi}}}\bigg(1+\frac{3\hat{H}^{4}\frac{d\hat{V}}{d\hat{\varphi}}}
{2\frac{d^{2}\hat{V}}{d\hat{\varphi}^{2}}}+\frac{d\hat{H}}{d\hat{t}}
\bigg)\Bigg]\hspace{2cm}\nonumber\\
\times\Bigg[\frac{-4\Big(\frac{d\hat{V}}{d\hat{\varphi}}\Big)
\Big(\frac{\kappa_{5}^{2}}{6}\rho_{\hat{\varphi}}\Big)}{6\Big(1+\frac{\kappa_{5}^{2}}{18}\rho_{\hat{\varphi}}\Big)
\Big(\rho_{\hat{\varphi}}+\hat{\lambda}+\frac{6\kappa_{4}^{2}}{\kappa_{5}^{4}}
\pm\frac{6}{\kappa_{5}^{2}}\,\sqrt{\frac{\kappa_{4}^{4}}{\kappa_{5}^{4}}+
\frac{\kappa_{4}^{2}}{3}\rho_{\hat{\varphi}}+
\frac{\kappa_{4}^{2}}{3}\hat{\lambda}-\frac{\kappa_{5}^{4}}{36}\hat{\lambda}^{2}-\frac{\mathcal{C}}{\hat{a}^{4}}}
\Big)}\Bigg]\Bigg[\frac{\frac{d\hat{V}}{d\hat{\varphi}}}{3\hat{H}^{4}}\Bigg]\,.\hspace{2cm}
\end{eqnarray}
\end{widetext}

The tensor perturbations amplitude of a given mode when leaving the
Hubble radius are given by
\begin{equation}
{\hat{A}_{T}}^{2}=\frac{4\kappa_{4}^{2}}{25\pi}\hat{H}^{2}\Bigg|_{\hat{k}=\hat{a}\hat{H}}\,.
\end{equation}
In our setup and within the slow-roll approximation, we find
\begin{eqnarray}
\hat{A}_{T}^{2}=\frac{4\kappa_{4}^{2}}{25\pi}\hat{V}\Bigg[\frac{\kappa_{4}^{2}}{3}
+\frac{1}{\hat{V}}\bigg(\frac{\kappa_{4}^{2}}{3}\hat{\lambda}
+\frac{2\kappa_{4}^{4}}{\kappa_{5}^{4}}\hspace{3.2cm}\nonumber\\
\pm\frac{2\kappa_{4}^{2}}{\kappa_{5}^{2}}\,\sqrt{\frac{\kappa_{4}^{4}}{\kappa_{5}^{4}}+
\frac{\kappa_{4}^{2}}{3}\hat{V}+\frac{\kappa_{4}^{2}}{3}\hat{\lambda}
-\frac{\kappa_{5}^{4}}{36}\hat{\lambda}^{2}}\,\bigg)\Bigg].\hspace{1.5cm}
\end{eqnarray}
The tensor spectral index is given by
\begin{equation}
\hat{n}_{T}=\frac{d \ln \hat{A}_{T}^{2}}{d \ln \hat{k}}
\end{equation}
which in Einstein frame, it takes the following form
\begin{eqnarray}
\hat{n}_{T}=-
\Bigg(\frac{\frac{d\hat{V}}{d\hat{\varphi}}}{3\hat{V}}\Bigg)\Bigg[\frac{\kappa_{4}^{2}}{3}
+\frac{1}{\hat{V}}\bigg(\frac{\kappa_{4}^{2}}{3}\hat{\lambda}
+\frac{2\kappa_{4}^{4}}{\kappa_{5}^{4}}\hspace{3cm}\nonumber\\
\pm\frac{2\kappa_{4}^{2}}{\kappa_{5}^{2}}\,\sqrt{\frac{\kappa_{4}^{4}}{\kappa_{5}^{4}}+
\frac{\kappa_{4}^{2}}{3}\hat{V}+
\frac{\kappa_{4}^{2}}{3}\hat{\lambda}-\frac{\kappa_{5}^{4}}{36}\hat{\lambda}^{2}}\,\bigg)\Bigg]^{-1}\hat{\Sigma}\,,\hspace{1.3cm}
\end{eqnarray}
where $\hat{\Sigma}$ is defined as
\begin{equation}
\hat{\Sigma}\equiv\frac{{\kappa_{{4}}}^{2}}{3}\bigg(\frac{\frac{d\hat{V}}{d\hat{\varphi}}}{\hat{H}^{2}}\bigg)\bigg(1\pm\frac
{{\kappa_{{4}}}^{2}}{{\kappa_{{5}}}^{2}} \frac {1}{\sqrt {{\frac
{{\kappa_{{4}}}^{4}}{{\kappa_{{5}}}^{4}}}+\frac{\kappa_{4}^{2}}{3}\hat{V}
+\frac{{\kappa_{{4}}}^
{2}}{3}\hat{\lambda}-\frac{\kappa_{5}^{4}}{36}\hat{\lambda}^{2}}}\bigg)\,.
\end{equation}

Finally, the tensor-to-scalar ratio in Einstein frame is given by
\begin{equation}
\hat{r}=\frac{\hat{A}_{T}^{2}}{\hat{A}_{s}^{2}}\simeq\frac{8\pi\kappa_{4}^{2}}{25}\frac{\exp
\Bigg(\int -\hat{{\cal{G}}}
d\hat{\varphi}\Bigg)}{\hat{{\cal{C}}}^{2}
\Big(\frac{d\hat{V}}{d\hat{\varphi}}\Big)^{2}\hat{k}}.
\end{equation}

Once again and similar to previous section, in which follows we
present an explicit example to see how our equations in Einstein
frame work.\\

\section{An explicit example: monomial case with $f\sim \varphi^{2}$ and $V\sim\varphi^{2m}$}

In this section, we use the same form of $f$ and $V$ defined in
equation (72) and (73). From equation (82), the potential in
Einstein frame takes the following form
\begin{equation}
\hat{V}=\frac{b}{2m}\frac{\varphi^{2m}}{(1+\kappa_{4}^{2}\xi\varphi^{2})^{2}}\,.
\end{equation}
With this form of the potential, we get the flat potential in the
large field regime in Einstein frame (see figure~\ref{fig:22}). As
the Jordan frame case, we study two types of potential: quadratic
potential with $m=1$ and quartic potential with $m=2$. In the large
$\varphi$ regime, the variation of $\hat{\varphi}$ versus $\varphi$
attains the following forms
\begin{equation}
\frac{d\hat{\varphi}}{d\varphi}=\frac{1}{\sqrt{\kappa_{4}^{2}\xi\varphi}}\,,\,\,
\varphi\cong\frac{\kappa_{4}^{2}\xi}{4}\hat{\varphi}^{2}
\hspace{3.1cm} m=1
\end{equation}

\begin{equation}
\frac{d\hat{\varphi}}{d\varphi}=\frac{1}{\varphi}\sqrt{\frac{1+6\xi}{\kappa_{4}^{2}\xi}}\,,\,\,
\varphi\cong\frac{1}{\sqrt{\kappa_{4}^{2}\xi}}\exp\frac{\sqrt{\kappa_{4}^{2}\xi}
\hat{\varphi}}{\sqrt{1+6\xi\kappa_{4}^{2}\xi}} \hspace{0.4cm}  m=2
\end{equation}

Now, in order to obtain the slow-roll parameters in Einstein frame,
we should rewrite these parameters in terms of the scalar field and
corresponding potential in Einstein frame
\begin{equation}
\hat{\epsilon}=\frac{1}{2{\kappa_{{4}}}^{2}}\Bigg(\frac{\partial
\hat{V}/\partial \hat{\varphi}}{\hat{V}}\Bigg)^{2}\times
\hat{{\cal{A}}}(\varphi)\,,
\end{equation}
and
\begin{equation}
\hat{\eta}=\frac{1}{{\kappa_{{4}}}^{2}}\Bigg(\frac{\partial^{2}\hat{V}/\partial
\hat{\varphi}^{2}}{\hat{V}}\Bigg)\times \hat{{\cal{B}}}(\varphi)\,,
\end{equation}
where by definition
\begin{widetext}
\begin{equation}
\hat{{\cal{A}}}(\varphi)=\frac{1
\pm\frac{\kappa_{4}^{2}}{\kappa_{5}^{2}}\bigg(\sqrt
{\frac{\kappa_{4}^{4}}{\kappa_{5}^{4}}+
\frac{\kappa_{4}^{2}}{3}\hat{V}+
\frac{\kappa_{4}^{2}}{3}\hat{\lambda}-\frac{\kappa_{5}^{4}}{36}\hat{\lambda}^{2}
}\,\,\bigg)^{-\frac{1}{2}} }{\Bigg(1
+\frac{1}{\hat{V}}\bigg(\hat{\lambda}+\frac{6\kappa_{4}^{2}}{\kappa_{5}^{4}}
\pm\frac{6}{\kappa_{5}^{2}}\,\sqrt{\frac{\kappa_{4}^{4}}{\kappa_{5}^{4}}+
\frac{\kappa_{4}^{2}}{3}\hat{V}+
\frac{\kappa_{4}^{2}}{3}\hat{\lambda}-\frac{\kappa_{5}^{4}}{36}\hat{\lambda}^{2}
}\bigg)\Bigg)^{2}}\,,
\end{equation}
\end{widetext}
and
\begin{widetext}
\begin{equation}
\hat{{\cal{B}}}(\varphi)=\left[1
+\frac{1}{\hat{V}}\Big(\hat{\lambda}+\frac{6\kappa_{4}^{2}}{\kappa_{5}^{4}}
\pm\frac{6}{\kappa_{5}^{2}}\,\sqrt{\frac{\kappa_{4}^{4}}{\kappa_{5}^{4}}+
\frac{\kappa_{4}^{2}}{3}\hat{V}+
\frac{\kappa_{4}^{2}}{3}\hat{\lambda}-\frac{\kappa_{5}^{4}}{36}\hat{\lambda}^{2}
}\Big)\right]^{-1}\,,
\end{equation}
\end{widetext}
where we defined
$\hat{\lambda}=\frac{1}{(1+\kappa_{4}^{2}f(\varphi))^{2}}\lambda$.
From equations (147) - (151), we obtain the slow-roll parameters in
the large field limit as
\begin{widetext}
\begin{equation}
\hat{\epsilon}=\left\{\begin{array}{ll}
\frac{2\left(1+\kappa_{4}^{2}\xi\varphi^{2}\right)^{2}}{\varphi^{2}\left(1+\kappa_{4}^{2}\xi\varphi\right)^{2}
\left(1+\kappa_{4}^{2}\xi\varphi^{2}+6\kappa_{4}^{2}\xi^{2}\varphi^{2}\right)}\left[\frac{1
\pm\frac{\kappa_{4}^{2}}{\kappa_{5}^{2}}\bigg(\sqrt
{\frac{\kappa_{4}^{4}}{\kappa_{5}^{4}}+
\frac{\kappa_{4}^{2}}{6}\frac{b\varphi^{2}}{(1+\kappa_{4}^{2}\xi\varphi^{2})^{2}}+
\frac{\kappa_{4}^{2}}{3}\hat{\lambda}-\frac{\kappa_{5}^{4}}{36}\hat{\lambda}^{2}
}\,\,\bigg)^{-\frac{1}{2}}}{\Bigg(1
+\frac{2(1+\kappa_{4}^{2}\xi\varphi^{2})^2}{b\varphi^{2}}\bigg(\hat{\lambda}+\frac{6\kappa_{4}^{2}}{\kappa_{5}^{4}}
\pm\frac{6}{\kappa_{5}^{2}}\,\sqrt{\frac{\kappa_{4}^{4}}{\kappa_{5}^{4}}+
\frac{\kappa_{4}^{2}}{6}\frac{b\varphi^{2}}{(1+\kappa_{4}^{2}\xi\varphi^{2})^{2}}+
\frac{\kappa_{4}^{2}}{3}\hat{\lambda}-\frac{\kappa_{5}^{4}}{36}\hat{\lambda}^{2}
}\bigg)\Bigg)^{2}}\right] \hspace{1cm} m=1\\\\
\frac{8}{\varphi^{2}\left(1+\kappa_{4}^{2}\xi\varphi^{2}+6\kappa_{4}^{2}\xi^{2}\varphi^{2}\right)}
\left[\frac{1\pm\frac{\kappa_{4}^{2}}{\kappa_{5}^{2}}\bigg(\sqrt{\frac{\kappa_{4}^{4}}{\kappa_{5}^{4}}+
\frac{\kappa_{4}^{2}}{12}\frac{b\varphi^{4}}{(1+\kappa_{4}^{2}\xi\varphi^{2})^{2}}+
\frac{\kappa_{4}^{2}}{3}\hat{\lambda}-\frac{\kappa_{5}^{4}}{36}\hat{\lambda}^{2}
}\,\,\bigg)^{-\frac{1}{2}} }{\Bigg(1
+\frac{4(1+\kappa_{4}^{2}\xi\varphi^{2})^2}{b\varphi^{4}}\bigg(\hat{\lambda}+\frac{6\kappa_{4}^{2}}{\kappa_{5}^{4}}
\pm\frac{6}{\kappa_{5}^{2}}\,\sqrt{\frac{\kappa_{4}^{4}}{\kappa_{5}^{4}}+
\frac{\kappa_{4}^{2}}{12}\frac{b\varphi^{4}}{(1+\kappa_{4}^{2}\xi\varphi^{2})^{2}}+
\frac{\kappa_{4}^{2}}{3}\hat{\lambda}-\frac{\kappa_{5}^{4}}{36}\hat{\lambda}^{2}
}\bigg)\Bigg)^{2}}\right]\hspace{2.3cm} m=2 \vspace{0.5cm}\\
\end{array}\right.
\end{equation}
\end{widetext}
and
\begin{widetext}
\begin{equation}
\hat{\eta}=\left\{\begin{array}{ll}
\frac{-2-2\kappa_{4}^{2}\xi\varphi^{2}}{(1+\kappa_{4}^{2}\xi\varphi^{2}+6\kappa_{4}^{2}\xi^{2}\varphi^{2})^{2}\varphi^{2}
(1+\kappa_{4}^{2}\xi\varphi)^{2}}\Big(-1+3\kappa_{4}^{4}\xi^{2}\varphi^{3}\\\\+12\kappa_{4}^{4}\xi^{3}\varphi^{3}+\kappa_{4}^{6}
\xi^{3}\varphi^{5}-3\kappa_{4}^{2}\xi\varphi^{2}-6\kappa_{4}^{2}\xi^{2}\varphi^{2}
+2\kappa_{4}^{2}\xi\varphi-2\kappa_{4}^{4}\xi^{2}\varphi^{4}
-18\kappa_{4}^{4}\xi^{3}\varphi^{4}\Big)\\\\ \times \Bigg[1
+\frac{2(1+\kappa_{4}^{2}\xi\varphi^{2})^2}{b\varphi^{2}}\Big(\hat{\lambda}+\frac{6\kappa_{4}^{2}}{\kappa_{5}^{4}}
\pm\frac{6}{\kappa_{5}^{2}}\,\sqrt{\frac{\kappa_{4}^{4}}{\kappa_{5}^{4}}+
\frac{\kappa_{4}^{2}}{6}(\frac{b\varphi^{2}}{(1+\kappa_{4}^{2}\xi\varphi^{2})^{2}})+
\frac{\kappa_{4}^{2}}{3}\hat{\lambda}-\frac{\kappa_{5}^{4}}{36}\hat{\lambda}^{2}
}\Big)\Bigg]^{-1}\hspace{2.3cm} m=1\vspace{0.5cm} \\\\\\
-4\frac{6\kappa_{4}^{4}\xi^{3}\varphi^{4}+2\kappa_{4}^{4}\xi^{2}\varphi^{4}-18\kappa_{4}^{2}\xi^{2}\varphi^{2}
-\kappa_{4}^{2}\xi\varphi^{2}-3}{\left(1
+\kappa_{4}^{2}\xi\varphi^{2}+6\kappa_{4}^{2}\xi^{2}\varphi^{2}\right)^{2}\varphi^{2}}\\\\
\times
\Bigg[1+\frac{4(1+\kappa_{4}^{2}\xi\varphi^{2})^2}{b\varphi^{4}}\Big(\hat{\lambda}+\frac{6\kappa_{4}^{2}}{\kappa_{5}^{4}}
\pm\frac{6}{\kappa_{5}^{2}}\,\sqrt{\frac{\kappa_{4}^{4}}{\kappa_{5}^{4}}+
\frac{\kappa_{4}^{2}}{12}\frac{b\varphi^{4}}{(1+\kappa_{4}^{2}\xi\varphi^{2})^{2}}+
\frac{\kappa_{4}^{2}}{3}\hat{\lambda}-\frac{\kappa_{5}^{4}}{36}\hat{\lambda}^{2}
}\Big)\Bigg]^{-1}
\hspace{2.6cm} m=2 \vspace{0.5cm}\\
\end{array}\right.
\end{equation}
\end{widetext}
Once $\hat{\epsilon}$ or $\hat{\eta}$ reach the unity, the
inflationary phase terminates.

\begin{figure*}
\flushleft\leftskip-5em{\includegraphics[width=2.5in]{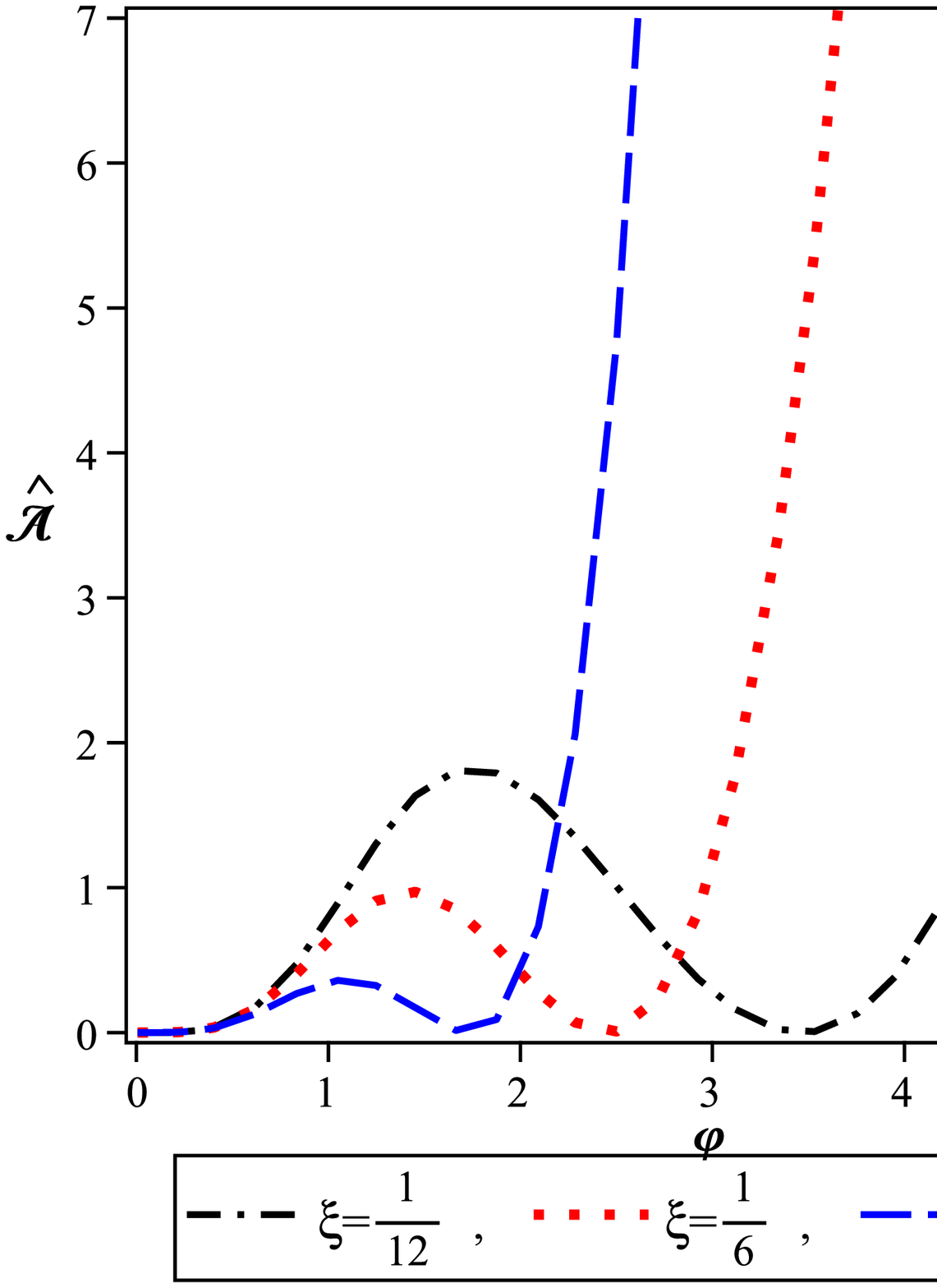}}\hspace{2.6cm}
{\includegraphics[width=2.5in]{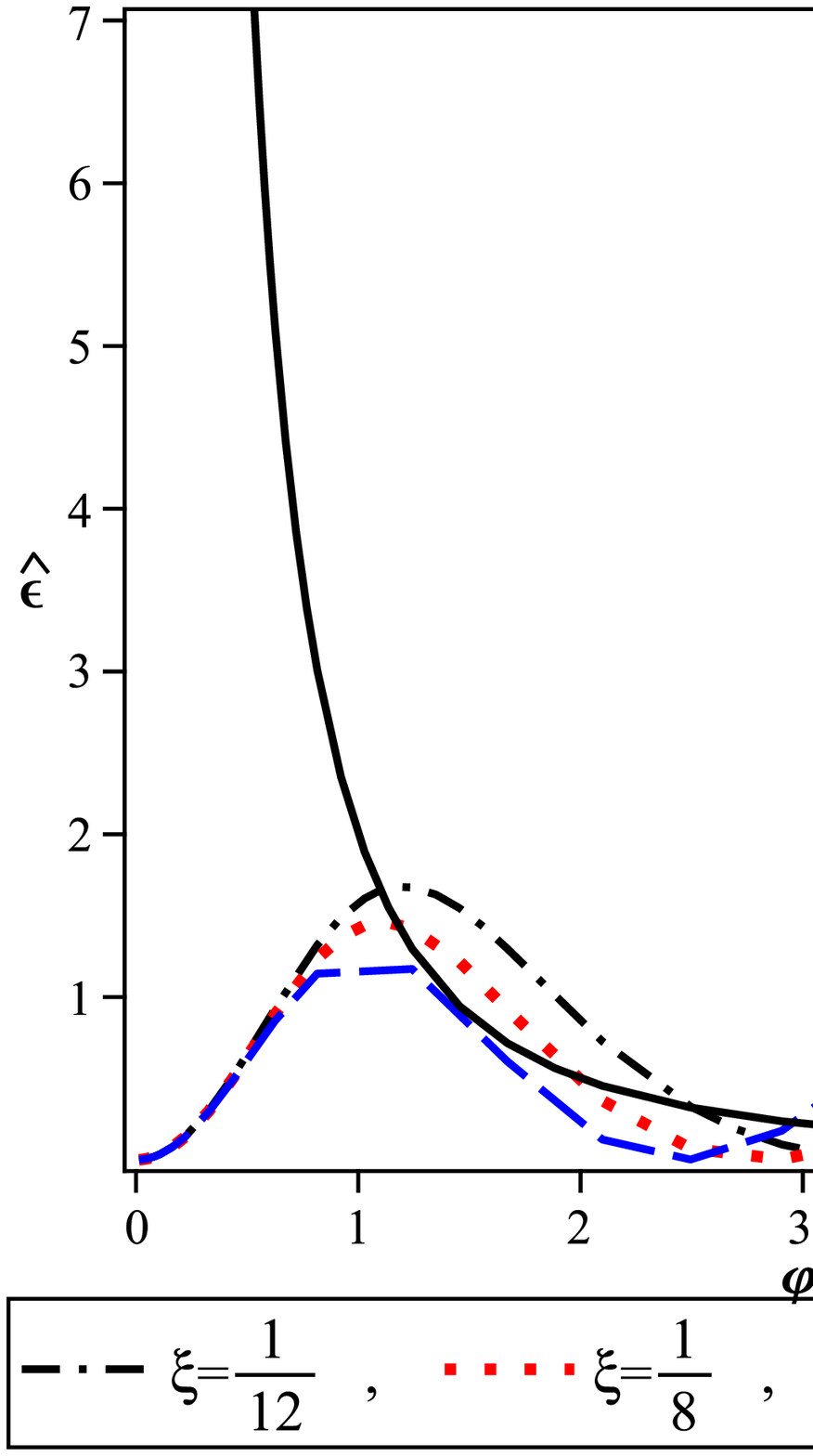}} \caption{\label{fig:11}The
evolution of the correctional factor $\hat{{\cal{A}}}$ (left panel)
and the first slow-roll parameter $\hat{\epsilon}$ (right panel)
versus the scalar field with a quadratic potential in Einstein
frame. The presence of the correctional factor, $\hat{{\cal{A}}}$,
causes the $\hat{\epsilon}$ to behave nearly as the standard 4D case
in the intermediate regime of the scalar field. In the large and
small scalar field regimes, it deviates from 4D behavior
drastically.}
\end{figure*}

\begin{figure*}
\flushleft\leftskip-8em{\includegraphics[width=2.5in]{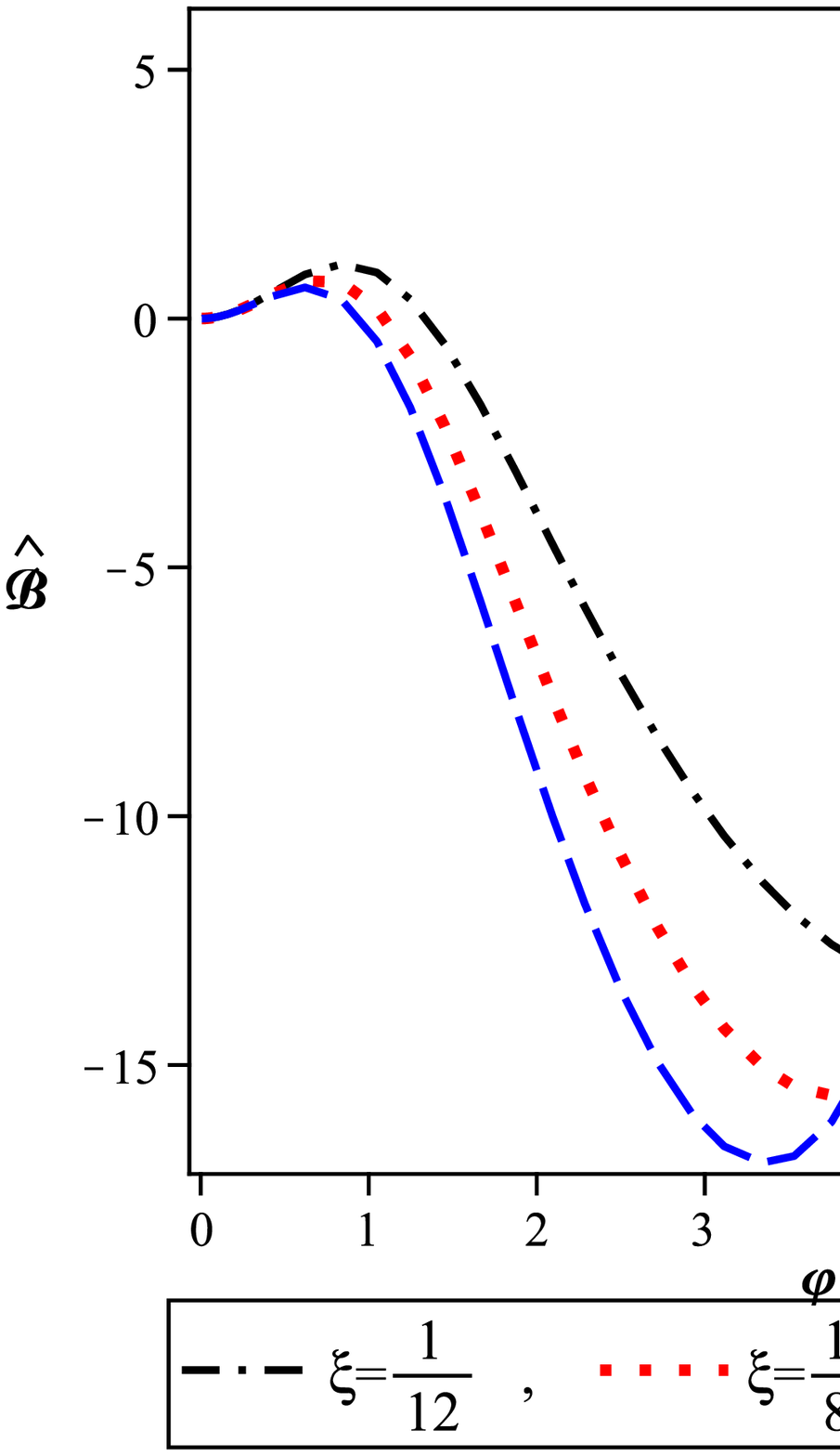}}\hspace{3.4cm}
{\includegraphics[width=2.5in]{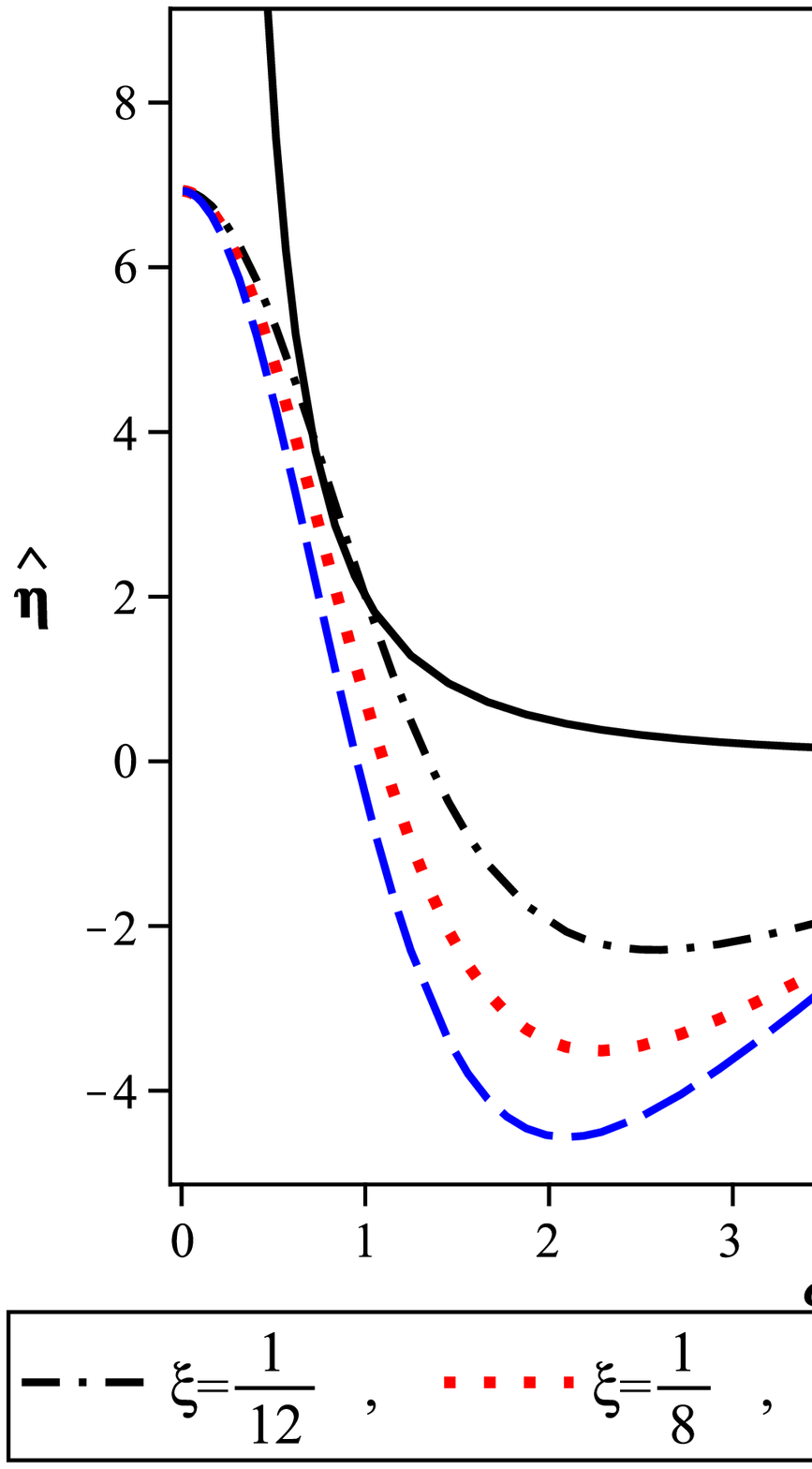}} \caption{\label{fig:12}The
evolution of the correctional factor $\hat{{\cal{B}}}$ (left panel)
and the second slow-roll parameter $\hat{\eta}$ (right panel) versus
the scalar field with a quadratic potential in Einstein frame. In
this case, there is a finite maximum value of $\hat{\eta}$ at
$\hat{\varphi}=0$.}
\end{figure*}

\begin{figure*}
\flushleft\leftskip-5em{\includegraphics[width=2.5in]{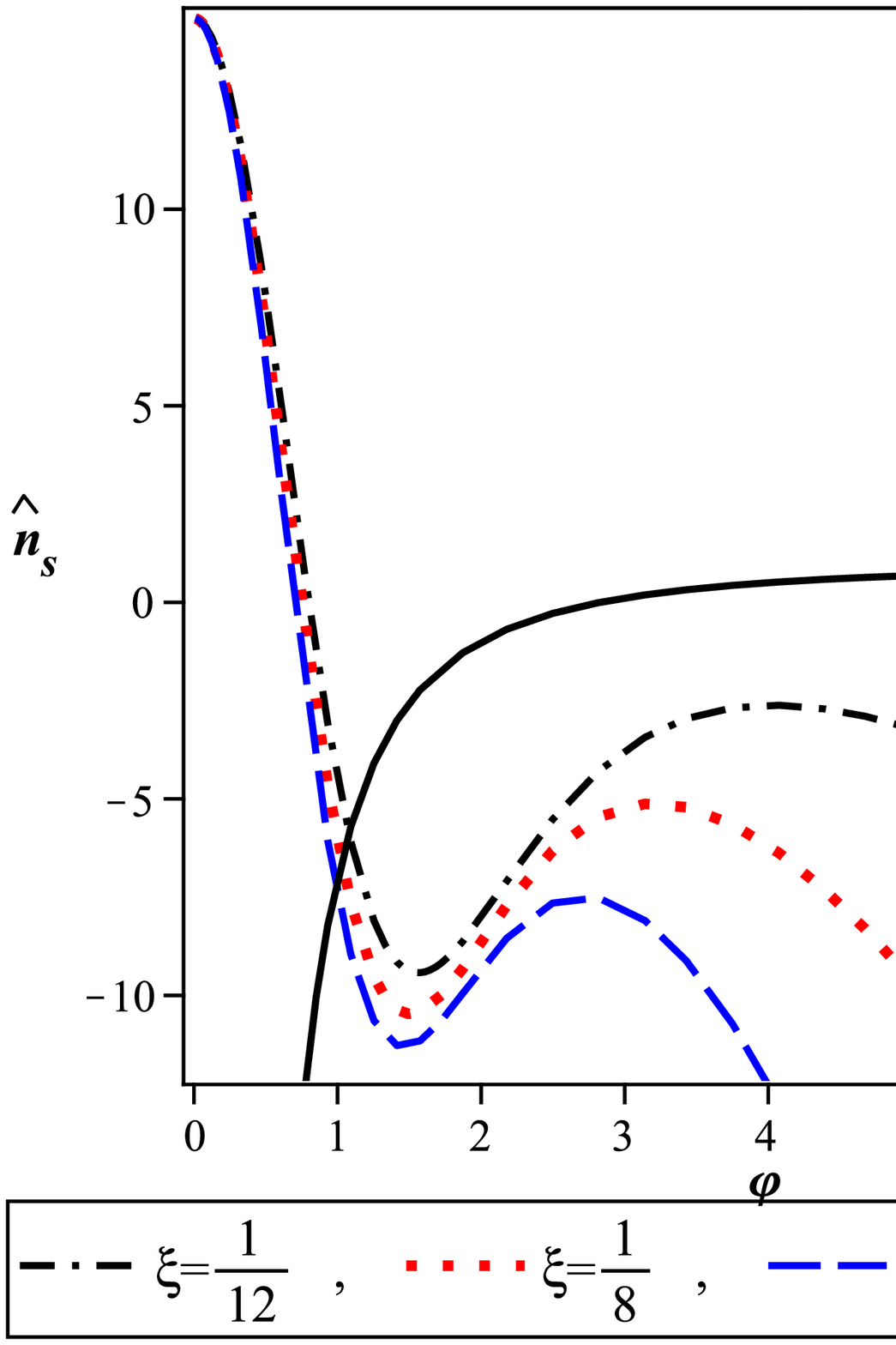}}\hspace{2.6cm}
{\includegraphics[width=2.5in]{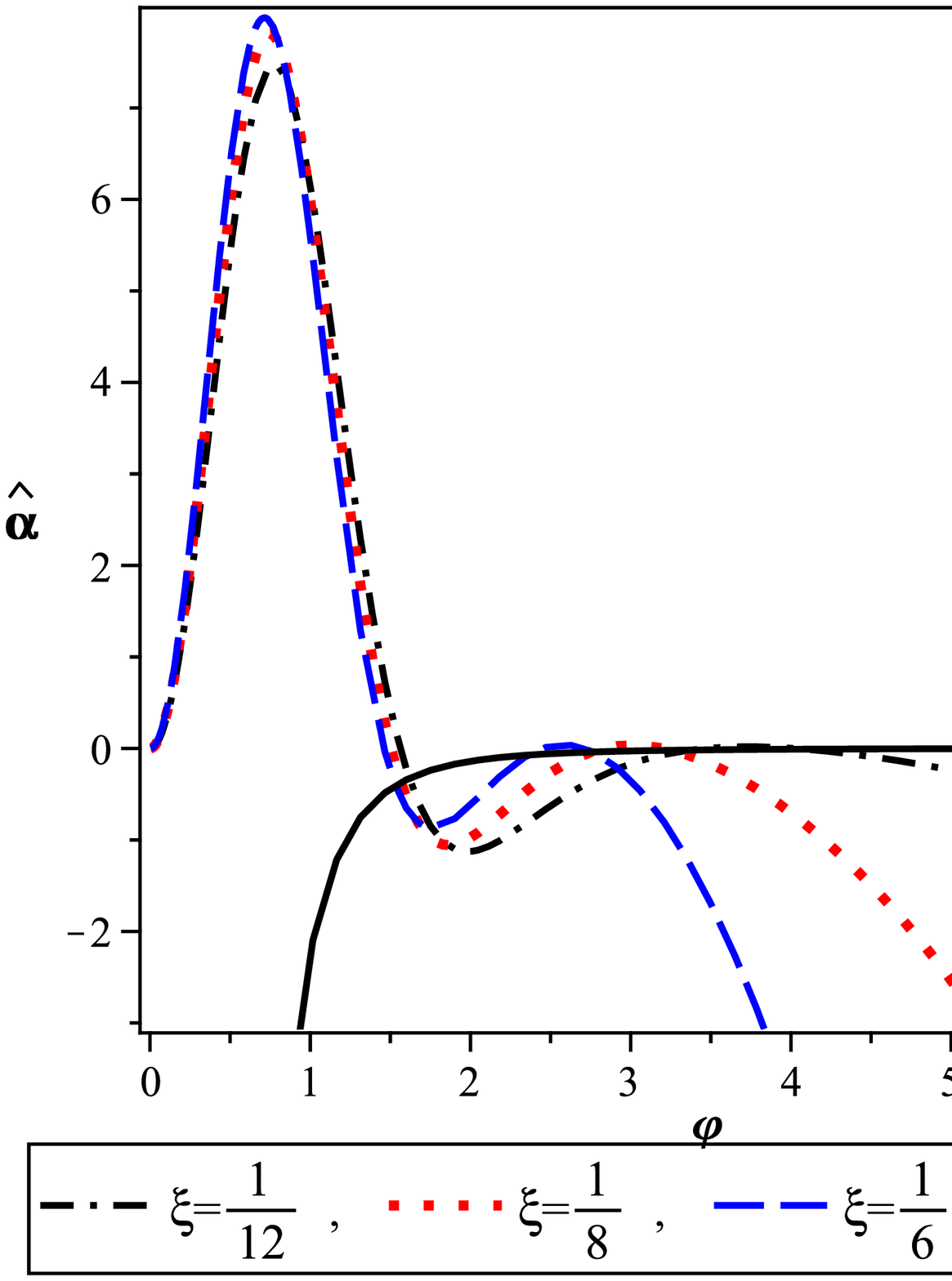}} \caption{\label{fig:13}The
evolution of the scalar spectral index (left panel) and the running
of the spectral index (right panel) versus the scalar field with a
quadratic potential in Einstein frame. In an intermediate field
regime, the behavior of $\hat{n}_{s}$ and $\hat{\alpha}$ are nearly
similar to the standard 4D behavior.}
\end{figure*}

\begin{figure}
\flushleft\leftskip-7em{\includegraphics[width=2.5in]{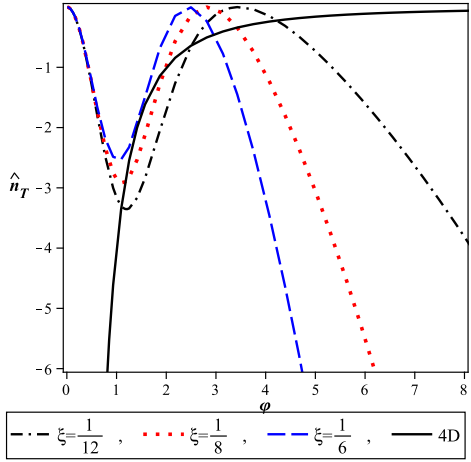}}\hspace{3cm}
\caption{\label{fig:14}The evolution of the tensor spectral index
versus the scalar field with a quadratic potential. $\hat{n}_{T}$ in
an intermediate field regime behaves similar to what it does in the
standard 4D case.}
\end{figure}

\begin{figure}
\flushleft\leftskip-7em{\includegraphics[width=2.5in]{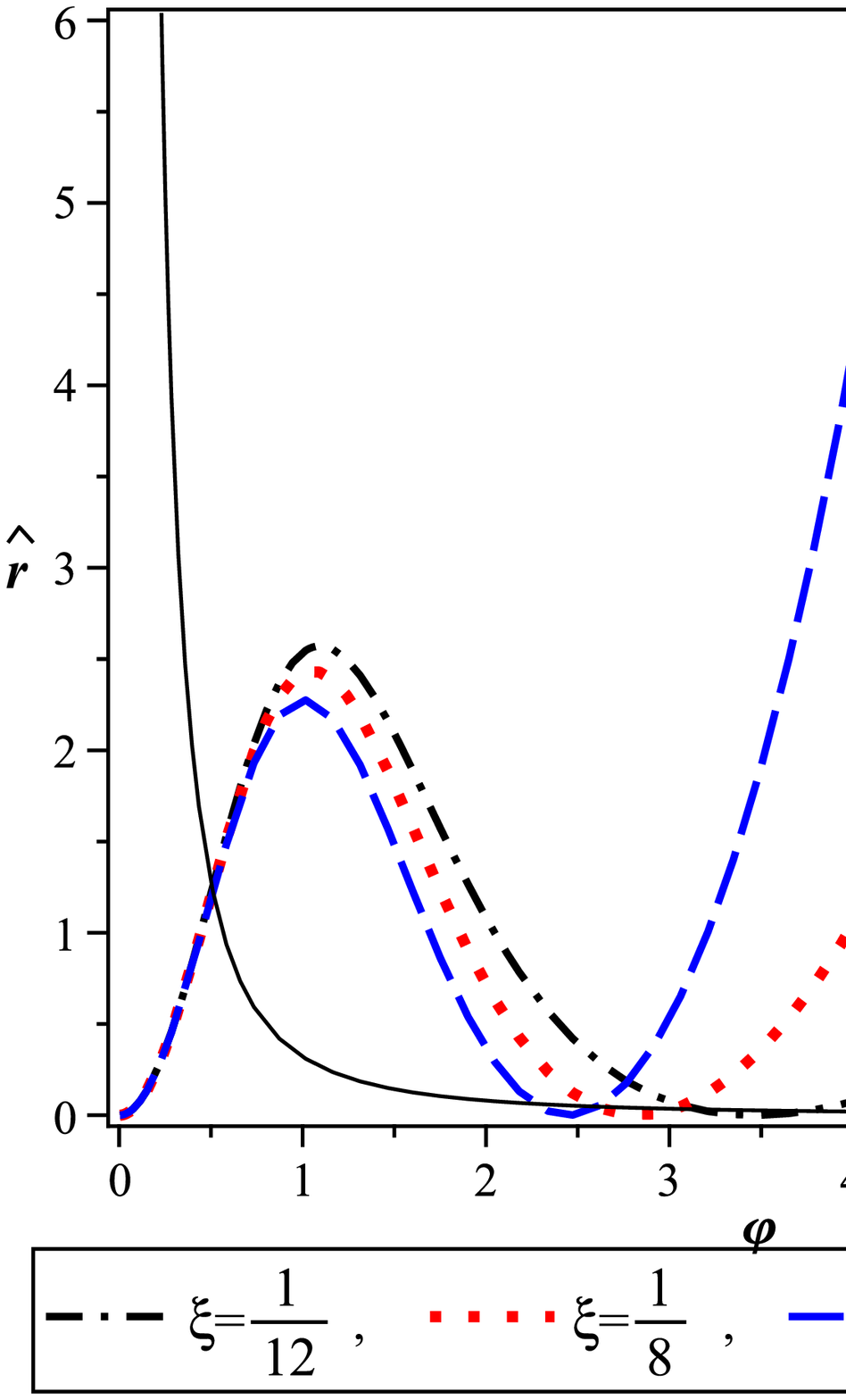}}\hspace{3cm}
\caption{\label{fig:15}The evolution of the tensor to scalar ratio
$r$ versus the scalar field with a quadratic potential.}
\end{figure}

\subsection{Quadratic Potential: $V(\varphi)=\frac{b}{2}\varphi^{2}$}

Similar to our analysis in Jordan frame, we firstly consider a
quadratic potential to analyze the outcome of the model in Einstein
frame. We show that with this potential in Einstein frame, there are
some differences with the Jordan frame case. These differences may
be a footprint of the physical non-equivalence of these two frames
in this braneworld setup (we return to this issue later). In
figure~\ref{fig:11} we depicted the behavior of the correctional
factor $\hat{{\cal{A}}}$ and $\hat{\epsilon}$ versus the scalar
field.

\begin{table*}
\caption{\label{tab:table3}The values of some inflation parameters
with a quadratic potential in Einstein frame at the time that
physical scales crossed the horizon.}
\begin{ruledtabular}
\begin{tabular}{ccccc}
$\xi$&$\hat{n}_{s}$&$\hat{r}$&$\hat{\alpha}$ \\ \hline\\
$\frac{1}{12}$& $0.9999999997$ &$2.690921210\times 10^{-68}$&-1.263758864$\times 10^{-216}$\\\\
$\frac{1}{8}$& $1.000000001$ &$1.387611488\times 10^{-68}$& -7.899897671$\times 10^{-217} $\\\\
$\frac{1}{6}$& $0.9999999991$ &$8.304922680\times 10^{-69}$& -5.265414757$\times 10^{-217}$\\
\end{tabular}
\end{ruledtabular}
\end{table*}

The left panel of this figure shows that as the scalar field
decreases, $\hat{{\cal{A}}}$ in two regimes (large and small scalar
field regimes) decreases. But there is an intermediate regime where
$\hat{{\cal{A}}}$ increases by reduction of the scalar field. The
behavior of $\hat{{\cal{A}}}$ affects the evolution of
$\hat{\epsilon}$. This can be seen in the right panel of
figure~\ref{fig:11}. This panel shows that only in intermediate
regime of the scalar field, $\hat{\epsilon}$ obeys nearly the 4D
behavior and in the large and small scalar field regimes, it
deviates from 4D behavior drastically. However, as we have shown in
the previous section, in Jordan frame with this type of potential
$\hat{\epsilon}$ in the large scalar field regime obeys the 4D
behavior. It should be noticed that there is a relative maximum for
$\hat{{\cal{A}}}$ and $\hat{\epsilon}$ which its value and location
depends on $\xi$. As $\xi$ increases, this maximum becomes smaller
and take places in smaller values of the scalar field.  This maximum
is larger for smaller values of $\xi$. Also, there is a minimum
value for the slow-roll parameter. As $\xi$ increases, this minimum
decreases and take places in larger values of the scalar field. We
note that for smaller $\xi$, the 4D behavior lasts in wider domain
of the scalar field values.

The behavior of the correctional factor $\hat{{\cal{B}}}$ and
$\hat{\eta}$ versus the scalar field is shown in
figure~\ref{fig:12}. Due to the effect of $\hat{{\cal{B}}}$, the
second slow-roll parameter in the large and small scalar field
regime deviates from the standard 4D behavior. But, in the
intermediate regime of the scalar field, $\hat{\eta}$ behaves
similar to what it does in 4D case. As $\xi$ decreases, this 4D
behavior lasts in wider domain of the scalar field values. Both
$\hat{\epsilon}$ and $\hat{\eta}$ can reach unity and therefore with
a quadratic potential in Einstein frame the inflation can ends
gracefully. We note that in contrast with Jordan frame case, with a
quadratic potential in Einstein frame, the second slow-roll
parameter can be negative in some values of the scalar field. Other
important parameters in an inflationary paradigm are the scalar and
tensor spectral indices. We have depicted the evolution of
$\hat{n}_{s}$ and $\hat{n}_{T}$ versus the scalar field in
figures~\ref{fig:13} (the left panel) and ~\ref{fig:14}
respectively. As these figures show, in an intermediate regime of
the scalar field these parameters decrease by reduction of the
scalar field as what they do in the standard four-dimensional model.
However, in the large and small scalar field regimes, $\hat{n}_{s}$
and $\hat{n}_{T}$ increase as the scalar field decreases. As $\xi$
decreases, the 4D behavior of $n_{s}$ and $n_{T}$ last in larger
domain of the scalar field. In the left panel of figure~\ref{fig:13}
we have shown the evolution of the running of the scalar spectral
index versus the scalar field. As this figure shows, in an
intermediate field regime, $\alpha$ decreases by reduction of the
scalar field similar to what it does in 4D case. This behavior
stopes at some value of the scalar field where $\alpha$ reaches a
relative minimum. Notice that, as $\xi$ decreases, the 4D behavior
of $\alpha$ lasts in wider domain of the scalar field.

The last parameter we consider is the tensor to scalar ratio,
$\hat{r}$. Figure~\ref{fig:15} shows the behavior of this parameter
versus the scalar field. As the scalar field decreases, $\hat{r}$
decreases until a minimum at some value of the scalar field is
reached and then it increases again. The increment of $\hat{r}$
stopes at some value of the scalar field and after that, $\hat{r}$
decreases again. This means that in an intermediate field regime,
the behavior of the tensor to scalar ratio in Einstein frame is
similar to the corresponding parameter in 4D case and in the large
and small field regime, it deviates from the standard 4D behavior
drastically. We note that as $\xi$ gets smaller, the 4D behavior of
$\hat{r}$ lasts in a wider domain of the scalar field.

Now, as the Jordan frame case, we proceed to calculate some
inflation parameters with a quadratic potential at the time of
horizon crossing. To find the value of the scalar field at the time
of horizon crossing, we treat as what we have done in the previous
section. We rewrite the Friedmann equation in high energy limit
($\hat{\rho}\gg\hat{\lambda}$) as follows
\begin{equation}
\hat{H}^{2}\simeq \Bigg(\frac{\kappa_{4}^{2}}{3}\hat{V}\Bigg)
\Bigg[1\pm\frac{2\kappa_{4}^{2}}{\kappa_{5}^{2}}\Bigg(\frac{\kappa_{4}^{2}}{3}\hat{V}\Bigg)^{-\frac{1}{2}}\,\Bigg]\,.
\end{equation}
So, the number of e-folds by using equation (99) can be expressed as
\begin{equation}
\hat{N}=-\int_{\hat{\varphi}_{hc}}^{\hat{\varphi}_{f}}
d\varphi\Bigg(\frac{d\hat{\varphi}}{d\varphi}\Bigg)^{2}
\Bigg(\frac{3\hat{V}}{d\hat{V}/d\varphi}\Bigg)\Bigg[\frac{\kappa_{4}^{2}}{3}
\pm\frac{2\kappa_{4}^{2}}{\kappa_{5}^{2}}\,\sqrt{\frac{\kappa_{4}^{2}}{3}}\,{\hat{V}}^{-\frac{1}{2}}\Bigg].
\end{equation}
In appendix {\bf C}, we have presented the solution of integral
(157) (where as before, we have assumed
$\hat{\varphi}_{hc}\gg\hat{\varphi}_{f}$). Then we found
$\hat{\varphi}_{hc}$ from that solution and by using equations
(139), (140) and (146) we find the values of the scalar spectral
index, its running and the tensor to scalar ratio at the time of
horizon crossing. Our analysis shows that although for different
values of $\xi$, the scalar spectral index is nearly scale
invariant, it is red-tilted for $\xi=\frac{1}{6}$ and
$\xi=\frac{1}{12}$ and blue-tilted for $\xi=\frac{1}{8}$. Also, by
reduction of $\xi$, the running of the spectral index increases and
the tensor to scalar ratio decreases. Table~\ref{tab:table3} shows
the value of $\hat{n_{s}}$, $\hat{r}$ and $\hat{\alpha}$ when the
physical scales crossed the horizon for three different values of
$\xi$. These values can be compared with recent observational data
as have been summarized in the last line of table~\ref{tab:table1}.

\subsection{Quartic Potential: $V(\varphi)=\frac{b}{4}\varphi^{4}$}

\begin{figure*}
\flushleft\leftskip-9em{\includegraphics[width=2.5in]{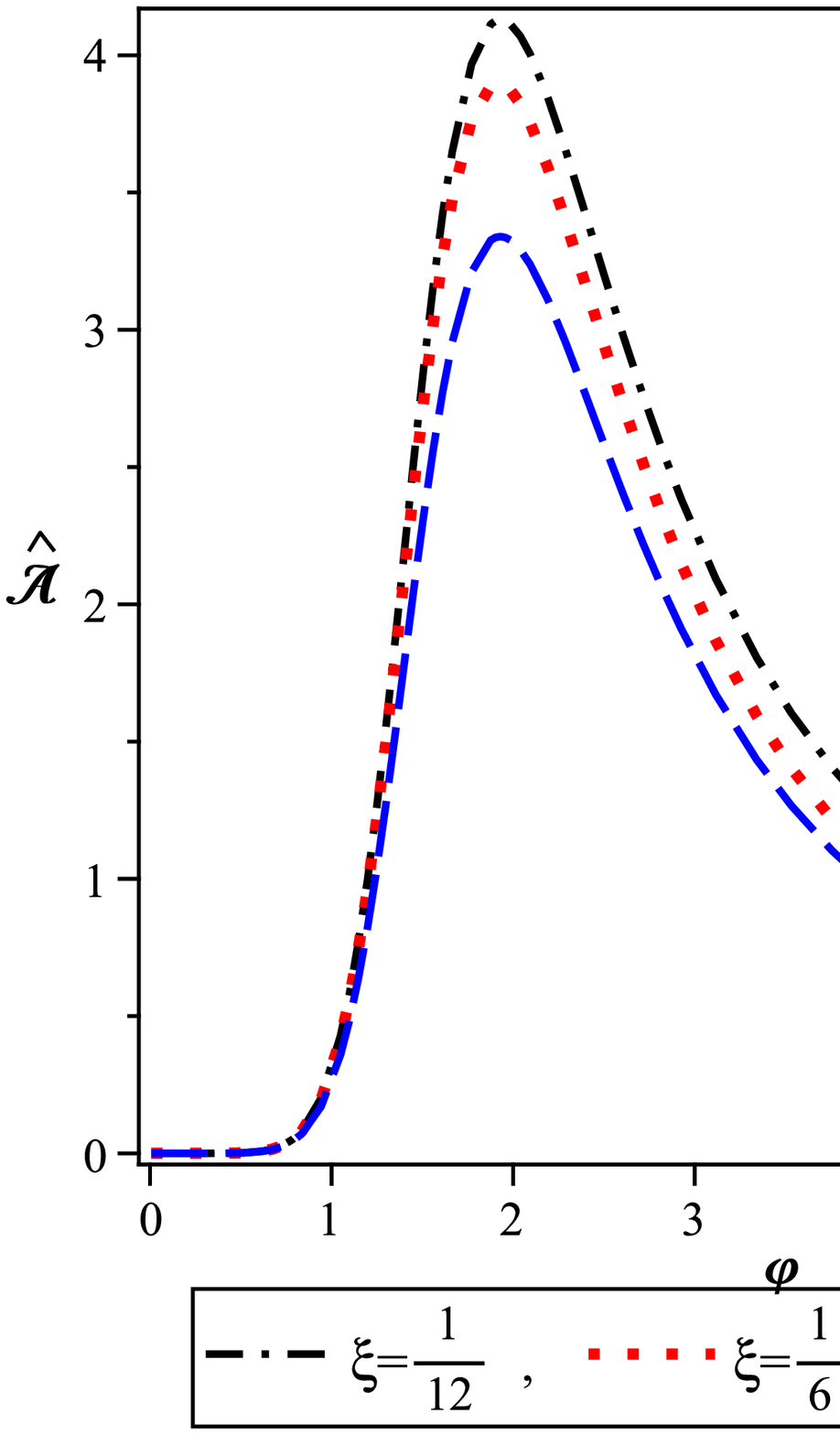}}\hspace{4.5cm}
{\includegraphics[width=2.5in]{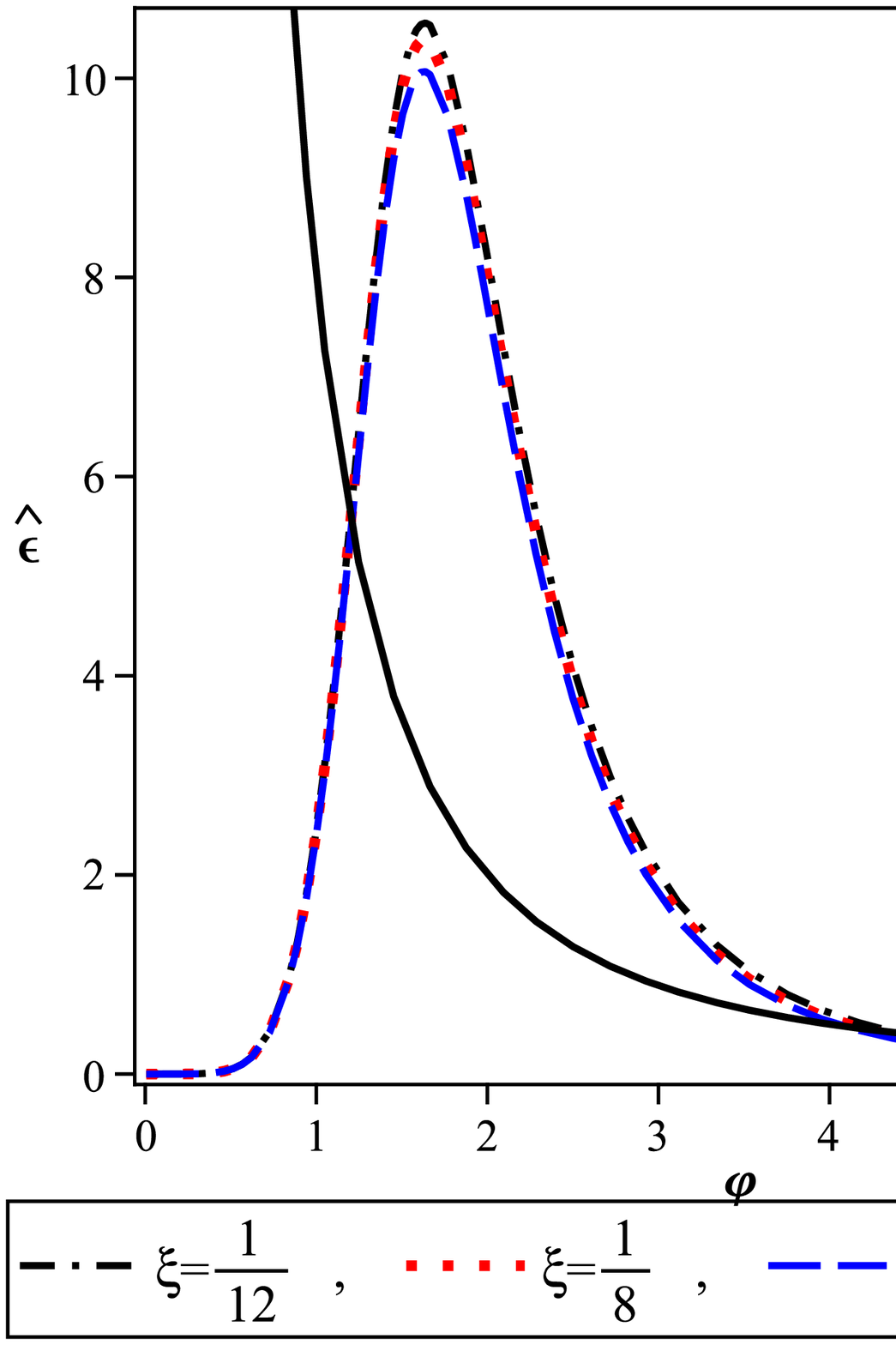}} \caption{\label{fig:16}The
evolution of the correctional factor $\hat{{\cal{A}}}$ (left panel)
and the first slow-roll parameter $\hat{\epsilon}$ (right panel)
versus the scalar field with a quartic potential in Einstein frame.
In the large scalar field regime, $\hat{\epsilon}$ behaves similar
to what it does in 4D case and in the small scalar field regime
deviates the 4D behavior.}
\end{figure*}

Now, as what we have done in Jordan frame, we analyze the model
parameter space with quartic potential in Einstein frame.
Figure~\ref{fig:16} shows the behavior of correctional factor,
$\hat{{\cal{A}}}$, and the slow roll parameter, $\hat{\epsilon}$,
versus the scalar field. As the scalar field decreases,
$\hat{{\cal{A}}}$ increases to a maximum value and then decreases.
As we have mentioned previously, the behavior of the correctional
factor affects the evolution of the corresponding inflation
parameters. Nevertheless, in spite of the presence of this factor,
in the large scalar field regime $\hat{\epsilon}$ behaves similar to
what it does in 4D case and increases by reduction of the scalar
field. It should be noticed that for larger values of $\xi$, the 4D
behavior of $\hat{\epsilon}$ lasts in wider domain of the scalar
field values.

\begin{figure*}
\flushleft\leftskip-8em{\includegraphics[width=2.5in]{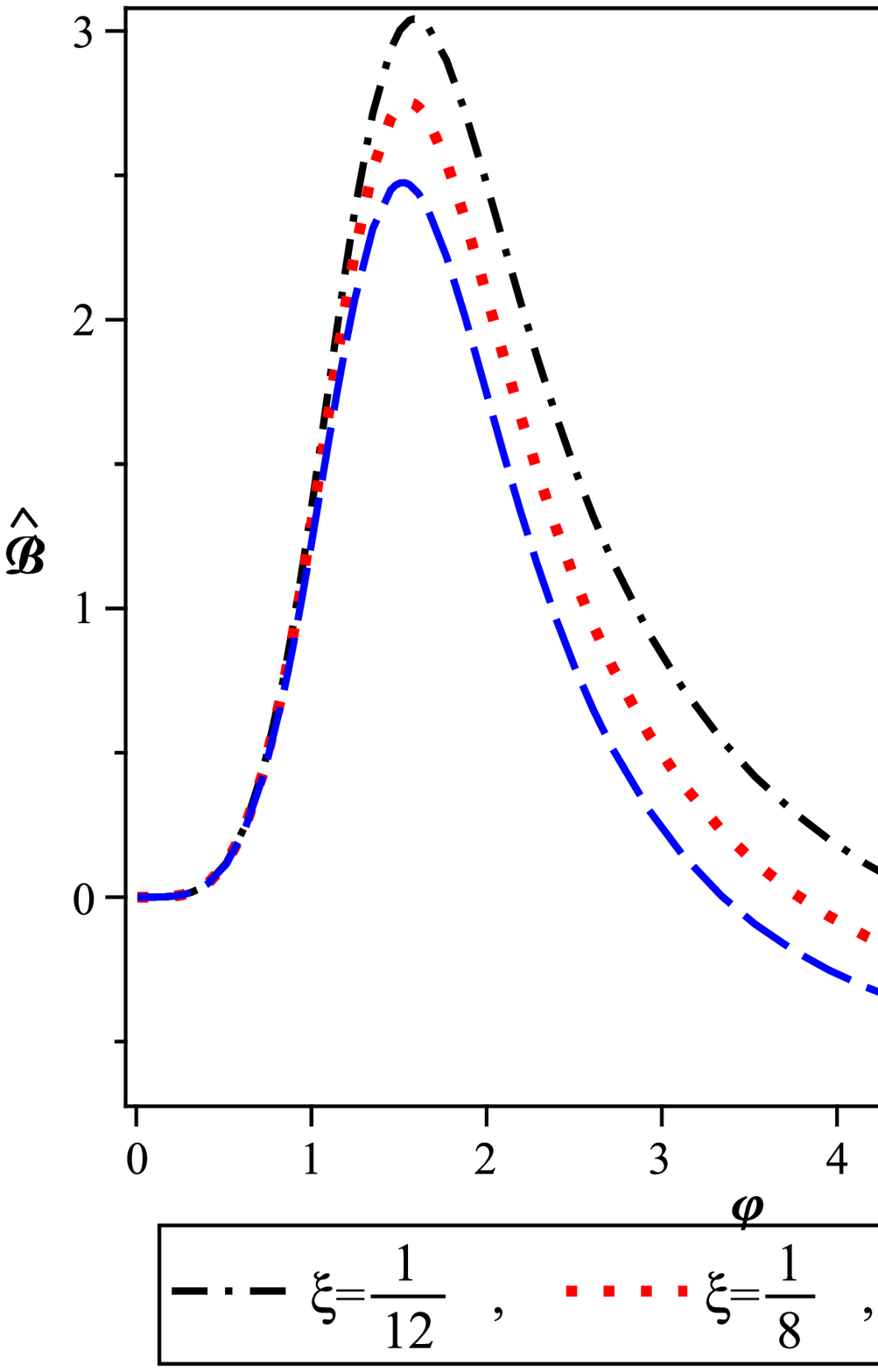}}\hspace{4cm}
{\includegraphics[width=2.5in]{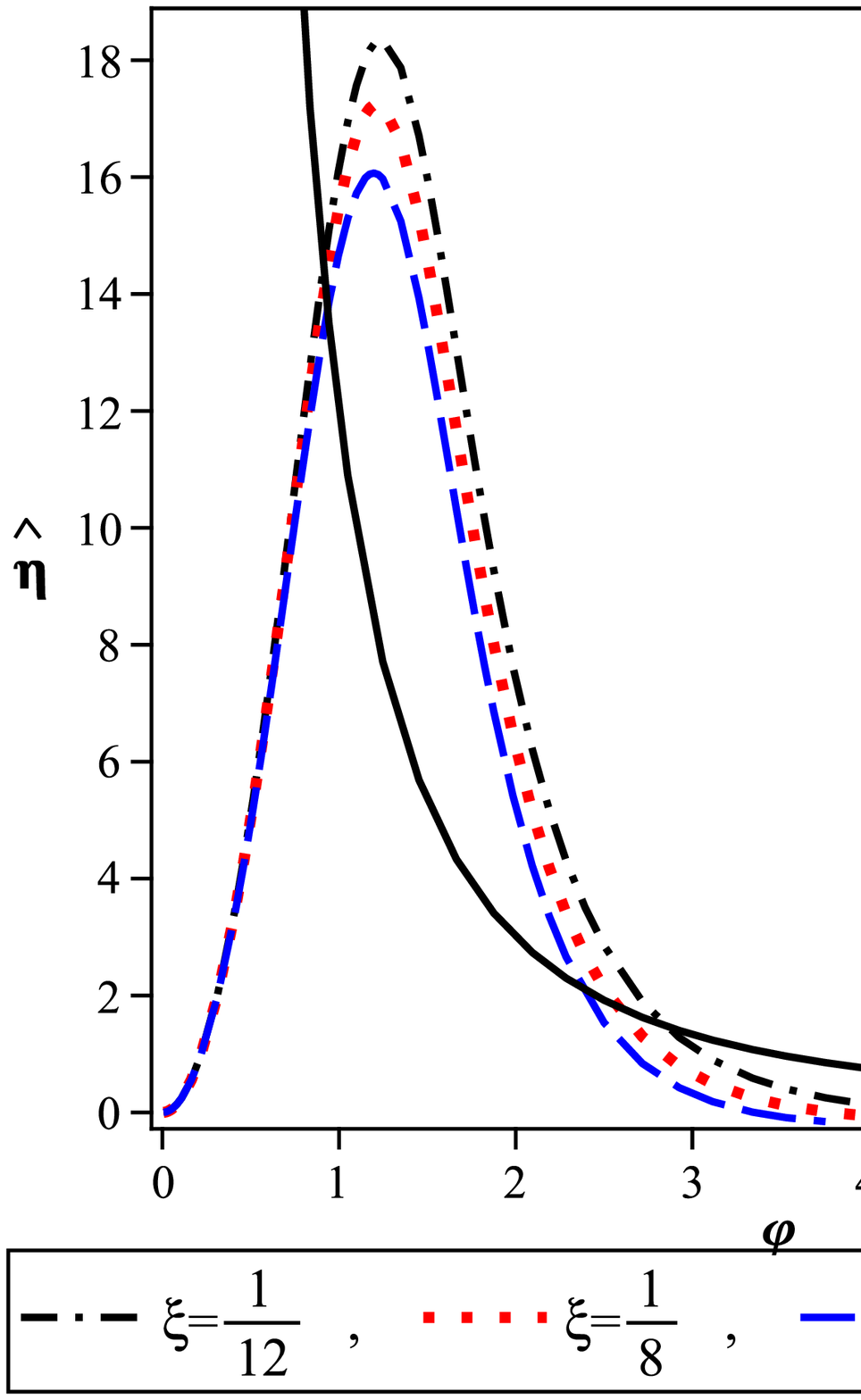}} \caption{\label{fig:17}The
evolution of the correctional factor $\hat{{\cal{B}}}$ (left panel)
and the second slow-roll parameter $\hat{\eta}$ (right panel) versus
the scalar field with a quartic potential in Einstein frame. The
effect of the correctional factor causes the $\hat{\eta}$ to follow
a behavior which deviates from the standard 4D behavior in the small
field regime.}
\end{figure*}

\begin{figure*}
\flushleft\leftskip-9em{\includegraphics[width=2.5in]{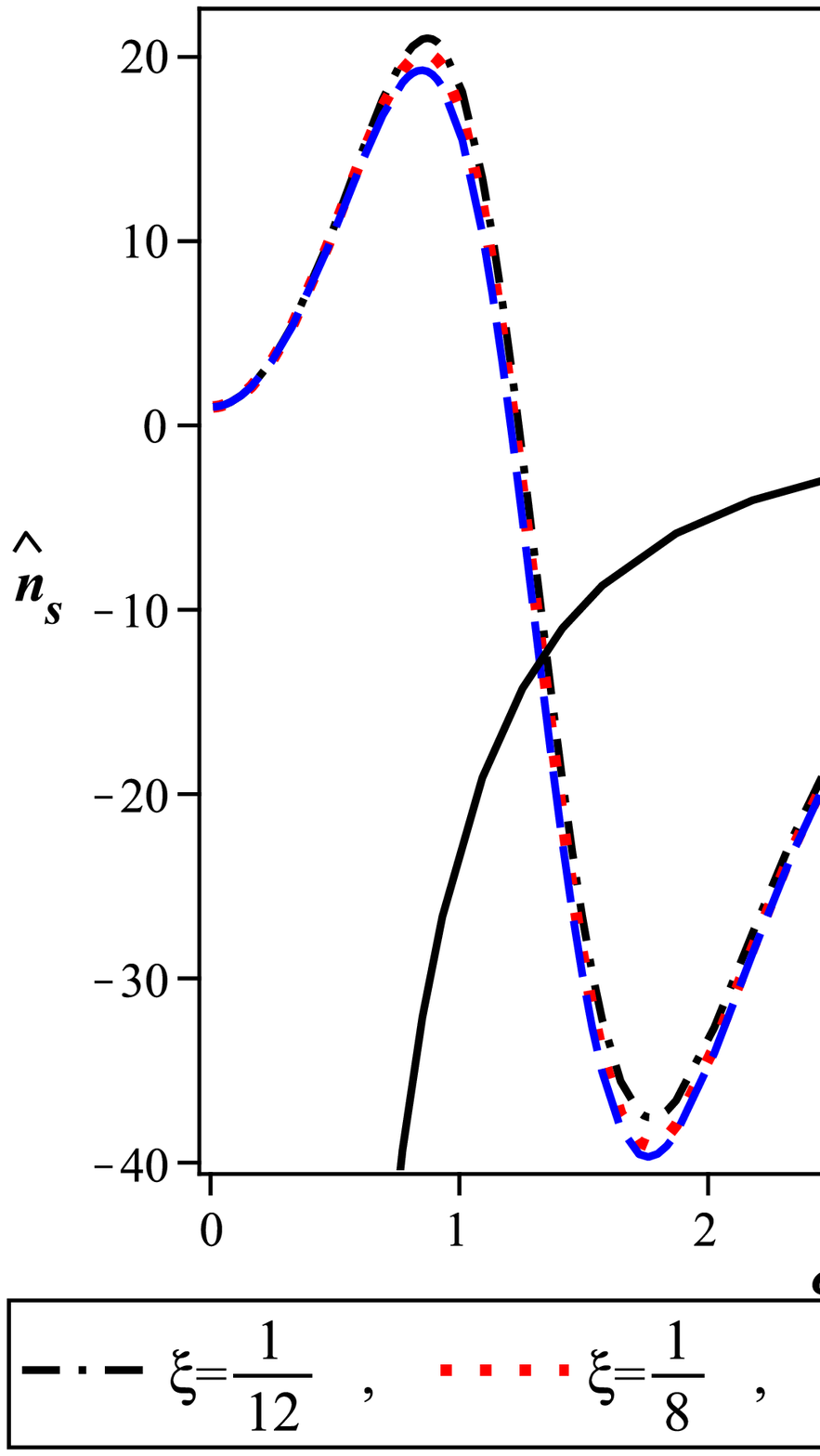}}\hspace{5cm}
{\includegraphics[width=2.5in]{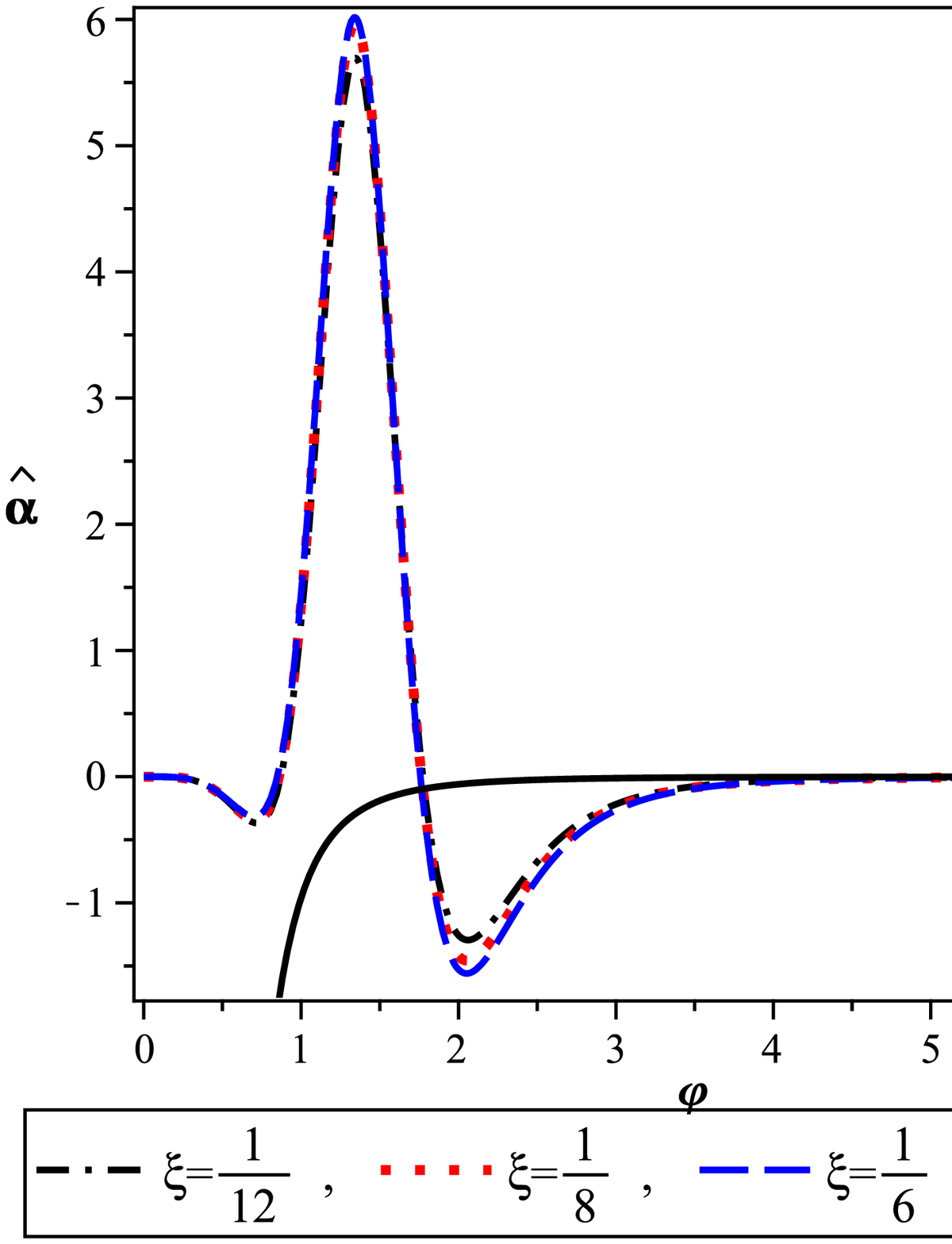}} \caption{\label{fig:18}The
evolution of the scalar spectral index (left panel) and the running
of the spectral index (right panel) versus the scalar field with a
quartic potential in Einstein frame. In the large and small scalar
field regime, the scalar spectral index and its running decrease by
reduction of the scalar field (as the 4D case).}
\end{figure*}

\begin{figure}
\flushleft\leftskip-9.5em{\includegraphics[width=2.5in]{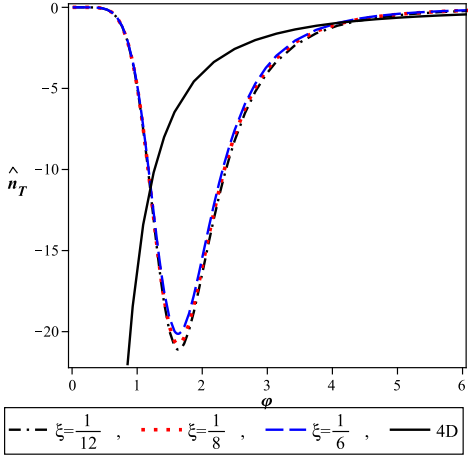}}\hspace{2cm}
\caption{\label{fig:19}The evolution of the tensor spectral index
versus the scalar field with a quartic potential in Einstein frame.
In the large scalar field regime, the tensor spectral index
decreases by reduction of the scalar field (as the 4D case).}
\end{figure}

\begin{figure}
\flushleft\leftskip-8.5em{\includegraphics[width=2.5in]{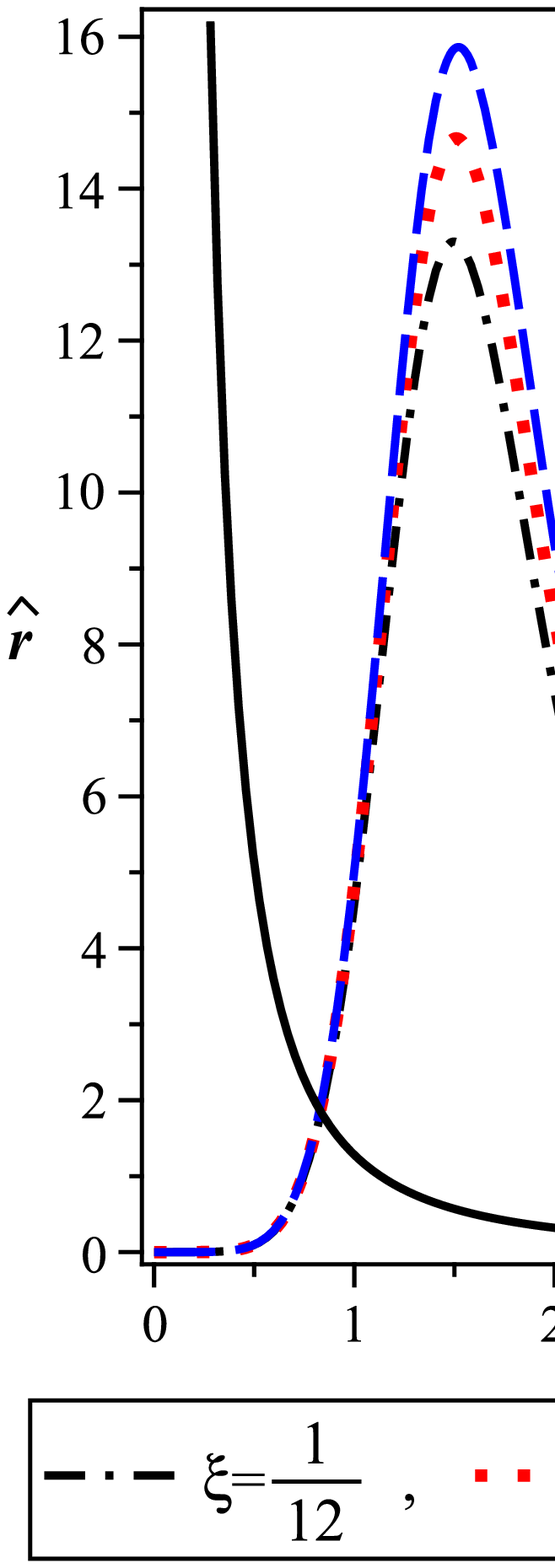}}\hspace{3cm}
\caption{\label{fig:20}The evolution of the tensor to scalar ratio
versus the scalar field with a quartic potential in Einstein frame.
The behavior of $\hat{r}$ in the large scalar field regime is
similar to the 4D behavior.}
\end{figure}

In figure~\ref{fig:17} we have depicted the behavior of
$\hat{{\cal{B}}}$ and $\hat{\eta}$ versus the scalar field. The
general behavior of $\hat{{\cal{B}}}$ and $\hat{\eta}$ is similar to
the behavior of $\hat{{\cal{A}}}$ and $\hat{\epsilon}$. In spite of
the effect of the correctional factor, the behavior of the slow-roll
parameter in the large scalar field regime is similar to the
behavior of the corresponding parameter in 4D case. In other words,
in large scalar field regime in this case, the braneworld nature of
the model can be neglected approximately. As $\xi$ becomes larger,
$\eta$ in larger region of the scalar field has the 4D behavior.
Also, in some values of the scalar field where $\hat{{\cal{B}}}$ is
negative, $\hat{\eta}$ has the negative values. Both
$\hat{\epsilon}$ and $\hat{\eta}$ can attain the unit value. So, the
graceful exit from the inflationary phase in this model is
guaranteed.
\begin{table*}
\caption{\label{tab:table4}The values of some inflation parameters
with a quartic potential in Einstein frame at the time that physical
scales crossed the horizon.}
\begin{ruledtabular}
\begin{tabular}{ccccc}
$\xi$&$\hat{n}_{s}$&$\hat{r}$&$\hat{\alpha}$ \\ \hline\\
$\frac{1}{12}$& $0.9999999985$ &$1.604411988\times10^{-20}$&-1.397561175$\times 10^{-162}$\\\\
$\frac{1}{8}$& $0.9999999995$ &$2.695140808\times10^{-19}$ & -7.310536449$\times 10^{-163}$ \\\\
$\frac{1}{6}$& $0.9999999995$ &$2.726926460\times10^{-19}$& -4.325996087$\times 10^{-163}$\\
\end{tabular}
\end{ruledtabular}
\end{table*}

In the left panel of figure~\ref{fig:18}, we have shown the
evolution of the scalar spectral index versus the scalar field. As
figure shows, in the large and small scalar field regimes,
$\hat{n}_{s}$ decreases by reduction of the values of the scalar
field similar to what it does in 4D case. In the intermediate regime
of the scalar field, this quantity deviates from the 4D behavior and
increases as the scalar field decreases. Figure~\ref{fig:19} shows
the evolution of the tensor spectral index versus the scalar field.
In the large scalar field regime, $\hat{n}_{T}$ evolves similar to
the evolution of the corresponding parameter in 4D model and
decreases by reduction of the scalar filed. But in the small scalar
field regime, it increases as the scalar field decreases. We note
that as $\xi$ increases, the 4D behavior of both scalar and tensor
spectral indices last in wider domain of the scalar field values.

In the right panel of figure~\ref{fig:18}, we have depicted the
evolution of the running of the scalar spectral index versus the
scalar field. Similar to other considered parameters in this
subsection, $\hat{\alpha}$ has the 4D behavior in the large scalar
field regime. For larger $\xi$, this 4D behavior lasts in larger
domain of the scalar field. By more reduction of the scalar field,
$\hat{\alpha}$ reaches a maximum and then deviates from the 4D
behavior. The last parameter which we consider is the tensor to
scalar ratio, $\hat{r}$ (figure~\ref{fig:20}). As the scalar field
decreases, $\hat{r}$ increases similar to what it does in 4D. The
increment of $r$ stopes at a maximum value which for larger $\xi$ is
smaller and take places in smaller values of the scalar field (this
means as $\xi$ increases, the 4D behavior of $\hat{r}$ lasts in
wider domain of the scalar field values). After that, it deviates
from 4D behavior and decreases by reduction of the scalar field.

Now we calculate some inflation parameters with a quartic potential
at the time of horizon crossing.  The Friedmann equation and the
number of e-folds are given by equations (156) and (157), but here
the potential is a quartic potential. The solution of integral (157)
with a quartic potential is presented in appendix {\bf D} (by
assuming $\hat{\varphi}_{hc}\gg\hat{\varphi}_{f}$). By finding
$\hat{\varphi}_{hc}$ from that solution, we obtain the value of the
scalar spectral index and the tensor to scalar ratio at the time of
horizon crossing. The results for three values of $\xi$ are shown in
table~\ref{tab:table4}.

With a quartic potential in Einstein frame, the value of
$\hat{n}_{s}$ at the time of horizon crossing is nearly scale
invariant and red-tilted. Also, the tensor to scalar ratio at the
time of horizon crossing decreases by reduction of $\xi$ and the
running of the spectral index increases by reduction of $\xi$.

\section{A special case}

The curvature perturbation in Einstein frame  $\hat{\zeta}$, remains
constant on large scales, but only so long as the condition
$\frac{\delta \hat{\varphi}}{(d\hat{\varphi})/(d\hat{t})}
=\frac{\delta\hat{\rho}_{eff}}{(d\hat{\rho}_{eff})/(d\hat{t})}$ (in
addition to the condition
$\frac{\delta\hat{p}_{eff}}{(d\hat{p}_{eff})/(d\hat{t})}
=\frac{\delta\hat{\rho}_{eff}}{(d\hat{\rho}_{eff})/(d\hat{t})}$) is
satisfied. This means that in this situation, the perturbations are
adiabatic \cite{Wan00}. In the warped DGP model and within the
slow-roll approximation, this condition is satisfied only when we
neglect the contribution of the dark radiation term in our analysis.
If we work in the large field regime, ($\hat{\varphi}\gg
\kappa_{4}^{-1}$ and $\hat{f}(\hat{\varphi})\gg \kappa_{4}^{-2}$)\,,
so this term is really negligible.\\

One can find the curvature perturbation on uniform density
hypersurfaces in terms of the scalar field fluctuations on spatially
flat hypersurfaces as follows
\begin{equation}
\hat{\zeta}=\frac{\hat{H}\delta
\hat{\varphi}}{d\hat{\varphi}/d\hat{t}}\,\,.
\end{equation}
Also the field fluctuations at Hubble crossing
($\hat{k}=\hat{a}\hat{H}$) in the slow-roll limit are given by
\cite{Lyt09,Maa00,Wan00}
\begin{equation}
<\delta
\hat{\varphi}^{2}>\simeq\bigg(\frac{\hat{H}}{2\pi}\bigg)^{2}\,.
\end{equation}
The scalar curvature perturbation amplitude of a given mode when
re-entering the Hubble radius is given by
\begin{equation}
\hat{A}_{s}^{2}=\frac{4<\hat{\zeta}^{2}>}{25}\,.
\end{equation}
So, in the slow-roll approximation, we find
\begin{widetext}
\begin{equation}
\hat{A}_{s}^{2}=\frac{9}{25\pi^{2}}\frac{\hat{V}^{3}}{(d\hat{V}/d\hat{\varphi})^{2}}\Bigg[\frac{\kappa_{4}^{2}}{3}
+\frac{1}{\hat{V}}\Big(\frac{\kappa_{4}^{2}}{3}\hat{\lambda}+\frac{2\kappa_{4}^{4}}{\kappa_{5}^{4}}
\pm\frac{2\kappa_{4}^{2}}{\kappa_{5}^{2}}\,\sqrt{\frac{\kappa_{4}^{4}}{\kappa_{5}^{4}}+
\frac{\kappa_{4}^{2}}{3}\hat{V}+
\frac{\kappa_{4}^{2}}{3}\hat{\lambda}-\frac{\kappa_{5}^{4}}{36}\hat{\lambda}^{2}
}\Big)\Bigg]^{3}\Bigg|_{\hat{k}=\hat{a}\hat{H}}\,.
\end{equation}
\end{widetext}
The scale-dependence of the perturbations is described
by the spectral index as
\begin{equation}
\hat{n}_{s}-1=\frac{d \ln \hat{A}_{s}^{2}}{d \ln \hat{k}}\,.
\end{equation}
The interval in wave number is related to the number of e-folds by
the relation $d \ln \hat{k}(\hat{\varphi})=d \hat{N}(\varphi)$, so
we obtain
\begin{widetext}
\begin{equation}
\hat{n}_{s}-1=-\Bigg(\frac{d\hat{V}/d\hat{\varphi}}{3\hat{V}}\Bigg)\Bigg[\frac{\kappa_{4}^{2}}{3}
+\frac{1}{\hat{V}}\Big(\frac{\kappa_{4}^{2}}{3}\hat{\lambda}+\frac{2\kappa_{4}^{4}}{\kappa_{5}^{4}}
\pm\frac{2\kappa_{4}^{2}}{\kappa_{5}^{2}}\,\sqrt{\frac{\kappa_{4}^{4}}{\kappa_{5}^{4}}+
\frac{\kappa_{4}^{2}}{3}\hat{V}+
\frac{\kappa_{4}^{2}}{3}\hat{\lambda}-\frac{\kappa_{5}^{4}}{36}\hat{\lambda}^{2}
}\Big)\Bigg]^{-1}\Upsilon\,,
\end{equation}
\end{widetext}
where the parameter $\Upsilon$ is defined as
\begin{eqnarray}
\Upsilon=-2\frac{d^{2}\hat{V}/d\hat{\varphi}^{2}}{d\hat{V}/d\hat{\varphi}}
+{\kappa_{{4}}}^{2}\frac{d\hat{V}/d\hat{\varphi}}{\hat{H}^{2}}\hspace{1.5cm}\nonumber\\
\times\bigg(1\pm\frac {{\kappa_{{4}}}^{2}}{{\kappa_{{5}}}^{2}} \frac
{1}{\sqrt {{\frac
{{\kappa_{{4}}}^{4}}{{\kappa_{{5}}}^{4}}}+\frac{{\kappa_{{4}}}^
{2}}{3}\hat{V}+\frac{{\kappa_{{4}}}^
{2}}{3}\hat{\lambda}-\frac{\kappa_{5}^{4}}{36}\hat{\lambda}^{2}}}\bigg)\,.
\end{eqnarray}
In terms of the slow-roll parameters, the spectral index becomes
\begin{equation}
\hat{n}_{s}-1=-6\hat{\epsilon}+2\hat{\eta}\,.
\end{equation}\\\\

The tensor perturbations amplitude of a given mode when leaving the
Hubble radius are given by
\begin{equation}
{\hat{A}_{T}}^{2}=\frac{4\kappa_{4}^{2}}{25\pi}\hat{H}^{2}\Bigg|_{\hat{k}=\hat{a}\hat{H}}\,\,.
\end{equation}
Therefore, in this warped DGP scenario and within the slow-roll
approximation in Einstein frame we find
\begin{widetext}
\begin{equation}
\hat{A}_{T}^{2}=\frac{4\kappa_{4}^{2}}{25\pi}\hat{V}\Bigg[\frac{\kappa_{4}^{2}}{3}
+\frac{1}{\hat{V}}\Big(\frac{\kappa_{4}^{2}}{3}\hat{\lambda}+\frac{2\kappa_{4}^{4}}{\kappa_{5}^{4}}
\pm\frac{2\kappa_{4}^{2}}{\kappa_{5}^{2}}\,\sqrt{\frac{\kappa_{4}^{4}}{\kappa_{5}^{4}}+
\frac{\kappa_{4}^{2}}{3}\hat{V}+
\frac{\kappa_{4}^{2}}{3}\hat{\lambda}-\frac{\kappa_{5}^{4}}{36}\hat{\lambda}^{2}
}\Big)\Bigg]\Bigg|_{\hat{k}=\hat{a}\hat{H}}\,\,.
\end{equation}
\end{widetext}

The tensor spectral index that is given by
\begin{equation}
\hat{n}_{T}=\frac{d \ln \hat{A}_{T}^{2}}{d \ln \hat{k}}\,,
\end{equation}
in our framework takes the following form
\begin{widetext}
\begin{equation}
\hat{n}_{T}=-\Bigg(\frac{d\hat{V}/d\hat{\varphi}}{3\hat{V}}\Bigg)\Bigg[\frac{\kappa_{4}^{2}}{3}
+\frac{1}{\hat{V}}\Big(\frac{\kappa_{4}^{2}}{3}\hat{\lambda}+\frac{2\kappa_{4}^{4}}{\kappa_{5}^{4}}
\pm\frac{2\kappa_{4}^{2}}{\kappa_{5}^{2}}\,\sqrt{\frac{\kappa_{4}^{4}}{\kappa_{5}^{4}}+
\frac{\kappa_{4}^{2}}{3}\hat{V}+
\frac{\kappa_{4}^{2}}{3}\hat{\lambda}-\frac{\kappa_{5}^{4}}{36}\hat{\lambda}^{2}
}\Big)\Bigg]^{-1}\Sigma\,,
\end{equation}
\end{widetext}
where $\Sigma$ is defined as
\begin{equation}
\Sigma=\kappa_{4}^{2}\frac{d\hat{V}/d\hat{\varphi}}{3\hat{H}^{2}}\left(
1\pm\frac{\kappa_{4}^{2}}{\kappa_{5}^{2}}{\frac {1}{\sqrt
{\frac{\kappa_{4}^{4}}{\kappa_{5}^{4}}+
\frac{\kappa_{4}^{2}}{3}\hat{V}+
\frac{\kappa_{4}^{2}}{3}\hat{\lambda}-\frac{\kappa_{5}^{4}}{36}\hat{\lambda}^{2}}}}\right).
\end{equation}
In terms of the slow-roll parameter $\hat{\epsilon}$, the tensor
(gravitational wave) perturbation finds the following expression
\begin{equation}
\hat{n}_{T}=-2\hat{\epsilon}\,.
\end{equation}

The ratio between the amplitudes of tensor and scalar perturbations
is given by
\begin{eqnarray}
\hat{r}=\frac{\hat{A}_{T}^{2}}{\hat{A}_{S}^{2}}\simeq \frac{4\pi
\kappa_{4}^{2}(d\hat{V}/d\hat{\varphi})^{2}}{9\hat{H}^{4}}\hspace{4.3cm}\nonumber\\
=\left(\frac{8\pi}{1
\pm\frac{\kappa_{4}^{2}}{\kappa_{5}^{2}}\bigg(\frac{\kappa_{4}^{4}}{\kappa_{5}^{4}}+
\frac{\kappa_{4}^{2}}{3}\hat{V}+
\frac{\kappa_{4}^{2}}{3}\hat{\lambda}-\frac{\kappa_{5}^{4}}{36}\hat{\lambda}^{2}
\,\,\bigg)^{-\frac{1}{2}} }\right)\,\hat{\epsilon}.\hspace{1cm}
\end{eqnarray}
So, the standard consistency condition between this ratio (i.e, the
relative amplitude of the two spectra) and the slow-roll parameter
$\hat{\epsilon}$ is modified by the factor in the parenthesis.\\

\subsection{large scalar field regime}

In the large scalar field regime, a quartic potential in Einstein
frame tends to a constant (see equation (147)). So, the brane
affects the standard form of the slow-roll parameters with a
constant factor. In the large scalar field regime, since the
denominator of the terms including $\hat{\lambda}$ and $\hat{a}$ are
negligible, so from equations (89), (95) and (96) we obtain the
slow-roll parameters in the large field limit as follows
\begin{eqnarray}
\hat{\epsilon}=\frac{4}{3\xi^{2}(1+1/(6\xi))}\left(\frac{1}{\kappa_{4}\varphi}\right)^{4}\hspace{4cm}\nonumber\\
\times\left[\frac{1
-\frac{\kappa_{4}^{2}}{\kappa_{5}^{2}}\bigg(\sqrt
{\frac{\kappa_{4}^{4}}{\kappa_{5}^{4}}+
\frac{b}{12\kappa_{4}^{2}\xi^{2}}}\,\,\bigg)^{-\frac{1}{2}}
}{\Bigg(1
+\frac{4\kappa_{4}^{4}\xi^{2}}{b}\bigg(\frac{6\kappa_{4}^{2}}{\kappa_{5}^{4}}
-\frac{6}{\kappa_{5}^{2}}\,\sqrt{\frac{\kappa_{4}^{4}}{\kappa_{5}^{4}}+
\frac{b}{12\kappa_{4}^{2}\xi^{2}}}\bigg)\Bigg)^{2}}\right]\,,\hspace{1cm}
\end{eqnarray}
and
\begin{eqnarray}
\eta=-\frac{4}{3\xi(1+1/(6\xi))}\left(\frac{1}{\kappa_{4}\varphi}\right)^{2}\hspace{3.6cm}\nonumber\\
\times\left[1
+\frac{4\kappa_{4}^{4}\xi^{2}}{b}\Big(\frac{6\kappa_{4}^{2}}{\kappa_{5}^{4}}
-\frac{6}{\kappa_{5}^{2}}\,\sqrt{\frac{\kappa_{4}^{4}}{\kappa_{5}^{4}}+
\frac{b}{12\kappa_{4}^{2}\xi^{2}}}\Big)\right]^{-1}\,.\hspace{1cm}
\end{eqnarray}

In order to find the values of $\hat{n}_{s}$ and $\hat{r}$ at the
time of horizon crossing, we should solve the integral (99) in the
large scalar field regime (where in this regime a quartic potential
in Einstein frame tends to a constant). The solution of the integral
is
\begin{eqnarray}
\hat{N}=\frac{9\kappa_{4}^{2}}{4}\bigg(\xi+\frac{1}{6}\bigg)\bigg(\varphi_{hc}^{2}-\varphi_{f}^{2}\bigg)\hspace{3.5cm}\nonumber\\
\times\Bigg[1
+\frac{4\kappa_{4}^{4}\xi^{2}}{b}\Big(\frac{6\kappa_{4}^{2}}{\kappa_{5}^{4}}
-\frac{6}{\kappa_{5}^{2}}\,\sqrt{\frac{\kappa_{4}^{4}}{\kappa_{5}^{4}}+
\frac{b}{12\kappa_{4}^{2}\xi^{2}}}\Big)\Bigg]\,.\hspace{0.7cm}
\end{eqnarray}
If we assume $\varphi_{hc}\gg\varphi_{f}^{2}$, then we find the
value of $\varphi_{hc}$ as
\begin{eqnarray}
\varphi=\frac{2\sqrt{\hat{N}}}{3\kappa_{4}}\left(\xi+\frac{1}{6}\right)^{-\frac{1}{2}}\hspace{4.7cm}\nonumber\\
\times\Bigg[1
+\frac{4\kappa_{4}^{4}\xi^{2}}{b}\Big(\frac{6\kappa_{4}^{2}}{\kappa_{5}^{4}}
-\frac{6}{\kappa_{5}^{2}}\,\sqrt{\frac{\kappa_{4}^{4}}{\kappa_{5}^{4}}+
\frac{b}{12\kappa_{4}^{2}\xi^{2}}}\Big)\Bigg]^{-\frac{1}{2}}\,.\hspace{0.7cm}
\end{eqnarray}
Now, $\hat{\epsilon}$ and $\hat{\eta}$ are defined as
\begin{equation}
\hat{\epsilon}=\frac{27(\xi+1/6)}{4\hat{N}^{2}}\Bigg[1
-\frac{\kappa_{4}^{2}}{\kappa_{5}^{2}}\bigg(\sqrt
{\frac{\kappa_{4}^{4}}{\kappa_{5}^{4}}+
\frac{b}{12\kappa_{4}^{2}\xi^{2}}}\,\,\bigg)^{-\frac{1}{2}}
\Bigg]\,,
\end{equation}
and
\begin{equation}
\hat{\eta}=-\frac{3}{\hat{N}\kappa_{4}^{2}}.
\end{equation}

\begin{figure}
{\includegraphics[width=2.5in]{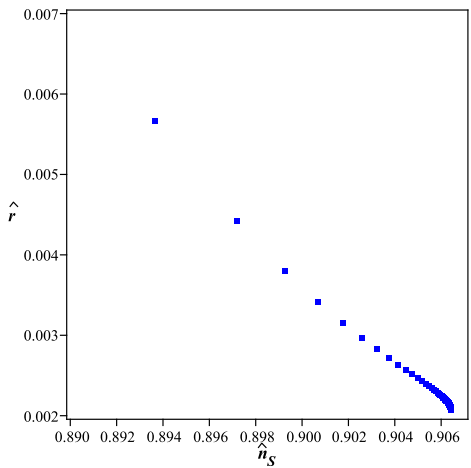}}\hspace{2cm}
\caption{\label{fig:21}The spectral index and the tensor to scalar
ratio for various values.}
\end{figure}
\begin{figure*}
{\includegraphics[width=2.5in]{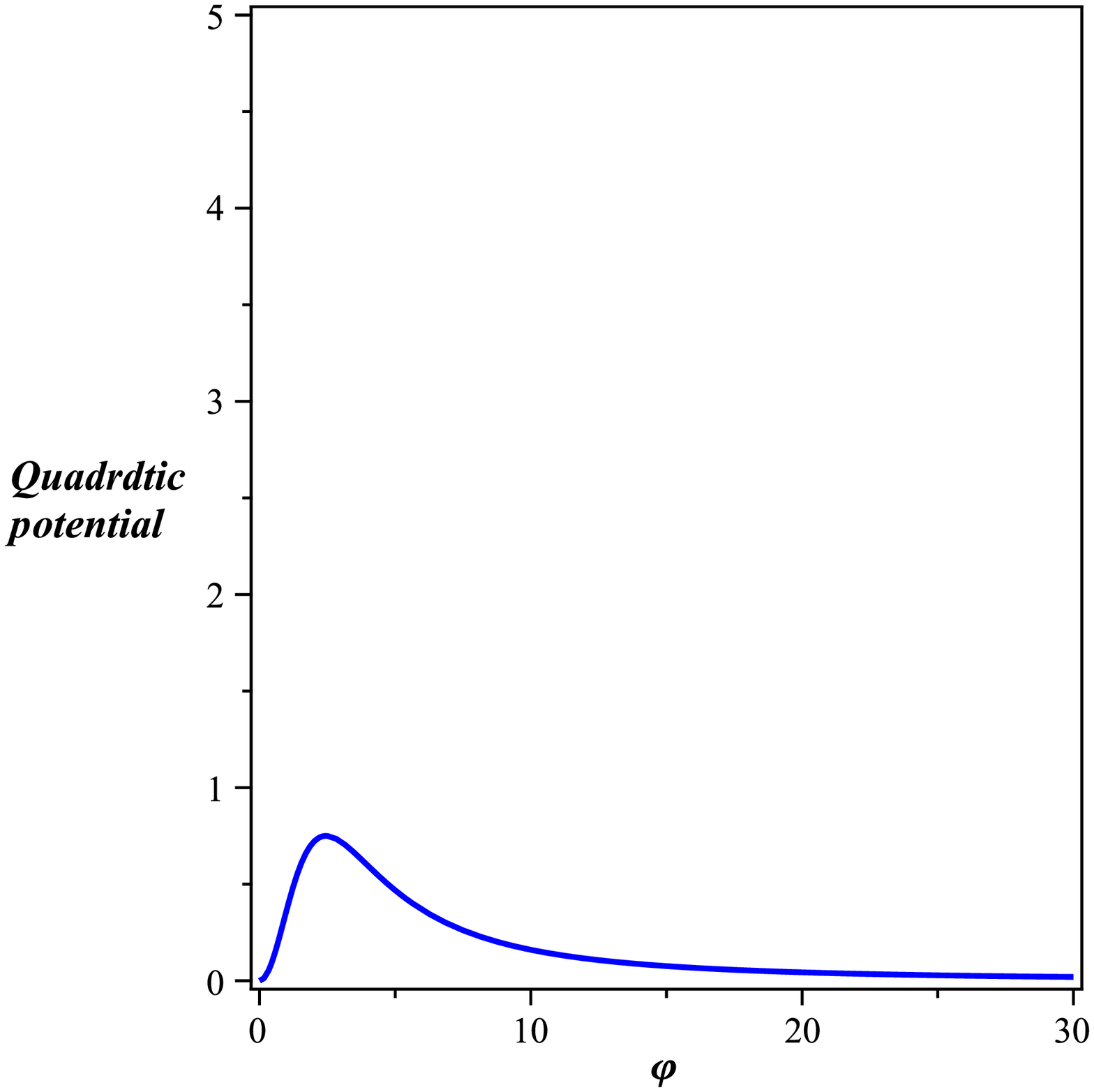}}\hspace{4cm}
{\includegraphics[width=2.5in]{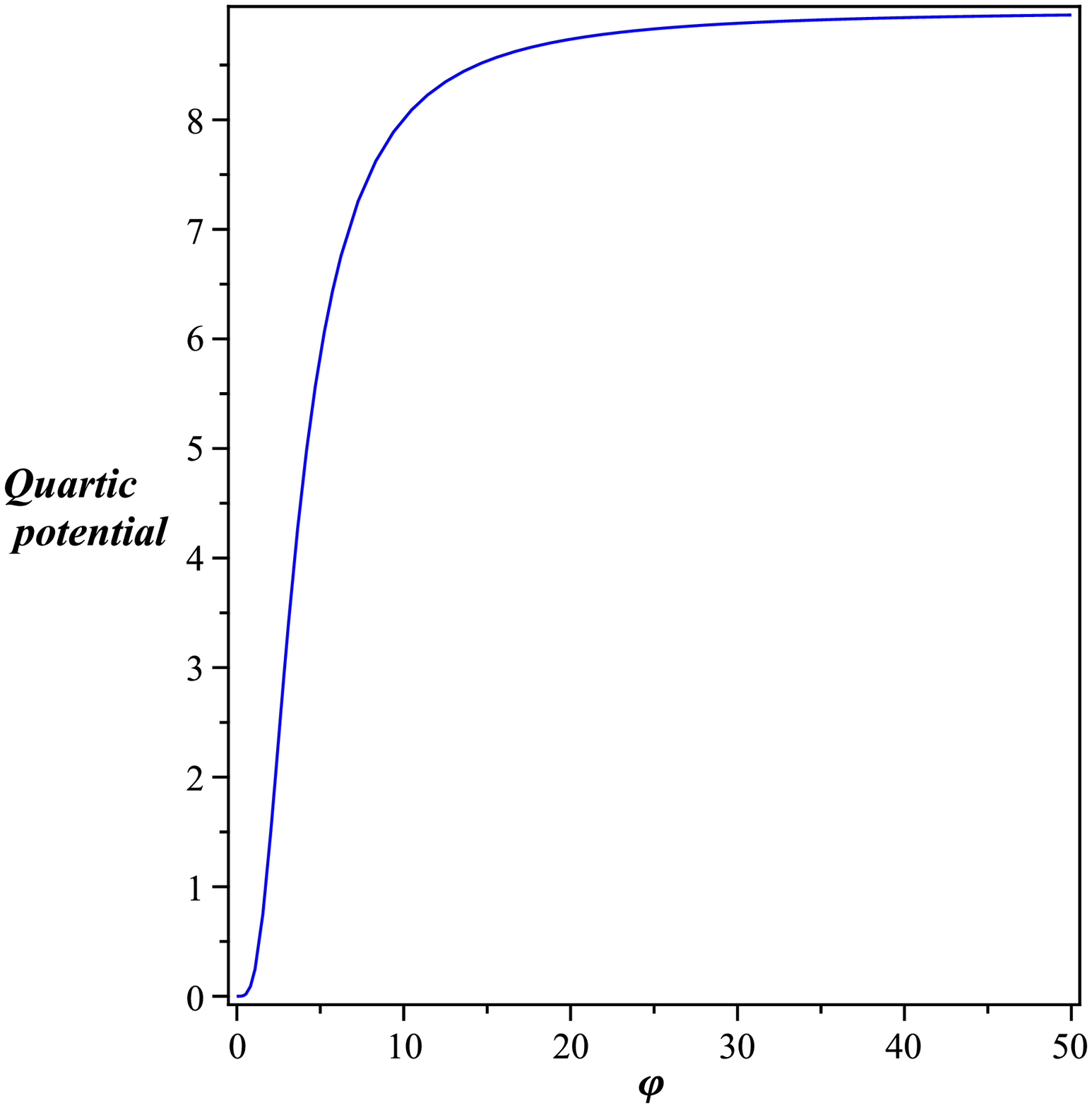}} \caption{\label{fig:22}The
quadratic and quartic potentials versus the scalar field in Einstein
frame. In this frame quadratic potential has a maximum and inflation
can occur just for those values of the scalar field located in the
left side of the maximum.}
\end{figure*}

From equations (139) and (146) and by using equations (177) and
(178) we find the scalar spectral index and the tensor to scalar
ratio at $\varphi=\varphi_{hc}$. In figure~\ref{fig:21} we have
depicted the scalar spectral index and tensor to scalar ratio in a
plot for various values of $\xi$ (we started with $\xi=\frac{1}{10}$
corresponding to the first point of the left hand side and then in
each step, we increased the value of $\xi$ by $\frac{1}{10}$). This
figure shows that as $\xi$ increases, the scalar spectral index
becomes larger but the tensor to scalar ratio gets smaller. By
increasing $\xi$, $\hat{n}_{s}$ and $\hat{r}$ tend to $0.906$ and
$0.002$ respectively (see a similar treatment for the non-minimal
Higgs boson as the inflaton in Einstein frame in \cite{Par08}).

It should be noticed that we don't consider the case with a
quadratic potential here, because this potential in the large scalar
field regime tends to zero and there is no inflation for the model
in this regime. It is due to the behavior of the quadratic potential
in Einstein frame (see figure~\ref{fig:22}). In Einstein frame,
there is a maximum for a quadratic potential that the slow-roll
conditions cannot be satisfied beyond it. But for a quartic
potential in Einstein frame the situation is different. In the large
scalar field regime, the quartic potential tends to a constant and
of course the slow-roll conditions can be satisfied.

\section{Conclusion}

In this paper we have studied the cosmological inflation on the
warped DGP braneworld, where a scalar field is non-minimally coupled
to the induced gravity term on the brane. We considered the warped
DGP setup since this braneworld scenario contains both UV and IR
modifications of the general relativity simultaneously. We have
studied the inflationary dynamics on the brane both in Jordan and
Einstein frame. We have calculated the inflation parameters and
perturbations in these two frames with details. In Jordan frame, the
brane world nature of the setup and the effects of the non-minimal
coupling between the scalar field and induced gravity on the brane
is manifest through the existence of some correctional factors in
slow-roll parameters. In Einstein frame, the effect of the
non-minimal coupling is implicit in the field equations and can be
manifested through the conformal transformation between two frames.

\begin{table*}
\caption{\label{tab:table5}Am analogy between Einstein an Jordan
frame.}
\begin{ruledtabular}
\begin{tabular}{ccccc}
\hspace{4cm}Jordan frame\hspace{6.8cm}Einstein frame \hspace{0.5cm} \\ \hline\\
\hspace{2.7cm}$\xi=\frac{1}{12}\hspace{1.5cm}\xi=\frac{1}{12}\hspace{1.4cm}\xi=\frac{1}{12}$
\hspace{2.8cm}$\xi=\frac{1}{12}\hspace{1.5cm}\xi=\frac{1}{12}\hspace{1.5cm}\xi=\frac{1}{12}$\\\\ \hline\\
\hspace{1.6cm}$n_{s}$\hspace{0.6cm} 0.9667051476 \quad 0.9653928105
\quad 0.9665654212\hspace{2cm}
0.9999999997 \quad 1.000000001 \quad 0.9999999991\\
\hspace{2.7cm}red-tilted \hspace{0.8cm} red-tilted \hspace{0.7cm} red-tilted \hspace{2.4cm} red-tilted \hspace{0.7cm} blue-tilted \hspace{0.7cm} red-tilted\\\\
\hspace{-0.4cm}$V\propto \varphi^{2}$\hspace{0.6cm}$r$
\hspace{0.61cm} 0.1575567389 \quad 0.2094435340 \quad
0.3137965930\hspace{2.5cm} $\sim 10^{-68}$
\hspace{0.8cm} $\sim 10^{-68}$\hspace{0.9cm} $\sim 10^{-69}$\\\\
\hspace{1.5cm}$\alpha$\hspace{0.95cm}$\sim
-10^{-51}$\hspace{0.8cm}$\sim -10^{-51}$\hspace{1.cm}$\sim
-10^{-51}$
\hspace{2.5cm}$\sim -10^{-216}$\hspace{0.8cm}$\sim -10^{-217}$\hspace{0.8cm}$\sim -10^{-217}$\\\\
\hline\\
\hspace{1.6cm}$n_{s}$\hspace{0.6cm} 1.000000000 \quad 0.9999999999
\quad 0.9999999996\hspace{2cm}
0.9999999985 \quad 0.9999999995 \quad 0.9999999995\\
\hspace{2.2cm}scale-invariant \hspace{0.5cm} red-tilted \hspace{0.7cm} red-tilted \hspace{2.6cm} red-tilted \hspace{0.7cm} red-tilted \hspace{0.7cm} red-tilted\\\\
\hspace{-0.35cm}$V\propto \varphi^{4}$\hspace{0.5cm}$r$
\hspace{1.1cm}$\sim 10^{-21}$\hspace{1.15cm}$\sim
10^{-20}$\hspace{1.15cm}$\sim 10^{-20}$\hspace{2.65cm} $\sim
10^{-19}$
\hspace{0.9cm} $\sim 10^{-21}$\hspace{1.2cm} $\sim 10^{-22}$\\\\
\hspace{1.6cm}$\alpha$\hspace{1.2cm}$\sim
-10^{-38}$\hspace{0.8cm}$\sim -10^{-38}$\hspace{1.cm}$\sim
-10^{-38}$ \hspace{2.4cm}$\sim -10^{-162}$\hspace{0.8cm}$\sim
-10^{-163}$\hspace{0.8cm}$\sim -10^{-163}$
\end{tabular}
\end{ruledtabular}
\end{table*}

The perturbations in these two frames are studied with details. The
adiabatic perturbations are generated if the inflaton field is the
only field in inflation period. But, if there is more than one
scalar field in a model or a scalar field interacts with other
fields such as the induced gravity on the brane, the isocurvature
perturbations are generated. In our case and in Jordan frame, the
presence of the non-minimal coupling between the inflaton field and
induced gravity on the brane and also the presence of the non-local
effects through the projection of the Weyl tensor on the brane lead
to a non-vanishing $\dot{\zeta}$ which affects the primordial
spectrum of perturbations. However, in Einstein frame (despite
implicit presence of the non-minimal coupling), isocurvature
perturbations are generated due to the presence of the non-local
effects through the projection of the Weyl tensor on the brane. If
we neglect this term in Friedmann equation, the perturbations become
adiabatic since neglecting the non-local effect leads to the
condition $\frac{\delta \hat{\varphi}}{(d\hat{\varphi})/(d\hat{t})}
=\frac{\delta\hat{\rho}_{eff}}{(d\hat{\rho}_{eff})/(d\hat{t})}$ to
be satisfied.

By adopting two types of potential ($V=\frac{b}{2m}\varphi^{2m}$;
$m=1,2$), we have performed numerical analysis of the model
parameters space in each case, the results of which are shown in
numerous tables and figures. We note that all of our numerical
analysis are done for normal branch of this DGP-inspired model which
is essentially ghost-free. In Jordan frame, both for quadratic and
quartic potential, all considered parameters ($\epsilon$, $\eta$,
$n_{s}$, $n_{T}$, $\alpha$ and $r$) in the large scalar field regime
evolve similar to what they do in 4D. In this frame, as $\xi$
becomes larger, the 4D behavior of these parameters lasts in a wider
domain of the scalar field values. By more reduction of the scalar
field, the evolution of the parameters deviate from the standard 4D
behavior. It seems that this deviation from the standard 4D behavior
is due to the presence of the tension term in the correctional
factors. Of course, with a quartic potential, the parameters
experience another standard 4D behavior in the small scalar field
regime. But, their values is very different from the values of the
corresponding parameters in 4D case.

In Einstein frame, the situation for two types of potentials is much
different. With a quartic potential in Einstein frame, the
considered parameters in the large scalar field regime have standard
4D behavior (similar to the quartic potential in Jordan frame). In
the small scalar field limit, the evolution of the parameters
deviate from standard 4D behavior. In this case, as $\xi$ increases,
the parameters mimic the standard 4D behavior in a relatively wider
domain of the scalar field values. Due to the shape of a quadratic
potential in Einstein frame, it is impossible to have inflation in
the large scalar field regime. But, when the scalar field is
confined to an intermediate regime, the slow-roll conditions can be
satisfied and the inflationary phase can be occurred. We note that
with a quadratic potential in Jordan frame, the inflationary phase
occurs in the large scalar field regime. In this case, the parameter
in the intermediate regime have the 4D behavior and in the large and
small scalar field regimes they deviate from the standard 4D
behavior considerably. Also, as $\xi$ increases, the 4D behavior of
all inflation parameters lasts in a wider domain of the scalar field
values. In general, with a quartic potential in both Jordan and
Einstein frame and with a quadratic potential just in Jordan frame,
by increasing of $\xi$ the 4D behavior of parameters lasts in larger
domain of the scalar field values. But, with a quadratic potential
in Einstein frame, 4D behavior lasts in a wider domain of the scalar
field for the smaller values of $\xi$.

We noticed that our analysis shows that although with a quadratic
potential in Jordan frame the slow-roll parameters are always
positive, with a quartic potential these parameters can be negative
for some values of the scalar field. Of course, in Einstein frame
both with quadratic and quatic potential, the slow-roll parameters
can get negative values.

We have also calculated some inflation parameters at the time that
physical scales had crossed the horizon. We note that for this
purpose, our analysis have been performed in the high energy limit
($\rho\gg\lambda$). Also, we have considered three values of $\xi$
in each case. The results of our analysis shows that in the warped
DGP model with a quadratic potential in Jordan frame, the scalar
perturbation is nearly scale invariant and red-tilted ($0.966\leq
n_{s}\leq 1$). In this case, the value of the tensor to scalar ratio
at the time of the horizon crossing is smaller than $0.24$ (except
for $\xi=\frac{1}{6}$ which has $r\simeq 0.31$ ). The running of the
scalar spectral index at the time of horizon crossing, is very close
to zero but it is negative as usual. So, with a quadratic potential
in Jordan frame, there is relatively good agreement between our
results and recent observation, specially for $\xi=\frac{1}{8}$ (the
result of WMAP+BAO+H$_{0}$ Mean data shows that $n_{s}=0.968 \pm
0.012$, $r<0.24(95 \% CL$ and $-0.022 \pm 0.020$). With a quartic
potential in Jordan frame, for $\xi=\frac{1}{6}$, the scalar
perturbation is quite scale invariant. But, for other $\xi$, it is
nearly scale invariant and red-tilted. With this potential, both the
running of the spectral index and the tensor to scalar ratio are
very close to zero. Then, we have found the value of $\hat{n}_{s}$,
$\hat{r}$ and $\hat{\alpha}$ at the horizon crossing time in
Einstein frame. With a quartic potential in Einstein frame, the
results are similar to the quartic potential in Jordan frame. The
scalar perturbation is nearly scale invariant and red-tilted. Also,
the running of the scalar perturbation and the tensor to scalar
ratio are very close to zero. With a quadratic potential,
$\hat{\alpha}$ and $\hat{r}$ are very close to zero too. With this
potential in Einstein frame, the scalar perturbation is nearly scale
invariant. But, for $\xi=\frac{1}{8}$, it is blue-tilted and for
other $\xi$, it is red-tilted (see table~\ref{tab:table5} which
summarizes all of these points). A careful inspection of these
results shows that by adopting a \emph{quadratic} potential and
working in \emph{Jordan frame}, the results of our analysis are more
reliable in comparison with recent observations. On the other hand,
as table~\ref{tab:table5} shows, the two frames are not equivalent
generally on physical ground. This is an important results. We
emphasize that the distinction between the various conformal frames
would be unphysical if one were dealing with conformal (Weyl)
gravity which is conformally invariant. Since general relativity is
not conformally invariant, our discussion entails the use of
compensator fields (like the dilaton) whose role is to basically
absorb the violations of conformal invariance. Inclusion of such
fields in our case helped us to address the comparative analysis of
cosmological perturbations in the Jordan and Einstein frames.

In section $8$ we have considered a special case where we have
neglected the dark radiation term in the Friedmann equation and
therefore the condition for adiabatic perturbation is satisfied. We
have considered the inflation parameters of the model in adiabatic
condition. Then we have repeated our analysis in the large scalar
field regime. Since in this regime, there was no inflationary phase
with a quadratic potential in Einstein frame, we have considered
only a quartic potential which is nearly constant in the large
scalar field regime. With this choice, the braneworld nature of the
model affects the standard form of the slow-roll parameters (and so,
other inflation parameters) with a constant factor. For the various
values of $\xi$, we have found the values of $\hat{n}_{s}$ and
$\hat{r}$ at the time of horizon crossing. The results are shown in
figure $21$. By increment of $\xi$, the scalar spectral index and
tensor to scalar ratio is saturated to $0.906$ and $0.002$
respectively.

\begin{acknowledgments}
This work has been supported financially by Research Institute for
Astronomy and Astrophysics of Maragha (RIAAM) under research project
No 1/2367.  We are very grateful to two anonymous referees for very
insightful comments and invaluable contribution in this work.
\end{acknowledgments}

\appendix

\section{Jordan Frame, m=1}
\begin{widetext}
\begin{eqnarray*}
N=-\frac{1}{4}\,{\frac {\kappa_{4}^{2} \left( 4\,\kappa_{4}^{2}\sqrt
{2}\sqrt {-\varphi_{hc}^{2} \left( -b+8\,{\xi}^{2}R-8\,\xi\,b
\right) }-\varphi_{hc}^{2}\kappa_{5}^{2}b
+8\,\varphi_{hc}^{2}\kappa_{5}^{2}{\xi}^{2}R-8\,\varphi_{hc}^{2}{
\kappa}^{2}\xi\,b \right) }{\kappa_{5}^{2} \left( \xi\,R-b \right)}}
\end{eqnarray*}
\end{widetext}

\section{Jordan Frame, m=2}
\begin{widetext}
\begin{eqnarray*}
N=\Big(1+16\xi\Big)\Big(\frac{\kappa_{4}^{4}}{\kappa_{5}^{2}\varphi_{hc}}\Big)\sqrt{\frac{b\,\varphi_{hc}^{4}
-16\,{\xi}^{2}\varphi_{hc}^{2}R+16\,\xi\,\varphi_{hc}^{4}b}{(b\,\varphi_{hc}^{2}-
16\,{\xi}^{2}R+16\,\xi\,\varphi_{hc}^{2}b)(b+16
\,\xi\,b)}}\hspace{9cm}\nonumber\\
\times\ln\Big({\frac {\varphi_{hc}\,b+16\,\xi\,\varphi_{hc}\,b
+\sqrt
{b\,\varphi_{hc}^{2}-16\,{\xi}^{2}R+16\,\xi\,\varphi_{hc}^{2}b}
\sqrt {b+16\,\xi\,b}}{\sqrt {b+16\,\xi\,b}}}\Big)\hspace{9cm}\nonumber\\
-\frac{\xi
R}{2}\Big(\frac{\kappa_{4}^{4}}{\kappa_{5}^{2}\varphi_{hc}}\Big)\sqrt{\frac{b\,\varphi_{hc}^{4}
-16\,{\xi}^{2}\varphi_{hc}^{2}R+16\,\xi\,\varphi_{hc}^{4}b}{(b\,\varphi_{hc}^{2}-16\,{\xi}^{2}R
+16\,\xi\,\varphi_{hc}^{2}b) b\,\xi^{2}\,R^{2}}}\hspace{11.3cm}\nonumber\\
\times\Bigg\{\ln\Big(-2b\,{\frac{\sqrt{b\,\xi\,R}\varphi_{hc}+16\,\sqrt
{ b\,\xi\,R}\varphi_{hc}\,\xi-16\,{\xi}^{2}R+\sqrt {\xi\,R}\sqrt
{b\,\varphi_{hc}^{2}-16\,{\xi}^{2}R+16\,\xi\,\varphi_{hc}^{2}b}
}{-\varphi_{hc}b+\sqrt{b\,\xi\,R}}}\,\,\Big)\hspace{5.5cm}\nonumber\\
-\ln\Big(2b\,{\frac{-\sqrt{b\,\xi\,R}\varphi_{hc}-16\,\sqrt{
b\,\xi\,R}\varphi_{hc}\xi-16\,{\xi}^{2}R+\sqrt {\xi\,R}\sqrt
{b\,\varphi_{hc}^{2}-16\,{\xi}^{2}R+16\,\xi\,\varphi_{hc}^{2}b}
}{\varphi_{hc}b+\sqrt {b\,\xi\,R}}}\,\,\Big)\Bigg\}\hspace{5.6cm}\nonumber\\
-\frac{1}{8}\,\kappa_{4}^{2}\varphi_{hc}^{2}-2\,\kappa_{4}^{2}\xi\,\varphi_{hc}^{2}-\frac{1}{8}\,{\frac
{\kappa_{4}^{2}\xi\,R\ln  \left( \xi\,R-b\,\varphi_{hc}^{2} \right)
}{b}}\hspace{4cm}
\end{eqnarray*}
\end{widetext}

\section{Einstein Frame, m=1}

\begin{widetext}
\begin{eqnarray*}
N=\Bigg[-\frac{1}{4}+54\kappa_{4}^{2}\xi^{4}\varphi_{hc}^{2}-\frac{3}{4}\xi
+9\xi^{2}-\frac{9}{2}\kappa_{4}^{2}\xi^{2}\varphi_{hc}^{2}
-\frac{1}{2}\kappa_{4}^{2}\xi\varphi_{hc}^{2}-18\kappa_{4}^{4}\xi^{4}\varphi_{hc}^{4}+27\kappa_{4}^{4}\xi^{5}\varphi_{hc}^{4}
-\frac{15}{4}\kappa_{4}^{4}\xi^{3}\varphi_{hc}^{4}\hspace{2.5cm}\nonumber\\
-\frac{1}{4}\kappa_{4}^{4}\xi^{2}\varphi_{hc}^{4}\Bigg]M-\Bigg[\frac{U}{4}-\frac{W}{4}-\frac{3}{2}W-12\kappa_{4}^{2}W
-\frac{1}{4}\kappa_{4}^{2}W+12\kappa_{4}^{2}U+\frac{3}{2}U
+\frac{1}{4}\kappa_{4}^{2}U\Bigg]E+\Bigg[396\sqrt{2}\frac{\kappa_{4}^{6}}{\kappa_{5}^{2}}\xi^{4}\varphi_{hc}^{5}\hspace{2.4cm}
\nonumber\\
+1188\sqrt{2}\frac{\kappa_{4}^{6}}{\kappa_{5}^{2}}\xi^{5}\varphi_{hc}^{5}
+\frac{7\sqrt{2}}{3}\frac{\kappa_{4}^2}{\kappa_{5}^{2}}\varphi_{hc}
+57\sqrt{2}\frac{\kappa_{4}^{6}}{\kappa_{5}^{2}}\xi^{3}\varphi_{hc}^{5}
+3\sqrt{2}\frac{\kappa_{4}^{6}}{\kappa_{5}^{2}}\xi^{2}\varphi_{hc}^{5}
+69\sqrt{2}\frac{\kappa_{4}^{4}}{\kappa_{5}^{2}}\xi^{2}\varphi_{hc}^{3}
+5\sqrt{2}\frac{\kappa_{4}^{4}}{\kappa_{5}^{2}}\xi\varphi_{hc}^{3}\hspace{3.1cm}\nonumber\\
+306\sqrt{2}\frac{\kappa_{4}^{4}}{\kappa_{5}^{2}}\xi^{3}\varphi_{hc}^{3}
+69\sqrt{2}\frac{\kappa_{4}^{4}}{\kappa_{5}^{2}}\xi^{2}\varphi_{hc}^{3}
+324\sqrt{2}\frac{\kappa_{4}^{4}}{\kappa_{5}^{2}}\xi^{4}\varphi_{hc}^{3}
+36\sqrt{2}\frac{\kappa_{4}^{2}}{\kappa_{5}^{2}}\xi^{2}\varphi_{hc}
+108\sqrt{2}\frac{\kappa_{4}^{2}}{\kappa_{5}^{2}}\xi^{3}\varphi_{hc}
+\frac{\sqrt{2}}{3}\frac{\kappa_{4}^{8}}{\kappa_{5}^{2}}\xi^{3}\varphi_{hc}^{7}\hspace{2.3cm}\nonumber\\
+54\sqrt{2}\frac{\kappa_{4}^{8}}{\kappa_{5}^{2}}\xi^{5}\varphi_{hc}^{7}
+180\sqrt{2}\frac{\kappa_{4}^{8}}{\kappa_{5}^{2}}\xi^{6}\varphi_{hc}^{7}
+7\sqrt{2}\frac{\kappa_{4}^{8}}{\kappa_{5}^{2}}\xi^{4}\varphi_{hc}^{7}
+216\sqrt{2}\frac{\kappa_{4}^{8}}{\kappa_{5}^{2}}\xi^{7}\varphi_{hc}^{7}
+648\sqrt{2}\frac{\kappa_{4}^{4}}{\kappa_{5}^{2}}\xi^{5}\varphi_{hc}^{3}
+19\sqrt{2}\frac{\kappa_{4}^{2}}{\kappa_{5}^{2}}\xi\varphi_{hc}\hspace{2.3cm}\nonumber\\
+1296\sqrt{2}\frac{\kappa_{4}^{6}}{\kappa_{5}^{2}}\xi^{6}\varphi_{hc}^{5}
\Bigg]Q-\Bigg[49\sqrt{2}\kappa_{4}^{2}\varphi_{hc}
(U+W)+\sqrt{2}\kappa_{4}^{4}\xi\varphi_{hc}^{3}
(U+W)+108\sqrt{2}\kappa_{4}^{4}\xi^{3}\varphi_{hc}^{3}(U+W)\hspace{3.3cm}
\nonumber\\
+18\sqrt{2}\kappa_{4}^{4}\xi^{2}\varphi_{hc}^{3}(U+W)
+216\sqrt{2}\kappa_{4}^{4}\xi^{4}\varphi_{hc}^{3}(U+W)\Bigg]Y
-\Bigg[1188\kappa_{4}^{4}\xi^{5}+1296\kappa_{4}^{4}\xi^{6}+396\kappa_{4}^{4}\xi^{4}
+57\kappa_{4}^{4}\xi^{3}\hspace{2.5cm}\nonumber\\
+3\kappa_{4}^{4}\xi^{2}
+3564\kappa_{4}^{6}\xi^{6}\varphi_{hc}^{2}+8424\kappa_{4}^{6}\xi^{7}\varphi_{hc}^{2}+7776\kappa_{4}^{6}\xi^{8}\varphi_{hc}^{2}
+738\kappa_{4}^{6}\xi^{5}\varphi_{hc}^{2}
+75\kappa_{4}^{6}\xi^{4}\varphi_{hc}^{2}+3\kappa_{4}^{6}\xi^{3}\varphi_{hc}^{2}\Bigg]DG\hspace{1.3cm}
\end{eqnarray*}
\end{widetext}

where

\begin{widetext}
\begin{eqnarray*}
U=\ln  \left( 4\,{\frac {\kappa_{4}^{2}\xi\, \left( 1-6\,\sqrt
{\kappa_{4}^{2}\xi} \varphi_{hc}\,\xi-\sqrt
{\kappa_{4}^{2}\xi}\varphi_{hc}+\sqrt {1+3\,\xi}\sqrt
{2+2\,\kappa_{4}^{
2}\xi\,\varphi_{hc}^{2}+12\,\kappa_{4}^{2}{\xi}^{2}\varphi_{hc}^{2}}
\right) }{\kappa_{4}^{2 }\xi\,\varphi_{hc}+\sqrt
{\kappa_{4}^{2}\xi}}} \right)\hspace{4.7cm}\nonumber\\
W=\ln  \left( -4\,{\frac {\kappa_{4}^{2}\xi\, \left( 1+6\,\sqrt
{\kappa_{4}^{2}\xi }\varphi_{hc}\,\xi+\sqrt
{\kappa_{4}^{2}\xi}\varphi_{hc}+\sqrt {1+3\,\xi}\sqrt
{2+2\,\kappa_{4}^
{2}\xi\,\varphi_{hc}^{2}+12\,\kappa_{4}^{2}{\xi}^{2}\varphi_{hc}^{2}}
\right) }{-\kappa_{4}^{2}\xi\,\varphi_{hc}+\sqrt
{\kappa_{4}^{2}\xi}}} \right)\hspace{4.3cm}\nonumber\\
D=\ln  \left( \sqrt {2\,\kappa_{4}^{2}\xi+12\,\kappa_{4}^{
2}{\xi}^{2}}\varphi_{hc}+\sqrt
{2+2\,\kappa_{4}^{2}\xi\,\varphi_{hc}^{2}+12\,\kappa_{4}^{2}{
\xi}^{2}\varphi_{hc}^{2}} \right)\hspace{8.6cm}\nonumber\\
M=\frac{\kappa_{4}^{3}\varphi_{hc}}{\left( 1+3\,\xi \right) \left(
6\,\xi+1 \right) \left( 1+\kappa_{4}^{2}\xi\,\varphi_{hc}^{2}
\right) ^{3} \left( {\frac
{1+\kappa_{4}^{2}\xi\,\varphi_{hc}^{2}+6\,\kappa_{4}^{2}{\xi}^{2}\varphi_{hc}^{2}}{
\left( 1+\kappa_{4}^{2}\xi\,\varphi_{hc}^{2} \right) ^{2}}} \right)
^{3/2}}\hspace{8.6cm}\nonumber\\
E=\frac{\kappa_{4}^{3}\sqrt
{2+2\,\kappa_{4}^{2}\xi\,\varphi_{hc}^{2}+12\,\kappa_{4}^{2}{\xi}^
{2}\varphi_{hc}^{2}}} {\left( 1+3\,\xi \right) ^{3/2}{\sqrt
{\kappa_{4}^ {2}\xi}} \left( 6\,\xi+1 \right) \left(
1+\kappa_{4}^{2}\xi\,\varphi_{hc}^{2} \right) ^{3} \left( {\frac
{1+\kappa_{4}^{2}\xi\,\varphi_{hc}^{2}+6\,\kappa_{4}^{
2}{\xi}^{2}\varphi_{hc}^{2}}{ \left(
1+\kappa_{4}^{2}\xi\,\varphi_{hc}^{2} \right) ^{2} }} \right)
^{3/2}}\hspace{7.4cm}\nonumber\\
Q=\frac{\kappa_{4}^{3}\sqrt
{2+2\,\kappa_{4}^{2}\xi\,\varphi_{hc}^{2}+12\,\kappa_{4}^{2}{\xi}^
{2}\varphi_{hc}^{2}} }{{\sqrt {\kappa_{4}^ {2}\xi}} \left( 6\,\xi+1
\right) ^{2} \left( 1+\kappa_{4}^{2}\xi\,\varphi_{hc}^ {2} \right)
^{4} \left( {\frac
{1+\kappa_{4}^{2}\xi\,\varphi_{hc}^{2}+6\,\kappa_{4}^{
2}{\xi}^{2}\varphi_{hc}^{2}}{ \left(
1+\kappa_{4}^{2}\xi\,\varphi_{hc}^{2} \right) ^{2} }} \right)
^{3/2}\left( 1+3\,\xi \right) \sqrt {{\frac {b\,\varphi_{hc}^{2}}{
\left( 1+\kappa_{4}^{2}\xi\,\varphi_{hc}^{2}
 \right) ^{2}}}}}\hspace{5.5cm}\nonumber\\
Y=\frac{\kappa_{4}^{3}\sqrt
{2+2\,\kappa_{4}^{2}\xi\,\varphi_{hc}^{2}+12\,\kappa_{4}^{2}{\xi}^
{2}\varphi_{hc}^{2}}} {\kappa_{5}^{2}\left( 1+3\,\xi \right)
^{3/2}{\sqrt {\kappa_{4}^ {2}\xi}} \left( 6\,\xi+1 \right)^{2}
\left( 1+\kappa_{4}^{2}\xi\,\varphi_{hc}^ {2} \right) ^{4} \left(
{\frac {1+\kappa_{4}^{2}\xi\,\varphi_{hc}^{2}+6\,\kappa_{4}^{
2}{\xi}^{2}\varphi_{hc}^{2}}{ \left(
1+\kappa_{4}^{2}\xi\,\varphi_{hc}^{2} \right) ^{2} }} \right)
^{3/2}\sqrt {{\frac {b\,\varphi_{hc}^{2}}{ \left(
1+\kappa_{4}^{2}\xi\,\varphi_{hc}^{2}
 \right) ^{2}}}}}\hspace{4.6cm}\nonumber\\
G=\frac{\kappa_{4}^{3}\sqrt
{2+2\,\kappa_{4}^{2}\xi\,\varphi_{hc}^{2}+12\,\kappa_{4}^{2}{\xi}^
{2}\varphi_{hc}^{2}}} {\left( 1+3\,\xi \right){\sqrt {\kappa_{4}^
{2}\xi}} \left( 6\,\xi+1 \right)\left(
2\,\kappa_{4}^{2}\xi+12\,\kappa_{4}^{2}{\xi}^{2} \right) ^{5/2}
\left( 1+\kappa_{4}^{2}\xi\,\varphi_{hc}^ {2} \right) ^{3} \left(
{\frac {1+\kappa_{4}^{2}\xi\,\varphi_{hc}^{2}+6\,\kappa_{4}^{
2}{\xi}^{2}\varphi_{hc}^{2}}{ \left(
1+\kappa_{4}^{2}\xi\,\varphi_{hc}^{2} \right) ^{2} }} \right)
^{3/2}}\hspace{4.7cm}
\end{eqnarray*}
\end{widetext}

\section{Einstein Frame, m=2}
\begin{widetext}
\begin{eqnarray*}
N=\frac {1}{64}\frac{\kappa_{4}^{2}}{{\left(6\,\xi+1\right)
^{5/2}\kappa_{5}^{2}\sqrt{
1+\kappa_{4}^{2}\xi\,\varphi_{hc}^{2}+6\,\kappa_{4}^{2}{\xi}^{2}\varphi_{hc}^{2}}\sqrt
{\xi} \sqrt {b}}}\Bigg[
\Big(288\,{\xi}^{2}\sqrt{b}\kappa_{5}{2}+96\,\kappa_{4}^{4}{\xi}^{2}-1152\,\kappa_{4}^{4}
{\xi}^{3}+12\,\kappa_{4}^{4}\xi\hspace{2.3cm}\nonumber\\
+\sqrt{b}\kappa_{5}{2}\Big)\Pi+\Big(576\sqrt{b}\kappa_{5}{2}\xi^{2}
+192\,\kappa_{4}^{4}\xi^{2}-2304\,\kappa_{4}^{4}{\xi}^{3}+24\,\kappa_{4}^{4}\xi
+2\sqrt{b}\kappa_{5}{2}\Big)\Theta
+\Big(576\,\kappa_{4}^{4}{\xi}^{4}\varphi_{hc}^{4}+192\,\kappa_{4}^{4}{\xi}^{3}\varphi_{hc}^{4}\hspace{1.9cm}\nonumber\\
+16\,\kappa_{4}^{4}{\xi}^{2}
\varphi_{hc}^{4}+1152\,\kappa_{4}^{2}{\xi}^{3}\varphi_{hc}^{2}+528\,\kappa_{4}^{2}{\xi}^{2}\varphi_{hc}^{2}
+56\,\kappa_{4}^{2}{\xi}\varphi_{hc}^{2}-13824{\xi}^{3}+192\xi+40
\Big)\Xi +\Big(-144\,\kappa_{4}^{4}{\xi}^{4}\varphi_{hc}^{4}\hspace{2.35cm}\nonumber\\
-48\,\kappa_{4}^{4}{\xi}^{3}\varphi_{hc}^{4}
-4\,\kappa_{4}^{4}{\xi}^{2}\varphi_{hc}^{4}-36\,\kappa_{4}^{2}{\xi}
^{2}\varphi_{hc}^{2}-6\,\kappa_{4}^{2
}{\xi}\varphi_{hc}^{2}-2\Delta-576{\xi}^{2}\Big)\Delta
 \Bigg]\hspace{6cm}
\end{eqnarray*}
\end{widetext}

where

\begin{widetext}
\begin{eqnarray*}
\Theta=\sqrt{1+\kappa_{4}^{2}\xi\,\varphi_{hc}^{2}
+6\,\kappa_{4}^{2}{\xi}^{2}\varphi_{hc}^{2}} \ln
 \left( \kappa_{4}\,\sqrt {\xi}\sqrt {6\,\xi+1}\varphi_{hc}+\sqrt {1+\kappa_{4}^{2}\xi\,\varphi_{hc}^{2}
 +6\,\kappa_{4}^{2}{\xi}^{2}\varphi_{hc}^{2}} \right)\hspace{3.63cm}\nonumber\\
\Pi=\sqrt{1+\kappa_{4}^{2}\xi\,
\varphi_{hc}^{2}+6\,\kappa_{4}^{2}{\xi}^{2}\varphi_{hc}^{2}}
\ln\left(2\right)\hspace{11cm}\nonumber\\
\Xi=\kappa_{4}^{5}{\xi}^{3/2}\varphi_{hc}\,\sqrt {6\,
\xi+1}\hspace{13cm}\nonumber\\
\Delta= \varphi_{hc}\,\kappa_{4}\,\sqrt {\xi}\sqrt {6\,\xi+1}\sqrt
{b}\kappa_{5}^{2}\hspace{12cm}
\end{eqnarray*}
\end{widetext}

\end{document}